\documentclass[tighten]{aastex6}

\usepackage{amsmath,bm}

\usepackage{graphicx}   
\usepackage{epstopdf}
\usepackage{soul}
\usepackage{hyperref}   
\usepackage{bm}
\usepackage{amsfonts}
\usepackage{natbib}
\usepackage{color}      
\usepackage{verbatim}

\newcommand{\Up}{{\bm u}^+}
\newcommand{\Um}{{\bm u}^-}
\newcommand{\Upm}{{\bm u}^\pm}
\newcommand{\Bp}{{\bm b}^+}
\newcommand{\Bm}{{\bm b}^-}
\newcommand{\Bpm}{{\bm b}^\pm}

\def\bk{{\bm b}_{\bm k}}

\def\bkpm{b^\pm_{\bm k}}
\def\ukpm{u^\pm_{\bm k}}
\def\ukp{u^+_{\bm k}}
\def\ukm{u^-_{\bm k}}

\def\uqp{u^+_{\bm q}}

\def\bkp{b^+_{\bm k}}
\def\bkm{b^-_{\bm k}}

\def\usk{u^{s_k}_{\bm k}}
\def\bsk{b^{s_k}_{\bm k}}
\def\bski{b^{\sigma_k}_{\bm k}}
\def\usp{u^{s_p}_{\bm p}}

\def\bspi{b^{\sigma_p}_{\bm p}}
\def\usq{u^{s_q}_{\bm q}}

\def\bsqi{b^{\sigma_q}_{\bm q}}
\def\bkps{b^{+*}_{\bm k}}
\def\bkms{b^{-*}_{\bm k}}
\def\Bkp{B^+_{{\bm k}_0}}
\def\Ukp{U^+_{{\bm k}_0}}
\def\Bkm{B^-_{{\bm k}_0}}
\def\Ukm{U^-_{{\bm k}_0}}
\def\Bpp{B^+_{{\bm p}_0}}
\def\Upp{U^+_{{\bm p}_0}}
\def\Bpm{B^-_{{\bm p}_0}}
\def\Upm{U^-_{{\bm p}_0}}

\newcommand{\Usp}{U^{s_{p_0}}_{{\bm p}_0}}
\newcommand{\Bspi}{B^{\sigma_{p_0}}_{{\bm p}_0}}
\newcommand{\Usk}{U^{s_{k_0}}_{{\bm k}_0}}
\newcommand{\Bski}{B^{\sigma_{k_0}}_{{\bm k}_0}}

\newcommand{\tuqp}{\tilde u^+_{\bm q}}
\newcommand{\tuqm}{\tilde u^-_{\bm q}}
\newcommand{\tbkp}{\tilde b^+_{\bm k}}
\newcommand{\tbkm}{\tilde b^-_{\bm k}}
\newcommand{\tbpp}{\tilde b^+_{\bm p}}
\newcommand{\tbpm}{\tilde b^-_{\bm p}}
\newcommand{\tbqp}{\tilde b^+_{\bm q}}
\newcommand{\tbqm}{\tilde b^-_{\bm q}}
\newcommand{\tusk}{\tilde u^{s_k}_{\bm k}}
\newcommand{\tusp}{\tilde u^{s_p}_{\bm p}}
\newcommand{\tusq}{\tilde u^{s_q}_{\bm q}}

\def\tbski{\tilde b^{\sigma_k}_{\bm k}}
\def\tbspi{\tilde b^{\sigma_p}_{\bm p}}
\def\tbsqi{\tilde b^{\sigma_q}_{\bm q}}
\def\tbskis{\tilde b^{\sigma_k*}_{\bm k}}
\def\tbspis{\tilde b^{\sigma_p*}_{\bm p}}
\def\tbsqis{\tilde b^{\sigma_q*}_{\bm q}}
\newcommand{\tusks}{\tilde u^{s_k*}_{\bm k}}
\newcommand{\tusps}{\tilde u^{s_p*}_{\bm p}}
\newcommand{\tusqs}{\tilde u^{s_q*}_{\bm q}}

\def\fup{f_u^+}
\def\fum{f_u^-}
\def\fbp{f_b^+}
\def\fbm{f_b^-}
\def\bk{\bm k}
\def\ba{\bm a}
\def\bb{\bm b}

\def\bp{\bm p}

\def\bq{\bm q}

\def\bx{\bm x}

\def\bomega{\bm \omega}
\def\bu{\bm u}
\def\hkm{{\bm h}^-_{\bm k}}
\def\hkp{{\bm h}^+_{\bm k}}

\def\hsk{{\bm h}^{s_k}_{\bm k}}
\def\hsp{{\bm h}^{s_p}_{\bm p}}

\newcommand{\be}{\begin{equation}}

\newcommand{\ee}{\end{equation}}

\renewcommand{\vec}[1]{\bm{#1}}
\newcommand{\fvec}[1]{\hat{\vec{#1}}}
\newcommand{\dt}{\partial_t}

\newcommand{\ml}[1]{}

\begin{document}
\vspace{-2em}
\begin{minipage}[l]{\textwidth}
Postprint version of the manuscript published in Astrophys. Journal {\bf 836}, 26 (2017) \\ 
\end{minipage}
\vspace{2em}
\title{Effects of magnetic and kinetic helicities on the growth of magnetic fields in laminar and turbulent flows by helical Fourier decomposition}

\author{Moritz Linkmann\altaffilmark{1,2}, Ganapati Sahoo\altaffilmark{1}, Mairi McKay\altaffilmark{2}, 
Arjun Berera\altaffilmark{2}, Luca Biferale\altaffilmark{1}}
\affiliation{\altaffilmark{1}Department of Physics \& INFN, University of Rome Tor Vergata, Via della Ricerca Scientifica 1, 00133 Rome, Italy. \\
\altaffilmark{2}School of Physics and Astronomy, University of Edinburgh, Peter Guthrie Tait Road, EH9 3FD, Edinburgh, UK.}
\date{\today}

\begin{abstract}

We present a numerical and analytical study of incompressible homogeneous
conducting fluids using a helical Fourier representation.  We analytically
study both small- and large-scale dynamo properties, as well as the inverse
cascade of magnetic helicity, in the most general minimal subset of interacting
velocity and magnetic fields on a closed Fourier triad. We mainly focus on the
dependency of magnetic field growth as a function of  the distribution of
kinetic and magnetic helicities among the three interacting wavenumbers.  By
combining  direct numerical simulations of the full magnetohydrodynamics 
equations with the helical Fourier decomposition we numerically confirm that in
the kinematic dynamo regime
the system develops a large-scale magnetic helicity with opposite sign compared
to the small-scale kinetic helicity, a sort of triad-by-triad
$\alpha$-effect in  Fourier space.  Concerning the small-scale perturbations,
we predict theoretically and confirm numerically that the largest instability
is achieved  for the magnetic component with the same helicity of the flow, in
agreement with the Stretch-Twist-Fold mechanism.  Vice versa, in presence of a
Lorentz feedback on the velocity, we  find that the inverse cascade of magnetic
helicity is mostly local if magnetic and kinetic helicities have opposite sign,
while it is more nonlocal and more intense if they have the same sign, as
predicted by the analytical approach.
\\
Our analytical and numerical results further
demonstrate the potential of the helical Fourier decomposition to elucidate
the entangled dynamics of magnetic and kinetic helicities both in fully
developed turbulence and in laminar flows. 
\\

\end{abstract}

\maketitle

\section{Introduction} \label{sec:intro}
Turbulent flows are ubiquitous on Earth and in the sky, 
e.g.~\citep{Frisch95,Pope00}.  
In many astrophysical and geophysical
cases \citep{Belenkaya09}, the fluid
is also conducting and one needs to control the entangled dynamics of velocity
and magnetic fields; this is the case of magnetohydrodynamic (MHD) 
turbulence \citep{Moffatt78,Biskamp03,Verma04}.
The spectrum of possible configurations 
and applications is  vast, depending on the presence of particular
external forcing mechanisms, mean flows or fields and boundaries 
\citep{Priest00,Priest14,Plihon14,Stieglitz01,Gailitis00,Gailitis01,Goodman02,Nornberg06,Frick10}. Here we
wish to address basic properties of all MHD configurations  
connected
with the interactions leading to the transfer of total
energy, magnetic helicity and kinetic helicity across scales. For MHD,
the presence of three inviscid invariants, total energy, magnetic and 
cross-helicity, makes the problem of predicting spectral properties difficult 
\citep{Iroshnikov64,Kraichnan65a,Matthaeus89,Goldreich95,Boldyrev05a,Boldyrev06}. 
Even the direction of the different
transfers is not completely under control, only empirical results exist 
\citep{Biskamp03}. Moreover, because of obvious applied and fundamental issues,
predicting or controlling the growth rate of a magnetic field is a key question,
connected to the famous dynamo problem 
\citep{Moffatt69,Krause80,Brandenburg03,Brandenburg05,Tobias13}. 
Since the magnetic energy is not conserved, the magnetic field may stretched, folded or advected 
even in absence of external input and dissipation.  
A huge amount of literature has been devoted to  
the identification of the key dynamical and statistical ingredients 
needed to promote or deplete such
a growth and to control the growth rate.  Magnetic and kinetic helicities are
among the key quantities that play a role in such a phenomenon:
\begin{align}
\label{eq:Hm}
H_m(t) &=\int_V d \bx \ \ba(\bx,t) \cdot \bb(\bx,t) \ , \\ 
\label{eq:Hk}
H_k(t) &=\int_V d \bx \ \bu(\bx,t) \cdot \bomega(\bx,t) \ ,  
\end{align}
where $\bu$, $\bb$, $\ba$ and $\bomega$ are  the velocity, the magnetic field,
the magnetic vector potential and the vorticity, respectively.  
The third ideal  
invariant of the MHD equations  
is given by cross-helicity:
\begin{align}
\label{eq:Hc}
H_c(t) &=\int_V d \bx \ \bu(\bx,t) \cdot \bb(\bx,t) \ ,  
\end{align}
that is connected to the degree of Alfv{\'e}nization of
the system, i.e. to the presence of waves traveling in the direction of the
mean global or local magnetic field \citep{Dobrowolny80,Biskamp93}. In this paper we will focus
on the importance of magnetic and kinetic helicities (helicities in short in
what follows) for the growth rate of a large-scale magnetic field, always
considering the case of almost vanishing cross helicity (see also the
concluding remarks about possible generalization of our work to include also
the latter). To be as simple as possible we will concentrate only on
 periodic and homogeneous conditions. The
analytical and numerical work is based on the
helical Fourier decomposition developed for Navier-Stokes equations by the
pioneering work of  
\citet{Waleffe92} and \citet{Constantin88}. This decomposition is exact and
has led to an important breakthrough in the understanding of the entangled
energy-helicity dynamics in three-dimensional Navier-Stokes turbulence
\citep{Waleffe92,Biferale13}. It has only recently been extended to MHD 
\citep{Lessinnes09,Linkmann16}, and it promises  to be a key tool also
 for problems where  the physics
is controlled by the interactions among magnetic or kinetic helical waves 
\citep{Cho11,Galtier03,Galtier05,Galtier14}.
Moreover, the helical Fourier basis is also the natural decomposition to be
used in numerical simulations, either to analyze the data or to perform
explicit numerical experiments by projecting the equations on a given subset
of Fourier modes, in order to highlight the physics of some particular interacting
waves. This procedure has already been carried out 
for the Navier-Stokes equations
with some surprising results \citep{Biferale12,Biferale13a,Sahoo15,Alexakis16a}  connected to  the discovery of a sub-class of (kinetic) helical modes
transferring energy backwards in a fully three-dimensional turbulent flow, i.e. the
identification of those Fourier interactions responsible for the energy
backscatter. 
Simulations of homogeneous magnetohydrodynamic turbulence in a periodic box without
a background magnetic field have been used as prototypical systems to study the
circumstances under which large-scale magnetic field growth occurs, such as the
$\alpha$-effect \citep{Steenbeck66,Brandenburg01} and the inverse cascade of magnetic
helicity 
\citep{Frisch75,Pouquet76,Balsara99,Brandenburg01,Alexakis06,Mueller12,Malapaka13}.
Concerns have been raised in the literature about the effectiveness of the
$\alpha$-effect in generating large-scale magnetic fields with strong
amplitude because of the detrimental feedback that fast-growing
small-scale magnetic fields have on the growth rate of the large-scale
magnetic field \citep{Vainshtein92,Cattaneo96}. This is the problem of
catastrophic $\alpha$-quenching, and much theoretical efforts have been made
in order to find dynamo models which are able to circumvent this problem.
In this paper we focus on the statistical and dynamical factors
that might promote or deplete the growth of a large-scale magnetic field. We do
this by using a systematic dissection of the three-dimensional MHD 
equations  in helical Fourier modes as pioneered by \citet{Linkmann16}. We first analyze 
the temporal evolution of three velocity and three magnetic
helical Fourier modes at wavenumbers, $k,p,q$,  the basic brick of any
quadratic non-linear transfer. Considering all helical combinations, there are 64 possible different 
subsets of closed dynamical systems that represent the minimal backbone of interacting modes conserving all inviscid invariants. 
A graphical representation of such a basic system is given
in Figure~\ref{fig:two-triad-system}.  
A stability analysis of a subfamily of these dynamical systems was
carried out by \citet{Linkmann16}, considering the most general equilibria. 
This led to the identification of linear instabilities that could be associated
with forward and inverse transfer of total energy and magnetic helicity. Here we 
restrict our attention to equilibria and instabilities that can be connected 
to cases of astrophysical interest, i.e. kinematic dynamo regimes 
and inverse cascade effects. 
 We further extend the analysis carried out by \citet{Linkmann16}  by specifying  
the magnetic field growth rates (if any) and providing clear predictions on the 
expected helical signatures of the dominant instabilities.      
This enables us to link specific dynamical properties connected 
 to the kinematic dynamo action and/or the 
inverse cascade of magnetic helicity with the geometrical structure of the triad (local versus non-local Fourier
interactions) and with its kinematic contents (helicities). Interestingly,  the
entangled dynamics of velocity field with the magnetic fields is already extremely rich at the
level of these most basic interactions. 
\\  
The second part of the paper is devoted to develop for the first time a
thoughtful numerical validation and benchmark of the previous theoretical
analysis by Direct Numerical Simulation (DNS) 
of the full MHD equations with and without a small-scale forcing on the magnetic field. 
Forcing the magnetic field allows us to switch from a situation where the magnetic field is
initially in the kinematic dynamo regime to a case where Lorentz force is
always acting at all scales, thanks to the strong injection of magnetic
fluctuations. We always analyze velocity and magnetic fluctuations in terms of
their helical Fourier components such as to be able to directly match the
theoretical predictions based on the simplified single-triad dynamics. We
changed the helical properties of the magnetic forcing (when applied)
to break the mirror symmetry with different injection mechanisms.  We study two
different configurations, 
with large- or small-scale injection of kinetic energy and
helicity, corresponding to turbulent or laminar regimes.  Furthermore, we also
present some ad hoc simulations by restricting the dynamics of the velocity
field to evolve only on modes with one given sign of  
helicity, which induces a strong breaking of the mirror symmetry already
at the level of the equations of motion.\\ 
The combined analytical and numerical analysis lead to the
following conclusions: (a) In presence of small-scale residual kinetic helicity
and with an initially weak 
magnetic field, the large-scale magnetic field develops a helical signature of opposite sign with respect to
the kinetic helicity, in agreement with the predictions of the classical $\alpha$-effect 
in presence of scale-separation. 
(b) In the same flow configuration as in point (a), 
but with a strong injection of small-scale magnetic fluctuations, the growing large-scale
magnetic field has the same sign of helicity as the injected helical small-scale magnetic fluctuations. 
This mechanism leads to an inverse cascade of magnetic helicity \citep{Frisch75}.
The two different behaviors are connected to two different
properties of the single-triad dynamics. In case (a) the growth is
$\alpha$-like as a result of the direct stretching of the  magnetic field lines by the helical
velocity field. In case (b) the growth is dominated by purely non-linear
effects that are due to the Lorentz force acting on the velocity field and to the stretching of
the magnetic field by the velocity. 

\section{Theoretical and analytical results} \label{sec:theory}
We consider the MHD equations for incompressible flow
\begin{align}
\label{eq:momentum}
\partial_t \vec{u}&= - \frac{1}{\rho}\nabla P -(\vec{u}\cdot \nabla)\vec{u}
 + \frac{1}{\rho}(\nabla \times \vec{b}) \times \vec{b} + \nu \Delta \vec{u}  \ , \\
\label{eq:induction}
\partial_t \vec{b}&= (\vec{b}\cdot \nabla)\vec{u}-(\vec{u}\cdot \nabla)\vec{b} + \eta \Delta \vec{b}\ , \\
\label{eq:incompr}
&\nabla \cdot \vec{u} = 0 \ \ \mbox{and} \ \  \nabla \cdot \vec{b} = 0 \ ,  
\end{align}
where $\vec{u}$ denotes the velocity field, $\vec{b}$ the magnetic induction
expressed in Alfv\'{e}n units, $\nu$ the kinematic viscosity, $\eta$ the
magnetic resistivity, $P$ the pressure and $\rho$ the density, which is set to
unity for convenience. In the limit of vanishing viscosity and resistivity, these equations conserve 
the magnetic helicity $H_m$, the cross-helicity
$H_c$, and the total energy
\begin{align}
\label{eq:Etotal}
E(t) &=\frac{1}{2}\int_V d \bx \ (|\bu(\bx,t)|^2 + |\bb(\bx,t)|^2) \ . 
\end{align}
We consider eqs.~\eqref{eq:momentum}-\eqref{eq:incompr} on a domain $[0,L)^3$ 
with periodic boundary conditions in order to assess scale-dependent dynamics by Fourier analysis. 
The Fourier transforms $\fvec{u}$ and $\fvec{b}$ of the velocity and magnetic field fluctuations evolve 
according to the following system of equations 
\begin{align}
\label{eq:Fmomentum}
(\partial_t+\nu k^2) \fvec{u}_{\bk}(t) = & \left(\mathbb{I} -\frac{\bk \otimes \bk}{k^2}\right) 
 \left [\sum_{\vec{k}+\vec{p}+\vec{q}=0} \left( -(i\vec{p}\times \fvec{u}_{\vec{p}}(t)^*
       \times \fvec{u}_{\bq}(t)^*+(i\vec{p}\times \fvec{b}_{\bp}(t))^*\times \fvec{b}_{\bq}(t)^*\right) \right] \ , \\
\label{eq:Finduction}
(\partial_t+\eta k^2) \fvec{b}_{\vec{k}}(t) = & \  i\vec{k} \times \sum_{\vec{k}+\vec{p}+\vec{q}=0} \fvec{u}_{\bp}(t)^* \times \fvec{b}_{\bq}(t)^* \ , 
\end{align}
where 
the asterisk denotes the complex conjugate. The inertial term
$(\vec{u}\cdot \nabla)\vec{u}$ in the momentum equation \eqref{eq:momentum} has
 been written in rotational form
$(\vec{u}\cdot \nabla)\vec{u} = (\nabla \times \vec{u}) \times \vec{u} + \nabla |\vec{u}|^2/2$ and
the pressure has been eliminated using the projector $P_{ij}=\delta_{ij} -k_ik_j/k^2$.
Owing to the statistical homogeneity of the vector field fluctuations, the
structure of the vector field couplings in Equations \eqref{eq:Fmomentum} and
\eqref{eq:Finduction} are expressed by triadic interactions of wavevectors,
a triad of wavevectors $\bk,\bp$ and $\bq$ being defined by requiring
$\bk + \bp + \bq = 0$.
In the following we study these triadic interactions with a view of extracting
information on the effects of helicities on the dynamics of the magnetic field. 
\\

\subsection{Helical decomposition}
Being solenoidal vector fields, the Fourier transforms $\fvec{u}_{\vec{k}}$ and
$\fvec{b}_{\vec{k}}$ are orthogonal to the wavevector $\bk$ and have only two
degrees of freedom. These can be expressed by projection onto circularly
polarized waves:
\begin{align}
\label{eq:basisu}
\fvec{u}_{\bk}(t)&=\ukp(t) \hkp + \ukm(t) \hkm=\sum_{s_k} \usk(t) \hsk \ , \\
\label{eq:basisb}
\fvec{b}_{\bk}(t)&=\bkp(t) \hkp + \bkm(t) \hkm=\sum_{s_k} \bsk(t) \hsk \ ,
\end{align}
where  $s_k = \pm$ and $\vec{h}^{\pm}_{\bk}$ are the two fully helical orthonormal eigenvectors of
the curl operator satisfying $i\bk \times \hsk = s_k k \hsk$. The above decomposition was first proposed 
for the three-dimensional Navier-Stokes case by 
\citet{Constantin88,Waleffe92} and later generalized for the MHD case by
\citet{Lessinnes09,Linkmann16}. 
Similar decompositions have also been exploited to build up simplified models 
of turbulence \citep{Lessinnes09,DePietro15}. Here we apply it, for the first 
time in a systematical way, to disentangle different basic transfers in the 
full MHD case. 
Since the helical basis vectors $\hsk$ are time independent, all information concerning the dynamics
of the system is contained in the helical coefficients $\usk(t)$ and $\bsk(t)$.

We can further decompose the kinetic and magnetic spectra in terms of their helical components as
\begin{align}
& E_{u}(k,t)  = E^+_u(k,t) + E^-_u(k,t);\qquad E_u^{\pm}(k,t) =  \frac{1}{2}\sum_{|\vec{k}|=k}  |\ukpm(t)|^2\\
& E_{b}(k,t)  = E^+_b(k,t) + E^-_b(k,t);\qquad E_b^{\pm}(k,t) =  \frac{1}{2}\sum_{|\vec{k}|=k}  |\bkpm(t)|^2.
\end{align}
The ideal invariants given by Equations \eqref{eq:Hm}, \eqref{eq:Hc} and \eqref{eq:Etotal} are conserved in single
triad interactions \citep{Lessinnes09,Lessinnes11} and can expressed as:
\begin{align}
&E(t) =\frac{1}{2}\sum_{\vec{k}} \left(|\ukp(t)|^2 + |\ukm(t)|^2 +|\bkp(t)|^2 + |\bkm(t)|^2  \right) \ , \\
&H_m(t) =\sum_{\vec{k}} \frac{1}{k} \left( |\bkp(t)|^2 - |\bkm(t)|^2 \right)  \ , \\ 
&H_c(t) =\sum_{\vec{k}} \mathcal{R}\left( \ukp(t) \bkps(t) + \ukm(t) \bkms(t) \right) \ , 
\end{align}
where $\mathcal{R}$ is the real part of a complex number.
Similarly, the kinetic helicity becomes
\be
H_k(t) = \sum_{\vec{k}} \fvec{u}(\vec{k},t)\fvec{\omega}(-\vec{k},t) =\sum_{\vec{k}} k \left(|\ukp(t)|^2 - |\ukm(t)|^2\right) \ ,
\ee
with $\fvec{\omega}(\vec{k},t)$ being the Fourier transform of the vorticity. For conciseness
we will drop the explicit notation of the 
time dependence of the helical coefficients and the corresponding energy spectra from now on.

\subsection{MHD-Helical dynamics} \label{sec:helical-dynamics}
\begin{figure}[h]
\center
\includegraphics[scale=0.15]{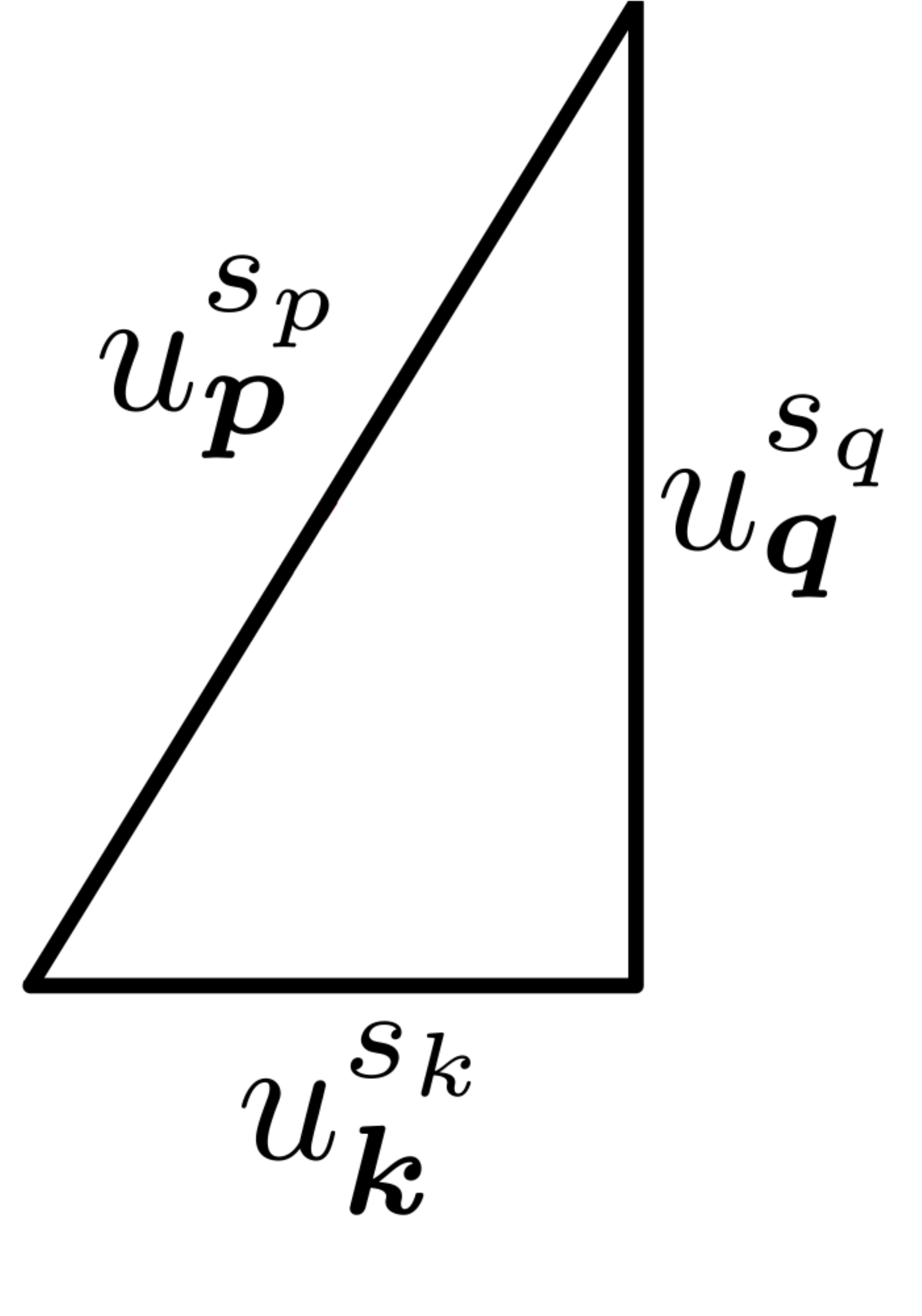} 
\includegraphics[scale=0.15]{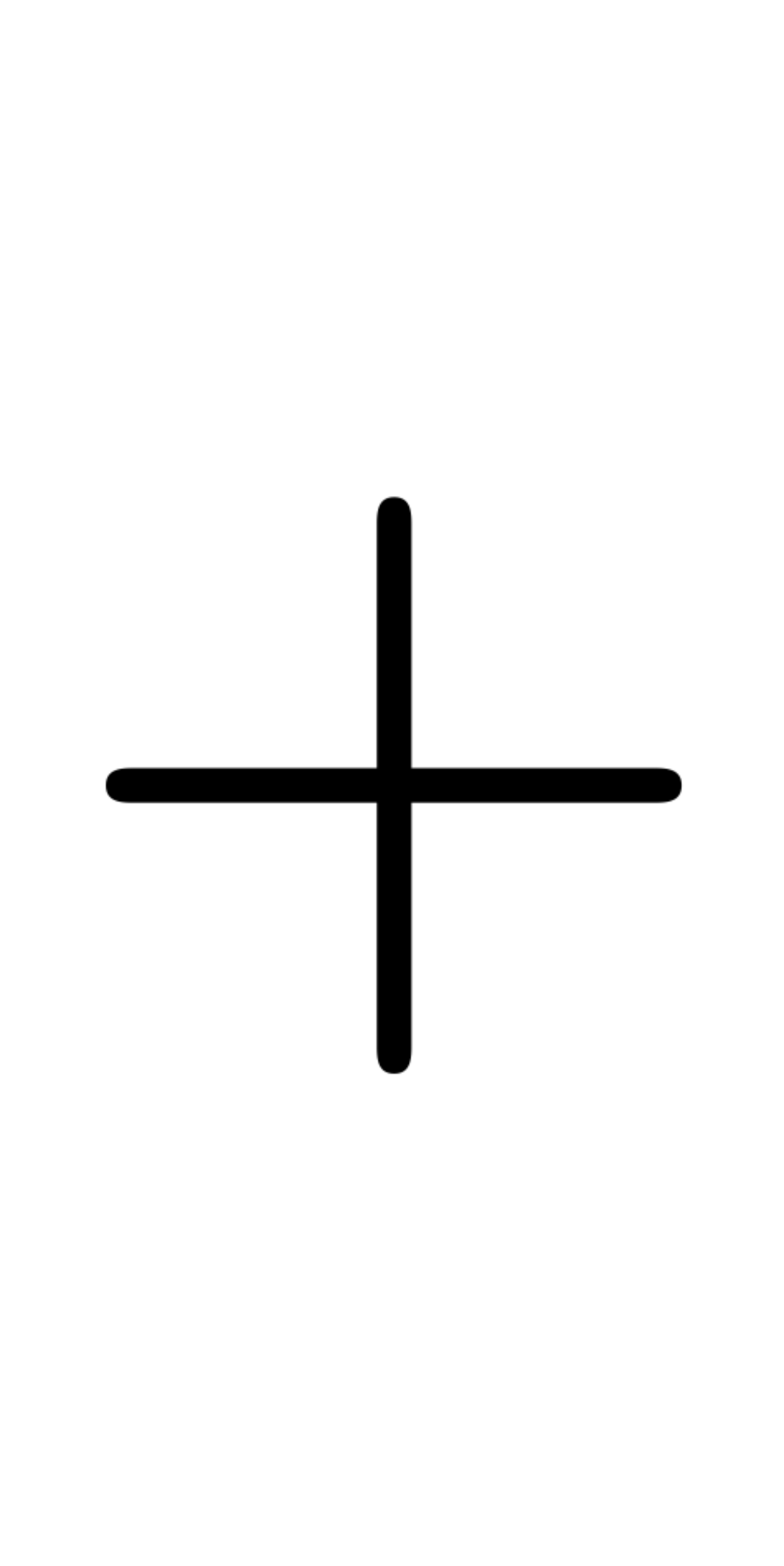}
\includegraphics[scale=0.15]{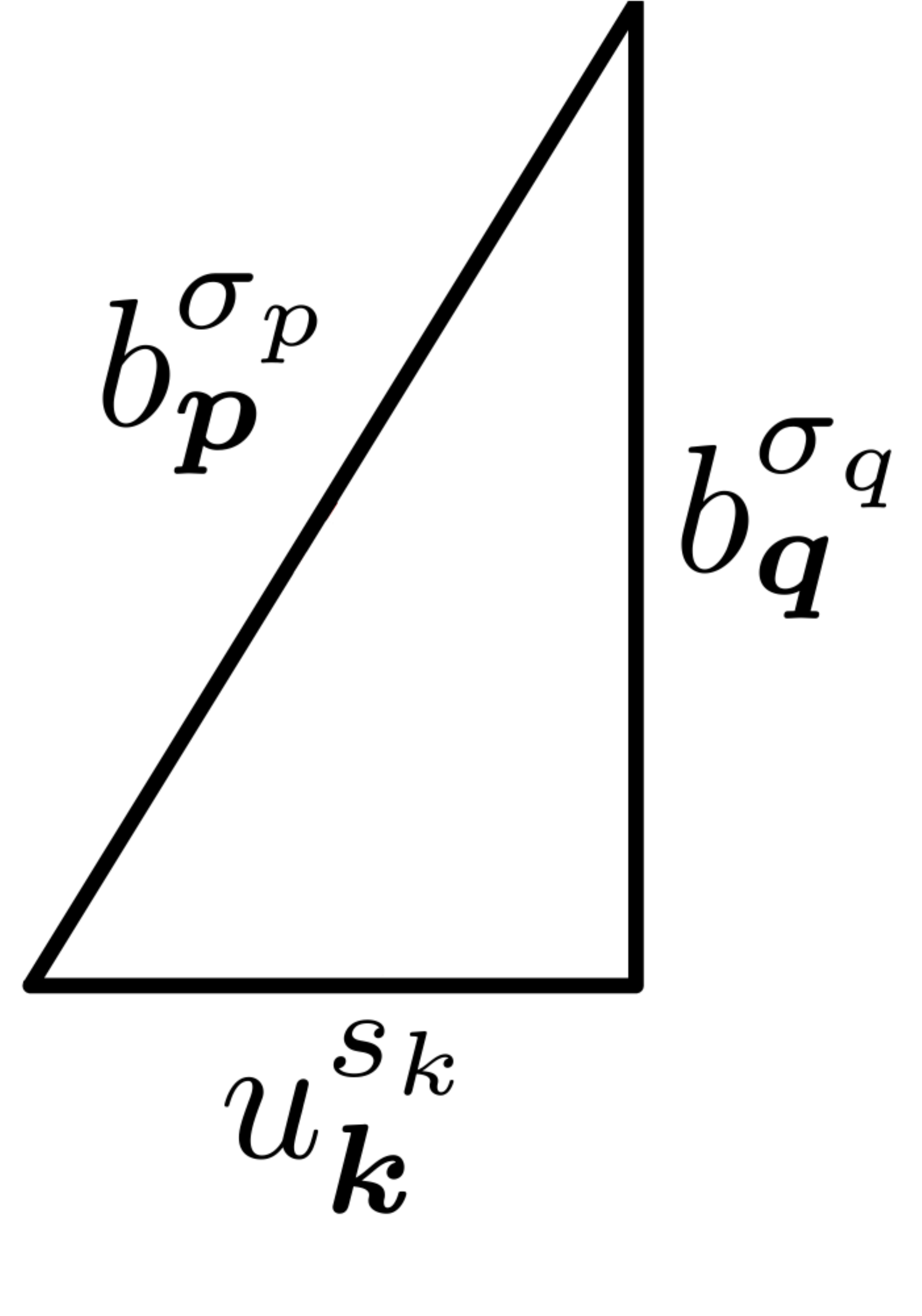}
\includegraphics[scale=0.15]{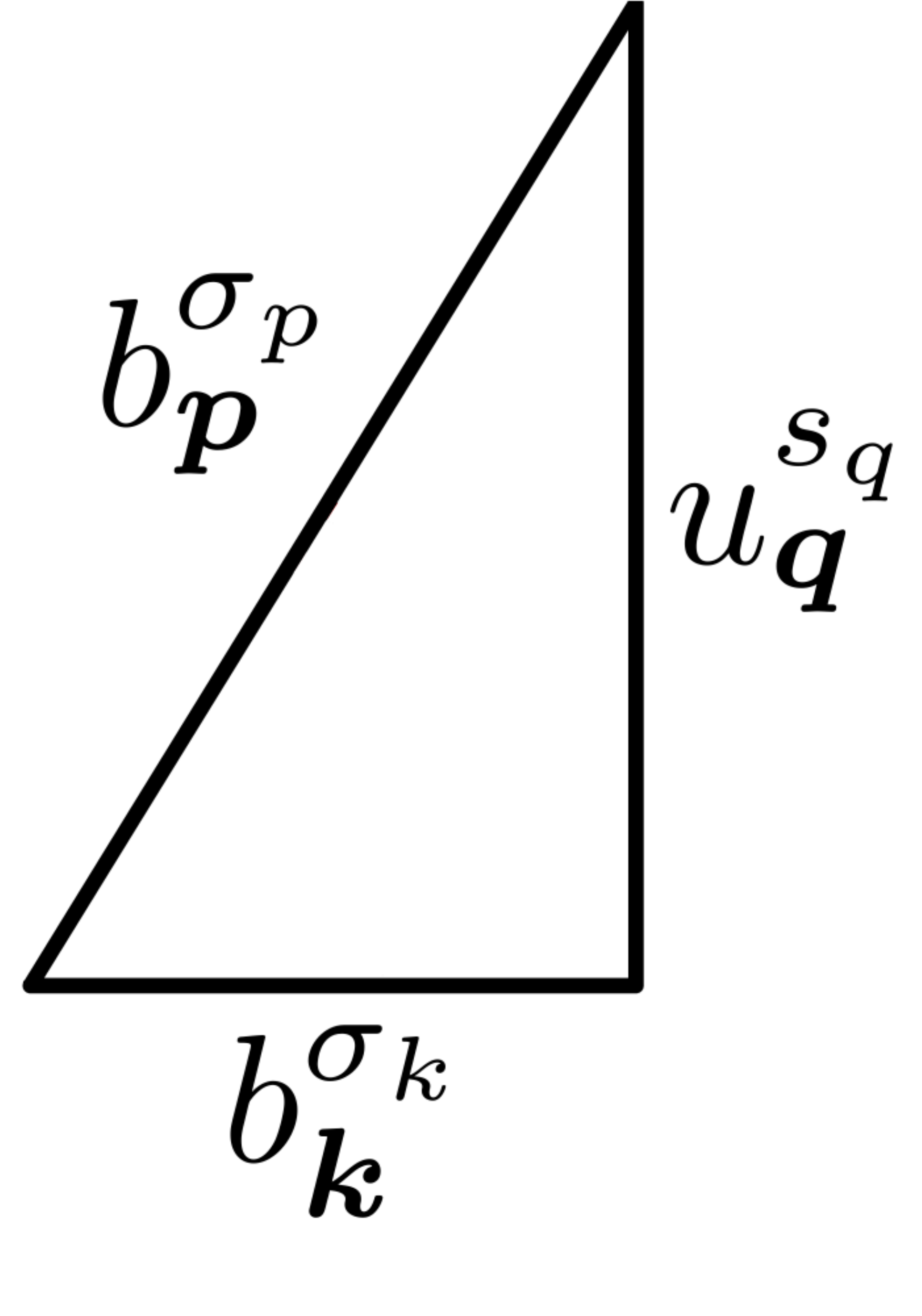}
\includegraphics[scale=0.15]{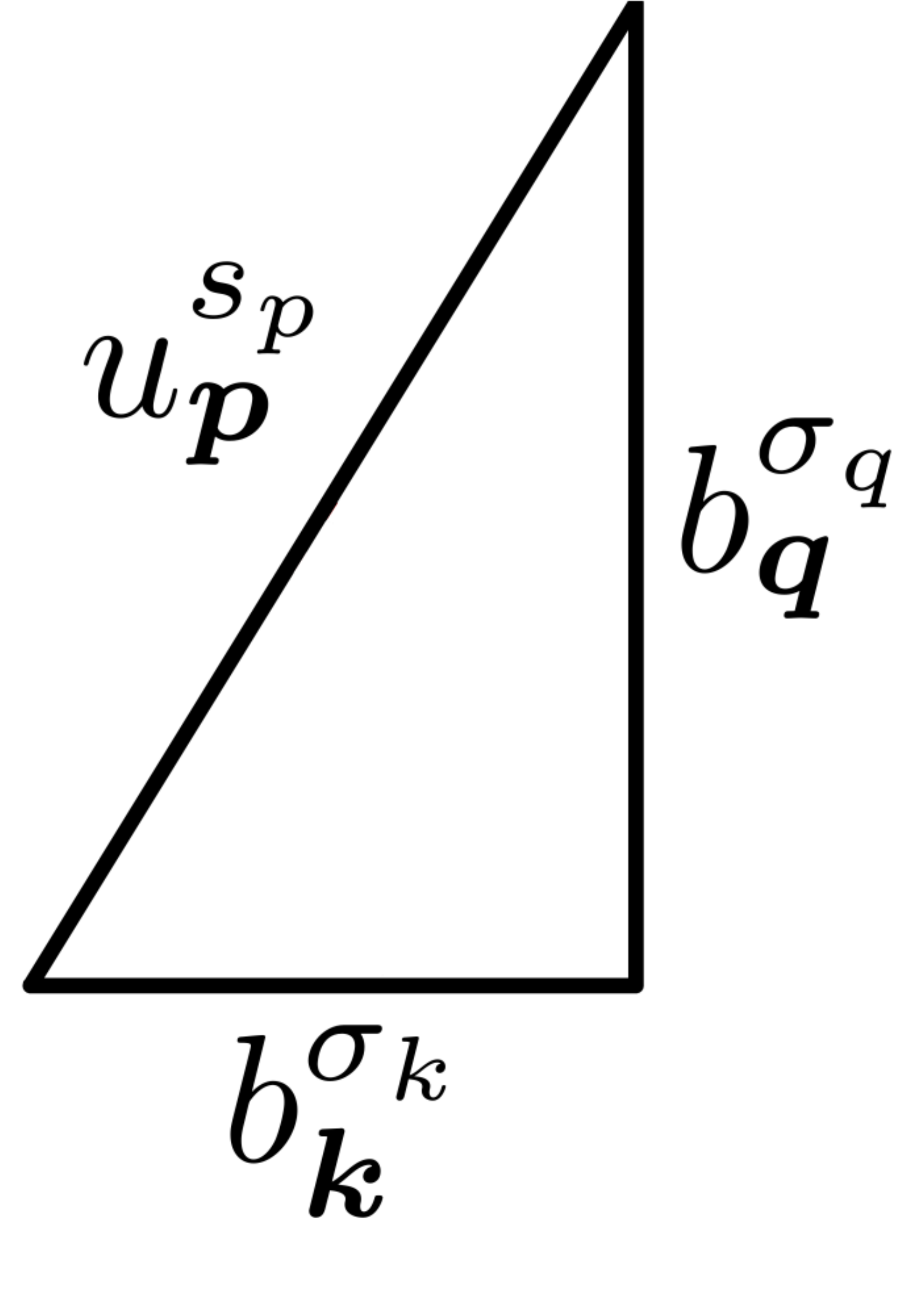}
\caption{Graphical representation of the minimal non-linear system consisting of the coupling 
among a  velocity  (left) 
and a magnetic-velocity (right) triadic interaction. 
The interaction of the magnetic field with the velocity field
corresponds to several terms in Equation \eqref{eq:basic-triads}, 
depending on whether the time evolution of the velocity field mode or the 
magnetic modes are considered.}
\label{fig:two-triad-system}
\end{figure}

Inserting the exact decompositions (\ref{eq:basisu}-\ref{eq:basisb}) into
(\ref{eq:Fmomentum}-\ref{eq:Finduction}) and taking the inner
product with $\hsk$,  we obtain the
helical Fourier version of the MHD equations \citep{Waleffe92,Lessinnes09}:
\begin{align}
\label{eq:MHD-helical}
(\partial_t+\nu k^2) {\usk}^* &= \frac{1}{2} \sum_{\vec{k}+\vec{p}+\vec{q}=0} 
\sum_{s_p, s_q} g^{IN}_{s_k s_p s_q} (s_p  p-s_q  q) \usp \usq 
-\sum_{\sigma_p, \sigma_q} g^{LF}_{s_k \sigma_p\sigma_q} (\sigma_p p-\sigma_q q) \bspi \bsqi \ , \nonumber \\ 
(\partial_t+\eta k^2) {\bski}^* & = 
  \frac{\sigma_k k}{2} \sum_{\vec{k}+\vec{p}+\vec{q}=0} 
\sum_{\sigma_p, s_q} g^{M1}_{\sigma_k\sigma_ps_q} \bspi \usq 
-\sum_{s_p, \sigma_q} g^{M2}_{\sigma_ks_p\sigma_q} \usp \bsqi
\end{align}
where the coupling coefficients $g^{IN}_{s_k s_p s_q}$ and  $g^{LF}_{s_k \sigma_p\sigma_q}$
originate from the inertial term and the Lorentz force in the momentum equation, respectively, while
$g^{M1}_{\sigma_k\sigma_ps_q}$ and $g^{M2}_{\sigma_ks_p\sigma_q}$ 
originate from the symmetrised induction equation. The coupling coefficients 
are given explicitly in
Equation \eqref{eqapp:gfactor} in Appendix \ref{app:stability}.
It is very important to realize that the non-linear triadic
interactions on the RHS of Equation \eqref{eq:MHD-helical}
can be  further decomposed into a subset of basic bricks consisting of three helical velocity  and 
magnetic modes at wavevectors $\bp,\bk,\bq$ such that $\bp+\bk+\bq=0$, 
characterized by one possible combination of chiral numbers 
 $(s_k,s_p,s_q) = (\pm,\pm,\pm)$ for the velocity modes and  
$(\sigma_k,\sigma_p,\sigma_q) = (\pm,\pm,\pm)$ for the magnetic modes as 
shown in Figure \ref{fig:two-triad-system}:
\begin{align}
\label{eq:basic-triads}
\dt {\usk}^* &=  g^{IN}_{s_k s_p s_q} (s_pp-s_qq)\:  \usp \usq  - g^{LF}_{s_k\sigma_p \sigma_q} (\sigma_pp-\sigma_qq)\:  \bspi \bsqi  \ , \nonumber \\
\dt {\usp}^* &=  g^{IN}_{s_k s_ps_q} (s_qq-s_kk)\:  \usq \usk - g^{LF}_{\sigma_ks_p\sigma_q}(\sigma_qq-\sigma_kk)\:  \bsqi \bski  \ , \nonumber \\
\dt {\usq}^* &=  g^{IN}_{s_k s_p s_q} (s_kk-s_pp)\:  \usk \usp - g^{LF}_{\sigma_k \sigma_ps_q} (\sigma_kk-\sigma_pp)\:  \bski \bspi  \ , \nonumber \\
\dt {\bski}^* &=  \sigma_k k\: \left(g^{M1}_{\sigma_k\sigma_p s_q} \bspi \usq - g^{M2}_{\sigma_ks_p \sigma_q} \usp \bsqi \right) \nonumber \ , \\
\dt {\bspi}^* &=  \sigma_p p\: \left(g^{M1}_{s_k \sigma_p \sigma_q} \bsqi \usk -  g^{M2}_{\sigma_k\sigma_p s_q} \usq \bski \right) \nonumber \ , \\
\dt {\bsqi}^* &=  \sigma_q q\: \left(g^{M1}_{\sigma_k s_p\sigma_q}\bski \usp -  g^{M2}_{s_k \sigma_p\sigma_q}\usk \bspi \right) \nonumber \ . \\
  \end{align}
Considering the symmetry of the original equations for a global change
of all positive helical waves into negative helical waves (mirror symmetry), 
we obtain 32 possible independent combinations of  chiral interactions
represented by Equation (\ref{eq:basic-triads}). 
Equation \eqref{eq:basic-triads} thus describes a minimal triadic interaction (MTI), where 
minimal refers to the smallest number of degrees of freedom necessary to represent the structure
of the quadratic couplings in the MHD equations.
It is crucial to realize that each of these
 32 MTI systems is closed in itself and that the energy,
the magnetic helicity and the cross-helicity are exactly preserved for each system.
A stability analysis  of the above eq.~\eqref{eq:basic-triads} shows that all
possible fixed points are given by non-zero entries at only one wavevector,
say at $\bp_0$ \citep{Linkmann16}.  We denote these equilibria by 
$(\Bpp,\Bpm,\Upp,\Upm)$.   Studying the stability of these equilibria  yields
information on possible energy or helicity 
transfers away from a given scale in
the system, as shown for the Navier-Stokes case by \citet{Kraichnan67,Waleffe92,Waleffe93}.
In the absence of the magnetic field, the structure of the system of
first-order ODEs describing a single triad  is very similar to the Euler
equations describing the torque-free rotation of a rigid body about its
principal axes of inertia, and the stability analysis was carried out by
\citet{Waleffe92}. The procedure remains conceptually the same for MHD with
added complications due to a larger number of coupled dynamical variables and
with the added value that one might gain some insight  into the basic physical
mechanisms governing kinematic dynamo action or into the effect of the Lorentz
force on the flow.  The stability analysis for the MHD case restricted  to {\it
homochiral} MTI system, i.e. 
interactions between the same chiral combination of
velocity and magnetic fields $s_p=\sigma_p,s_q=\sigma_q,s_k=\sigma_k$,
has been performed  by \citet{Linkmann16}.
In this paper we perform a few important steps forward with respect to the
analysis of \citet{Linkmann16}.
First, we will clarify under which circumstances the analysis by \citet{Linkmann16}
is sufficient, i.e. where a distiction between homo- and heterochiral triads is relevant
and where it is not required.
Second, we will extend the results concerning kinematic dynamo action and the 
inverse magnetic helicity cascade in order to deliver qualitatively testable predictions. 
Finally, in
Section \ref{sec:simulations} we carry out a direct benchmark of all
predictions by performing  high-resolution  DNS of the full MHD equations 
and exploiting the helical decomposition to rationalize the numerical results.

\subsubsection{Stability analysis of a general MTI system}
Let us first fix the notation. We always consider the following order for wavenumbers: $$ k< q< p.$$
 As a result,
whenever we wish to address the growth of perturbations at small wavenumbers (large scales), 
we take the equilibria for magnetic or velocity field 
at the largest wavenumber (smallest scale) $p_0= |\bp_0|$. Vice versa, 
for the growth of perturbations at large wavenumbers (small scales),
we fix the equilibria at the smallest wavenumber (largest scale) $k_0= |\bk_0|$. 
Up to first order, 
the evolution equations of the large-scale perturbations, $\tusk$, $\tusq$, $\tbski$ and $\tbsqi$ are 
\begin{equation}
\label{eq:basic-triads-equilibria}
Large \, Scale \ \ \begin{cases}
\dt \tusks & =  g^{IN}_{s_ks_{p_0}s_q}\: (s_pp_0-s_qq)\: \Usp \tusq  
              - g^{LF}_{s_k\sigma_{p_0}} \sigma_q \:(\sigma_{p_0} p_0- \sigma_q q)\: \Bspi \tbsqi \ , \nonumber \\
\dt \tusqs & =  g^{IN}_{s_ks_{p_0}s_q}\:(s_kk-s_{p_0}p_0)\: \tusk \Usp 
             -  g^{LF}_{\sigma_k \sigma_{p_0}s_q}\:(\sigma_kk-\sigma_{p_0}p_0)\: \tbski \Bspi \ , \nonumber \\
\dt \tbskis & =  \sigma_kk\: \left( g^{M1}_{\sigma_k\sigma_{p_0} s_q} \Bspi \tusq 
                                  - g^{M2}_{\sigma_ks_{p_0}\sigma_q} \Usp \tbsqi \right) \nonumber \ , \\
\dt \tbsqis & =  \sigma_qq\: \left( g^{M1}_{\sigma_ks_{p_0}\sigma_q} \tbski \Usp 
                                  - g^{M2}_{s_k\sigma_{p_0}\sigma_q} \tusk \Bspi \right) \ , 
\end{cases}
\end{equation}
while for the evolution of small-scale perturbations $\tusp$, $\tusq$, $\tbspi$ and $\tbsqi$ we have
\begin{equation}
\label{eq:basic-triads-equilibria2}
Small \, Scale \ \ \begin{cases}
\dt \tusps & =   g^{IN}_{s_{k_0}s_ps_q}\:(s_qq-s_{k_0}k_0)\: \tusq \Usk  - g^{LF}_{\sigma_{k_0}s_p\sigma_q}\:(\sigma_qq-\sigma_{k_0}k_0)\: \tbsqi \Bski \ , \nonumber \\
\dt \tusqs & =   g^{IN}_{s_{k_0}s_ps_q}\:(s_{k_0}k_0-s_pp)\: \Usk \tusp -  g^{LF}_{\sigma_{k_0}\sigma_ps_q}\:(\sigma_{k_0}k_0-\sigma_pp)\: \Bski \tbspi \ , \nonumber \\
\dt \tbspis & =  \sigma_pp\: \left( g^{M1}_{s_{k_0}\sigma_p\sigma_q} \tbsqi \Usk - g^{M2}_{\sigma_{k_0}\sigma_p\sigma_q} \tusq \Bski \right) \nonumber \ , \\
\dt \tbsqis & =  \sigma_qq\: \left( g^{M1}_{\sigma_{k_0}s_p\sigma_q} \Bski \tusp -   g^{M2}_{s_{k_0}\sigma_p\sigma_q} \Usk \tbspi \right) \ . 
\end{cases}
\end{equation}
In the following we will address only fully kinetic helical 
equilibria $(B^{\pm}_{\bp_0}=0, U^-_{\bp_0}=0, U^+_{\bp_0} \ne 0)$ or fully magnetic helical 
equilibria  $(U^{\pm}_{\bp_0}=0, B^-_{\bp_0}=0, B^+_{\bp_0} \ne 0)$.
In order to assess the linear stability of a given equilibrium solution of each single set of
coupled first-order ODEs describing an individual MTI system, a further differentiation in time is
required, leading to four coupled second-order ODEs. The procedure is
explained in full detail in the papers by \citet{Waleffe92} for purely inertial
dynamics and \citet{Linkmann16} for MHD.
Here we will consider 
the stability properties of a given equilibrium for each of the  32 MTI 
 closed sub-systems for both large- or small-scale equilibria,
and we discuss 
the stability properties of the equilibrium as a function of the
helical (magnetic or kinetic) content of the equilibrium and the perturbations.
The  exponential growth of a given perturbation is associated with
transfers from the equilibrium mode into the perturbing modes.
For example, kinematic
dynamo action can be represented by the growth of a magnetic perturbation starting from a  mechanical equilibrium given by
$(B^\pm_{\bp_0} = U^-_{\bp_0}=0,  U^+_{\bp_0} \ne 0)$, while the feedback  of the Lorentz force on the
fluid can be described  by the dynamics of the velocity perturbation starting from a magnetic helical equilibrium
$(U^\pm_{\bp_0} = B^-_{\bp_0}=0,  B^+_{\bp_0} \ne 0)$.

\section{Summary of predictions} \label{sec:summary-theory}
\subsection{Large scale kinematic dynamo \label{sec:lsdynamo}}
\begin{figure}[h]
\center
\includegraphics[scale=0.25]{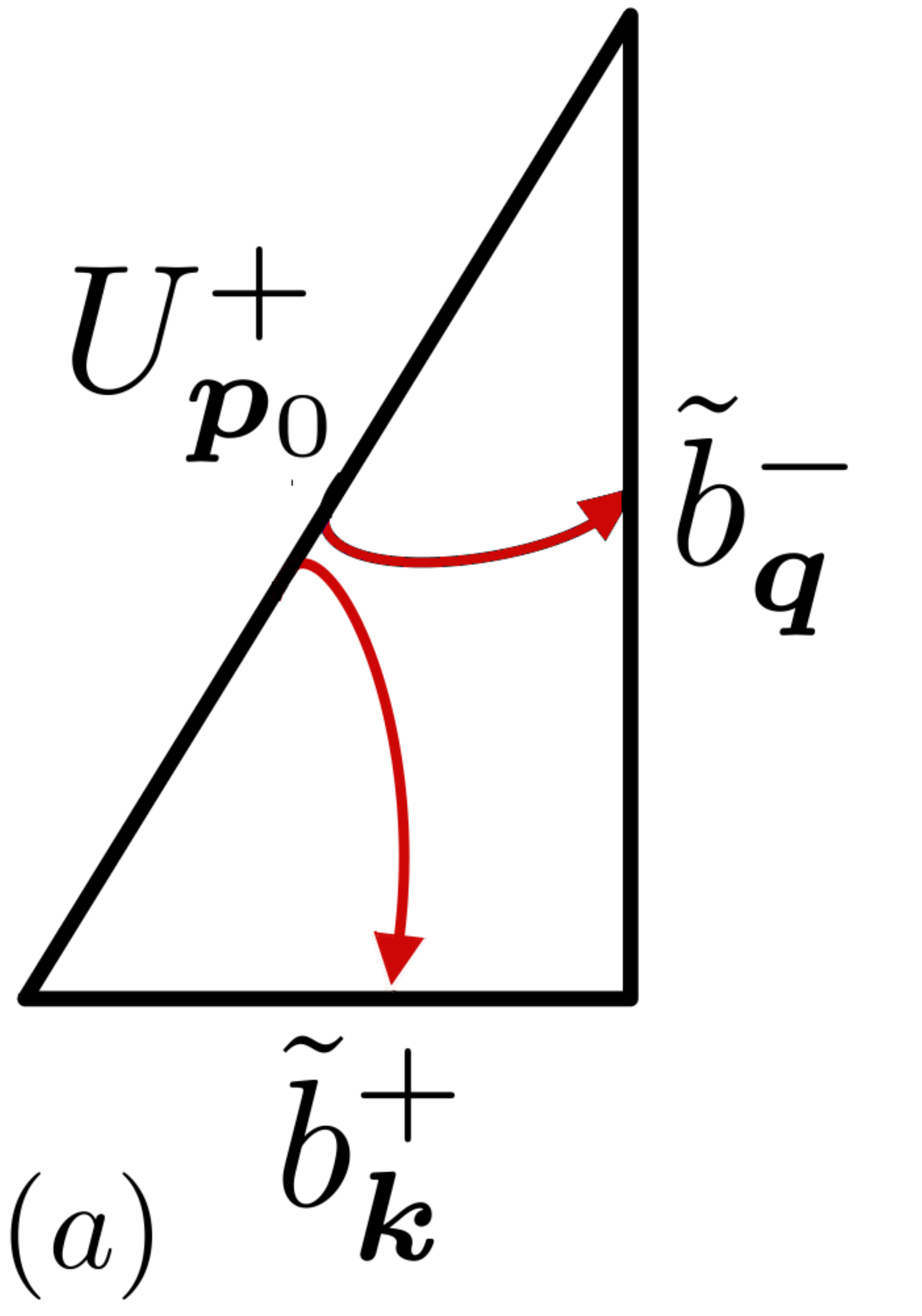} 
\includegraphics[scale=0.25]{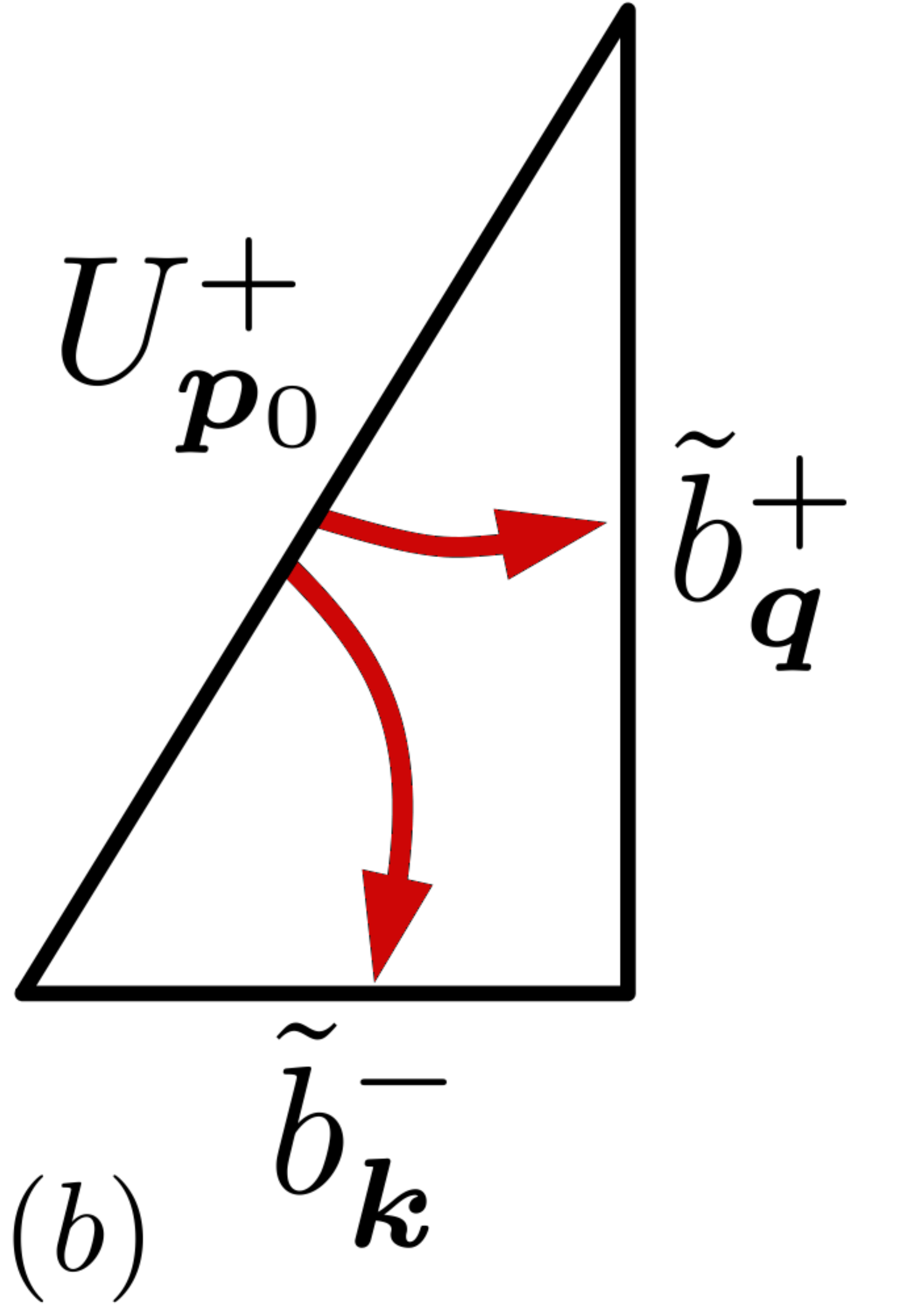} 
\caption{Summary of large-scale dynamo processes $D2: \ \Upp \xrightarrow{\tbqm} \tbkp$ (a) and $ D3: \ \Upp \xrightarrow{\tbqp} \tbkm$ (b). 
The thickness of the arrows indicates
the qualitative value of the growth rate.}
\label{fig:ls-kinematic-triads}
\end{figure}

In order to study large-scale kinematic dynamo action restricted to each one of the 32 MTI system separately, 
we start from a fully helical mechanical equilibrium at the smallest scale, that is subject to magnetic perturbation at larger scales. 
Following the notation introduced in the previous section, this corresponds to an equilibrium at wavevector $\bp_0$ 
given by $(\Bpp=0,\Bpm =0, \Upp \neq 0, \Upm =0)$ that is subject to magnetic perturbations $\tbkp$, $\tbkm$, $\tbqp$ and $\tbqm$,
where $p_0 > q > k$. 
Given the particular equilibrium, one can show from Equation \eqref{eq:basic-triads} 
that it is sufficient to consider 4 classes only: as a result of the choice of the equilibrium,
the 32 MTI systems reduce to 16, which is further reduced to 4 as we only consider the 
evolution of the magnetic perturbations. Thus the only possibilities are given by all 
combinations  $(\sigma_k, s_{p_0}=+, \sigma_q)$, and we recover the set of MTIs
connected to kinematic dynamo action by \citet{Linkmann16}. 
Note that the distinction between homo- and heterochiral MTI systems is not relevant here, 
as only magnetic perturbations to a mechanical equilibrium are considered. 
Even if mechanical perturbations were considered, the evolution of the magnetic perturbations 
does not depend on the helicity content of the mechanical perturbations as they do not 
couple. In the following, for the sake of completeness 
 we first recapitulate the known results on triadic dynamo action 
\citep{Linkmann16} and subsequently extend the analysis in order to supply qualitative 
predictions concerning the growth rates.   
The evolution equations of the perturbations for each of the 4 individual 
MTI systems are \\
\begin{minipage}[c]{.25\textwidth}
  \centering
\includegraphics[scale=0.1]{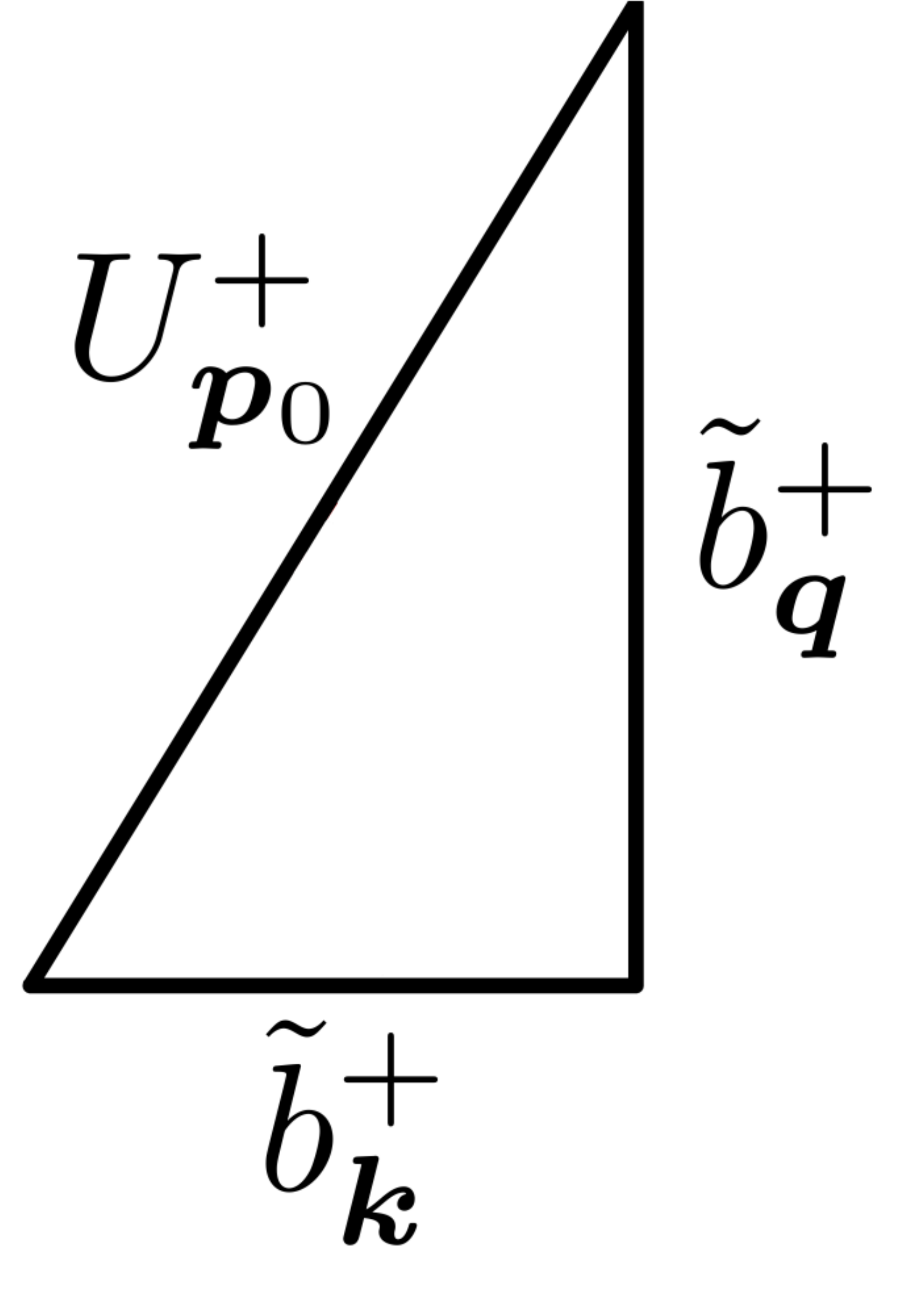} \\ 
\includegraphics[scale=0.1]{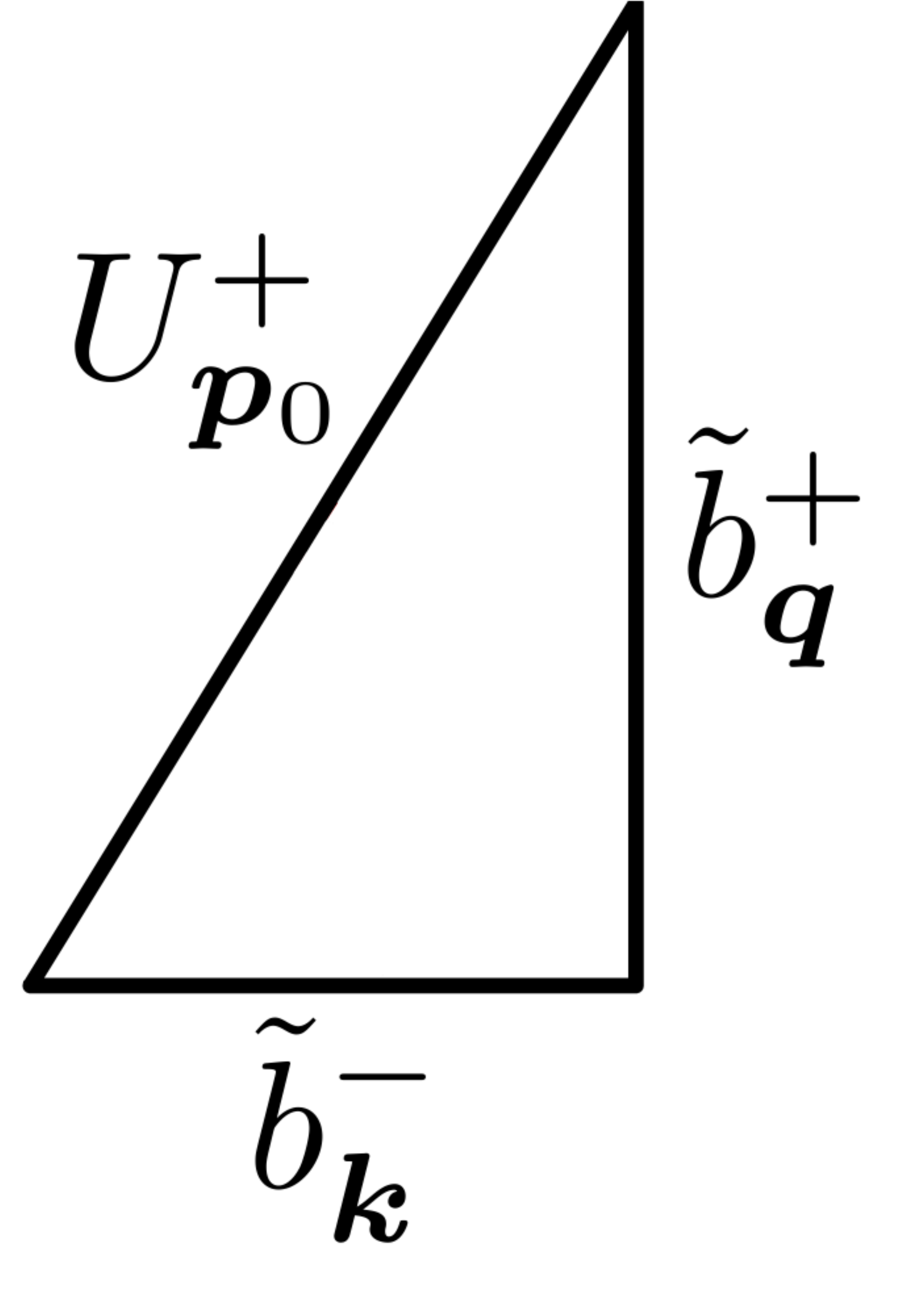} \\ 
\includegraphics[scale=0.1]{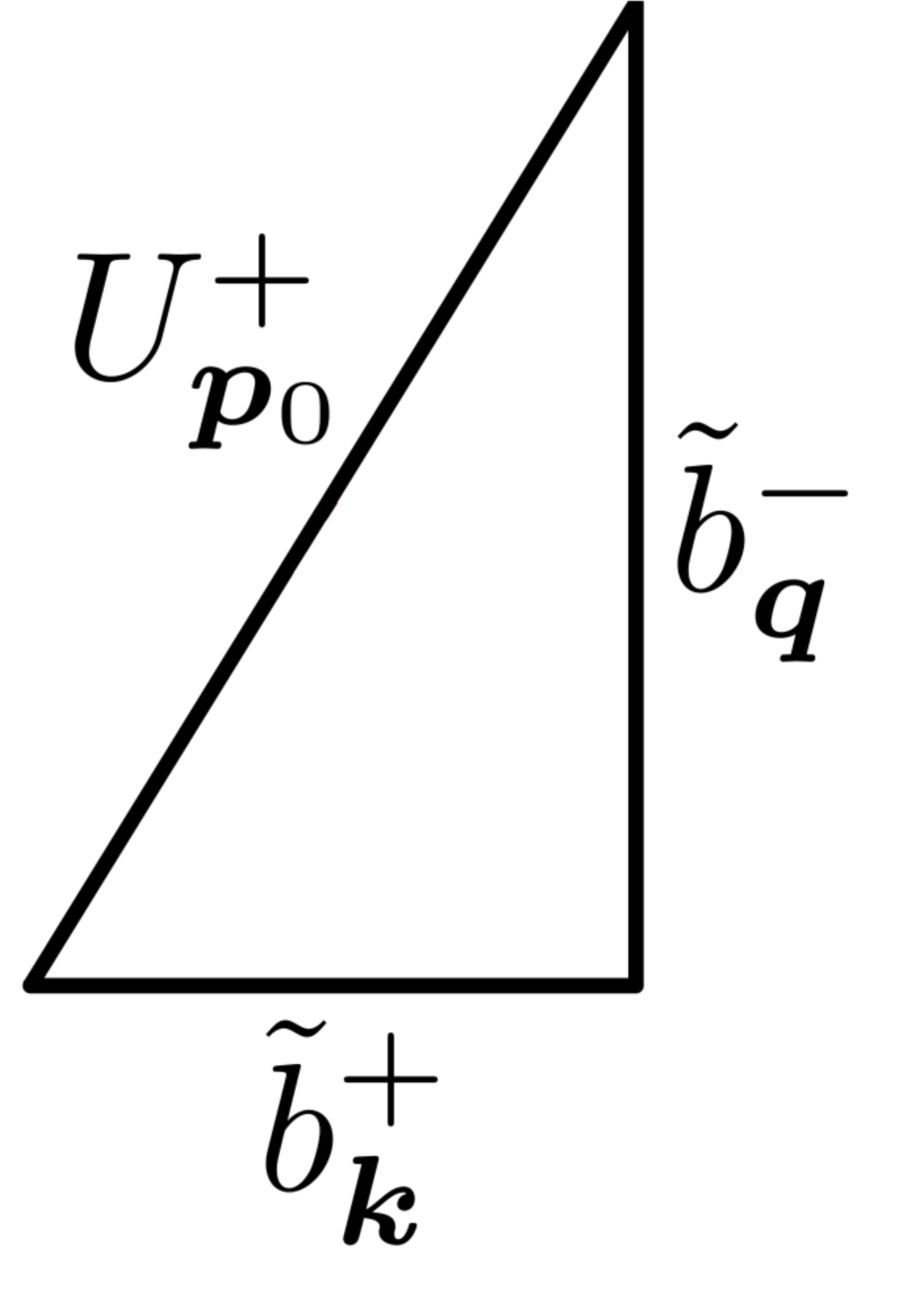} \\ 
\includegraphics[scale=0.1]{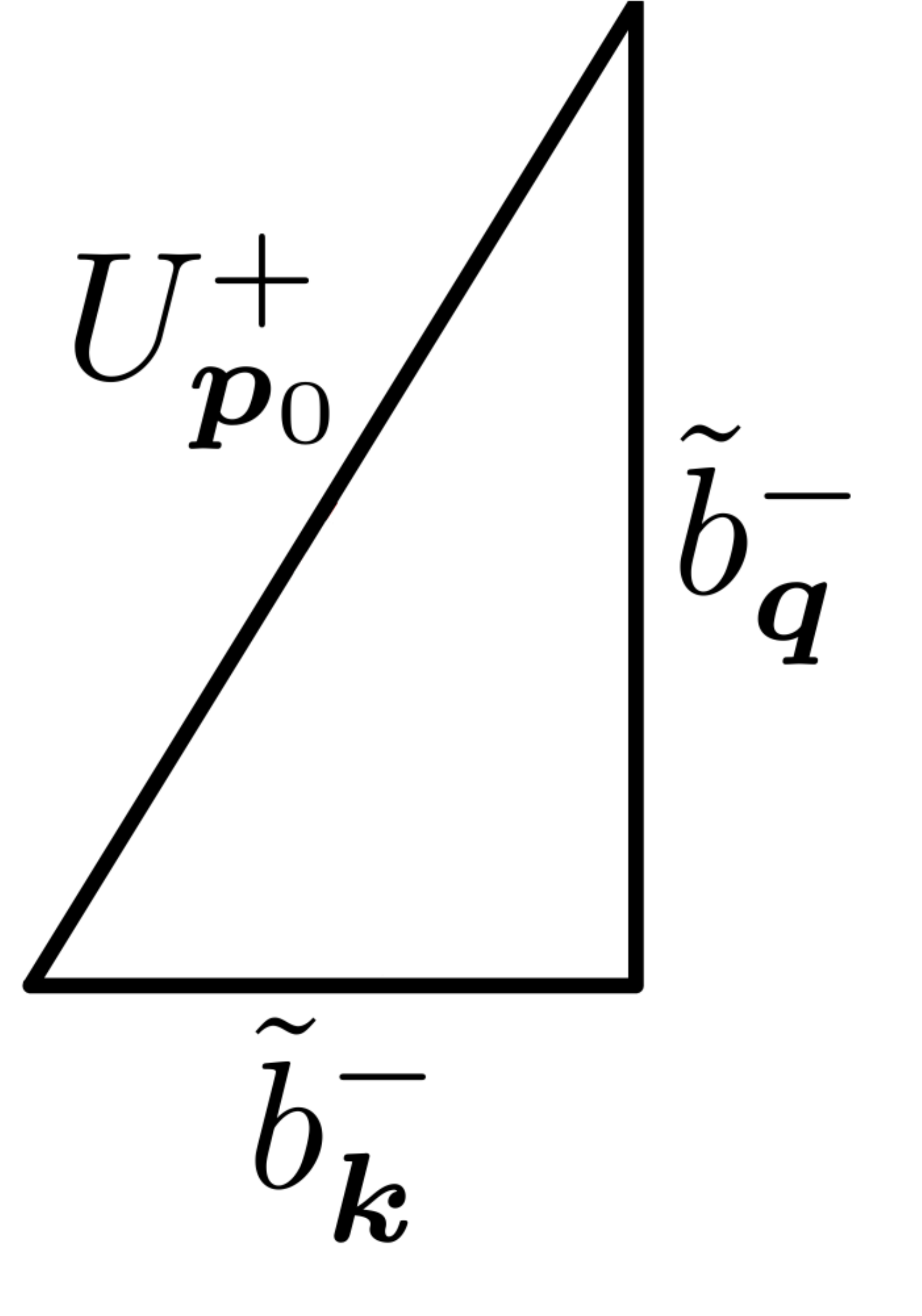} \\ 
\end{minipage}
\begin{minipage}[c]{.75\textwidth}
\begin{align}
& D1\qquad 
\begin{cases}
\dt |\tbkp|^2 = -k\: g^{M2}_{+++} \: \tbkp \Upp \tbqp  + \mbox{c.~c.} \ ,  \\
\dt |\tbqp|^2 = q\: g^{M1}_{+++} \: \tbqp \tbkp \Upp  + \mbox{c.~c.} \ ,   
\end{cases} \\
\nonumber \\
\nonumber \\
& D2\qquad 
\begin{cases}
\dt |\tbkp|^2 = -k\: g^{M2}_{++-}  \:\tbkp \Upp \tbqm  + \mbox{c.~c.} \ , \\ 
\dt |\tbqm|^2 = -q\: g^{M1}_{++-}  \:\tbqm \tbkp \Upp  + \mbox{c.~c.} \ ,  
\end{cases} \\ 
\nonumber \\
\nonumber \\
& D3\qquad 
\begin{cases}
\dt |\tbkm|^2 = k\: g^{M2}_{-++} \: \tbkm \Upp \tbqp + \mbox{c.~c.} \ ,  \\
\dt |\tbqp|^2 = q\: g^{M1}_{-++} \: \tbqp \tbkm \Upp + \mbox{c.~c.} \ ,  
\end{cases} \\
\nonumber \\
\nonumber \\
& D4\qquad 
\begin{cases}
\dt |\tbkm|^2 = k\: g^{M2}_{-+-} \: \tbkm \Upp  \tbqm + \mbox{c.~c.} \ ,  \\
\dt |\tbqm|^2 = -q\: g^{M1}_{-+-} \: \tbqm \tbkm \Upp+ \mbox{c.~c.} \ ,  \
\end{cases}
\end{align}
\end{minipage}
where c.~c.~denotes the complex conjugate, and we have written the evolution
equations in terms of the magnetic energy in order to highlight the structure of 
the different helical interactions D1-D4.
The second-order equations governing magnetic field growth 
at $\bk$ and $\bq$ become for this particular equilibrium case 
\begin{align} 
\label{eq:bpluskin1-ls} 
& D1\qquad 
\begin{cases}
\partial_t^2 \tbkp = -g^{M2}_{+++}\:g^{M1*}_{+++}\: kq |\Upp|^2  \tbkp \ , \\ 
\partial_t^2 \tbqp = -g^{M1}_{++-}\:g^{M2*}_{++-}\: kq |\Upp|^2  \tbqp \ ,  
\end{cases} \\
\nonumber \\
\label{eq:bpluskin2-ls} 
& D2\qquad 
\begin{cases}
\partial_t^2 \tbkp = g^{M2}_{++-}\:g^{M1*}_{++-}\: kq |\Upp|^2  \tbkp \qquad  \qquad \Upp \xrightarrow{\tbqm} \tbkp \ , \\ 
\partial_t^2 \tbqm = g^{M1}_{++-}\:g^{M2*}_{++-}\: kq |\Upp|^2  \tbqm \qquad \qquad  \Upp \xrightarrow{\tbkp} \tbqm \ , 
\end{cases} \\
\nonumber \\
\label{eq:bminuskin1-ls} 
& D3\qquad 
\begin{cases}
\partial_t^2 \tbkm = g^{M2}_{-++}\:g^{M1*}_{-++}\: kq |\Upp|^2 \tbkm \qquad \qquad  \Upp \xrightarrow{\tbqp} \tbkm \ , \\
\partial_t^2 \tbqp = g^{M1}_{-++}\:g^{M2*}_{-++}\: kq |\Upp|^2 \tbqp \qquad \qquad  \Upp \xrightarrow{\tbkm} \tbqp \ , 
\end{cases} \\
\nonumber \\
\label{eq:bminuskin2-ls} 
& D4\qquad 
\begin{cases}
\partial_t^2 \tbkm = -g^{M2}_{-+-}\:g^{M1*}_{-+-}\: kq |\Upp|^2 \tbkm \ , \\ 
\partial_t^2 \tbqm = -g^{M1}_{-+-}\:g^{M2*}_{-+-}\: kq |\Upp|^2 \tbqm \ .
\end{cases}
\end{align}
From the definition of the coupling coefficients 
 given by Equation \eqref{eqapp:gfactor} 
and from the structure of the products shown 
in Equation \ref{eqapp:gsquare} in Appendix \ref{app:stability} it is immediate to realize that
the prefactors in front of the fluctuating fields
in Equations \eqref{eq:bpluskin1-ls}-\eqref{eq:bminuskin2-ls}
are positive for classes D2 and D3 while they are 
negative for classes D1 and D4. 
Triadic dynamo action is therefore described by processes D2 and D3.
The arrows in Equations \eqref{eq:bpluskin2-ls} and \eqref{eq:bminuskin1-ls}  
indicate the transfer direction associated with the two fields
and the superscripts indicate the `catalyzer' mode.  
In both processes the magnetic perturbations are of opposite helicity, 
therefore we state the first observation: \\
(i) {\it  Magnetic field 
perturbations of mutually opposite helicity are necessary to enable dynamo action} \citep{Linkmann16}.
 \\ \noindent
Having summarised the known results, we now proceed to a qualitative analysis of the 
growth rates. 
It is important to stress that at the largest scale, $1/k$, the magnetic fluctuation with 
 helicity opposite to the one of the  stretching velocity field  grows faster, because the
ratio between the growth rates for processes D2 and D3 is given by
\be 
\label{eq:ls-ratio}
\left(\frac{g^{M2}_{++-}\:g^{M1*}_{++-}}{g^{M2}_{-++}\:g^{M1*}_{-++}}\right)^{1/2}= \frac{|k+p_0-q|}{|-k+p_0+q|} < 1 \ , 
\ee
as can be seen from Equation \eqref{eqapp:gorder-ls} in Appendix \ref{app:stability}. 
Therefore this is leading to an  $\alpha$-like 
dynamo process, and we can make the second important observation: \\
(ii) {\it The main instability leading to large-scale dynamo action is of $\alpha$-type}. \\ \noindent 
Moreover, 
we notice that the growth becomes
more and more of $\alpha$-type if the geometry of the triad is 
strongly non-local, i.e. $k << p_0 \simeq q$, 
because then $(g^{M2}_{++-}\:g^{M1*}_{++-}/g^{M2}_{-++}\:g^{M1*}_{-++})^{1/2} \to 0$, as can be seen from Equation \eqref{eq:ls-ratio}.   
Our predictions for the large-scale dynamo are summarized in 
Figure \ref{fig:ls-kinematic-triads}.

\subsection{Small scale kinematic dynamo \label{sec:ssdynamo}}
\begin{figure}[h]
\center
\includegraphics[scale=0.25]{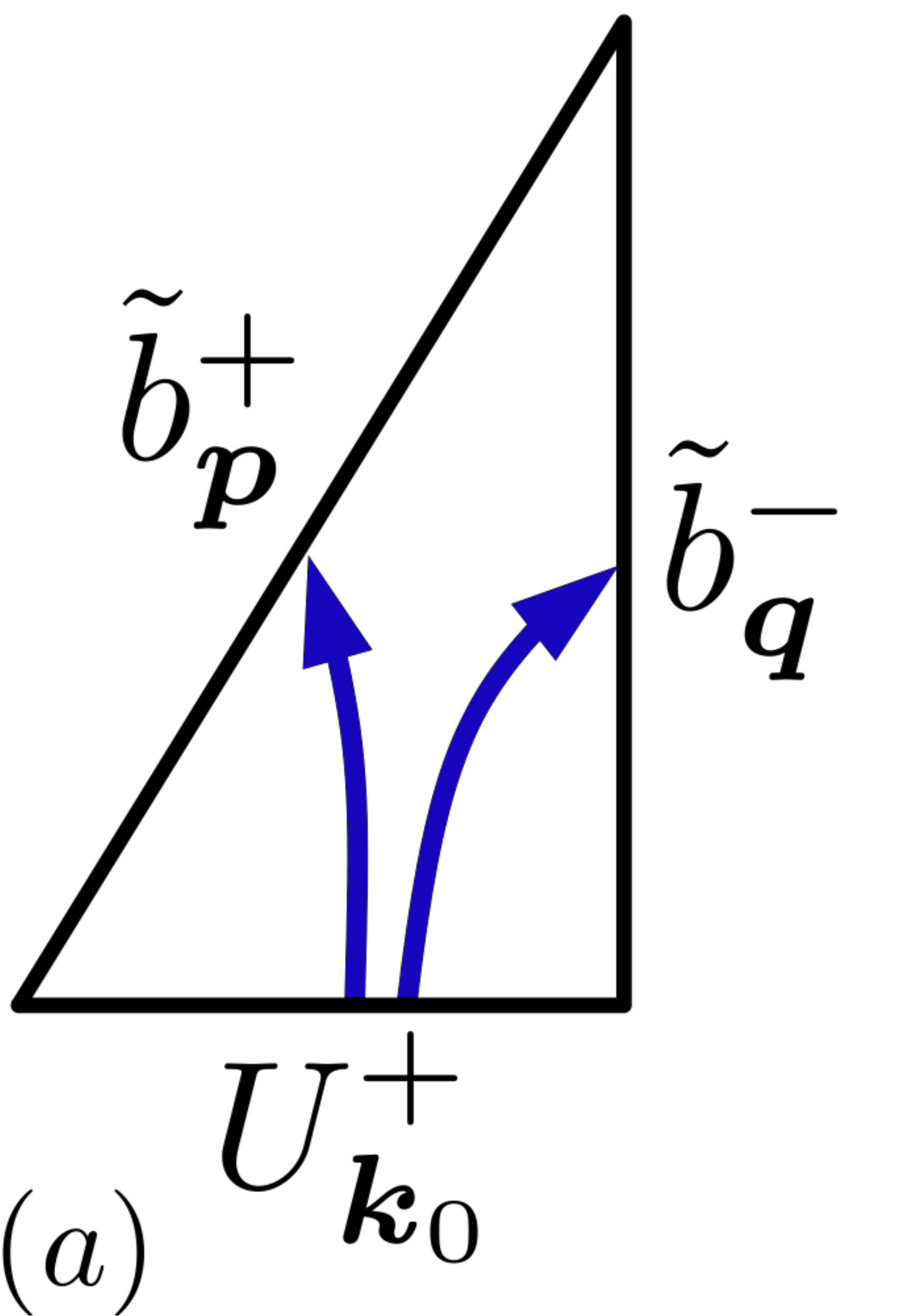} 
\includegraphics[scale=0.25]{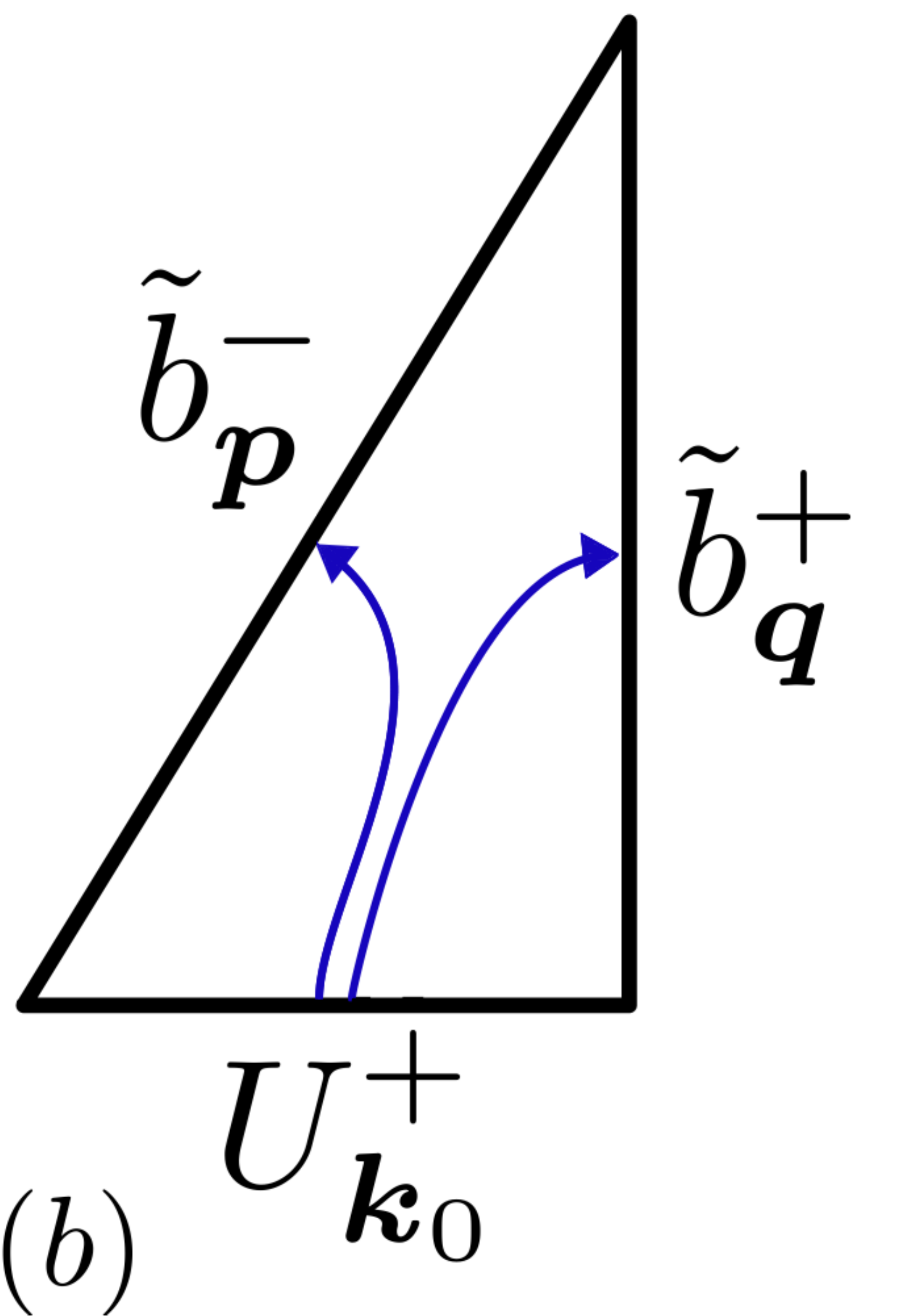} 
\caption{Summary of small-scale dynamo processes $D2: \ \Ukp \xrightarrow{\tbqm} \tbpp$ (a) and $D3: \ \Ukp \xrightarrow{\tbqp} \tbpm$ (b).  
The thickness of the arrows indicates the qualitative value of the growth rate.}
\label{fig:ss-kinematic-triads}
\end{figure}

We now consider a mechanical equilibrium $(\Bkp=0,\Bkm =0, \Ukp \neq 0, \Ukm)$ at the largest scale, 
which corresponds to the smallest wavenumber $k_0$ and study magnetic perturbations
$\tbpp$, $\tbpm$, $\tbqp$ and $\tbqm$ at smaller scales, i.e. at larger wavenumbers with $p > q > k_0$.
The evolution of the magnetic energy for each of the four individual MTI systems is 
\begin{align}
& D1\qquad 
\begin{cases}
\label{eq:basic-triads-ssdynamo1}
\dt |\tbpp|^2& = -p \:g^{M1}_{+++} \: \tbpp \tbqp \Ukp  + \mbox{c.~c.} \ ,  \\
\dt |\tbqp|^2& = q \:g^{M2}_{+++} \: \tbqp \Ukp \tbpp  + \mbox{c.~c.} \ ,  
\end{cases} \\
\nonumber \\
\label{eq:basic-triads-ssdynamo2}
& D2\qquad 
\begin{cases} 
\dt |\tbpp|^2& = -p \: g^{M1}_{++-}  \:\tbpp \tbqm \Ukp  + \mbox{c.~c.} \ ,  \\
\dt |\tbqm|^2& = -q \: g^{M2}_{++-}  \:\tbqm \Ukp \tbqm  + \mbox{c.~c.} \ ,  
\end{cases} \\
\nonumber \\
\label{eq:basic-triads-ssdynamo3}
& D3\qquad 
\begin{cases}
\dt |\tbpm|^2& = p \: g^{M1}_{+-+} \: \tbpm \tbqp \Ukp + \mbox{c.~c.} \ ,  \\
\dt |\tbqp|^2& = q \: g^{M2}_{+-+} \: \tbqp \Ukp \tbpm + \mbox{c.~c.} \ ,  
\end{cases} \\
\nonumber \\
\label{eq:basic-triads-ssdynamo4}
& D4\qquad 
\begin{cases}
\dt |\tbpm|^2& = p \: g^{M1}_{+--} \: \tbpm  \tbqm \Ukp + \mbox{c.~c.} \ ,  \\
\dt |\tbqm|^2& = -q \: g^{M2}_{+--} \: \tbqm \Ukp  \tbpm + \mbox{c.~c.} \ , 
\end{cases}
\end{align}
from which we derive the second-order equations
\begin{align} \label{eq:bpluskin1-ss} 
D1\qquad & 
\begin{cases}
\partial_t^2 \tbpp& = - g^{M1}_{+++}\:g^{M2*}_{+++}\: pq |\Ukp|^2  \tbpp \ , \\ 
\partial_t^2 \tbqp& = - g^{M2}_{+++}\:g^{M1*}_{+++} \:pq |\Ukp|^2  \tbqp \ ,  
\end{cases} \\
\nonumber \\
\label{eq:bpluskin2-ss} 
D2\qquad & 
\begin{cases}
\partial_t^2 \tbpp& =  g^{M1}_{++-}\:g^{M2*}_{++-}\: pq |\Ukp|^2  \tbpp \qquad \qquad \Ukp \xrightarrow{\tbqm} \tbpp \ , \\ 
\partial_t^2 \tbqm& =  g^{M2}_{++-}\:g^{M1*}_{++-}\: pq |\Ukp|^2  \tbqm \qquad \qquad \Ukp \xrightarrow{\tbpp} \tbqm \ , 
\end{cases} \\
\nonumber \\
\label{eq:bminuskin1-ss} 
D3\qquad & 
\begin{cases}
\partial_t^2 \tbpm &= g^{M1}_{+-+}\:g^{M2*}_{+-+}\: pq |\Ukp|^2 \tbpm \qquad \qquad \Ukp \xrightarrow{\tbqp} \tbpm \ , \\
\partial_t^2 \tbqp &=  g^{M2}_{+-+}\:g^{M1*}_{+-+}\: pq |\Ukp|^2 \tbqp \qquad \qquad \Ukp \xrightarrow{\tbpm} \tbqp \ , 
\end{cases} \\
\nonumber \\
\label{eq:bminuskin2-ss} 
D4\qquad & 
\begin{cases}
\partial_t^2 \tbpm &= - g^{M1}_{+--}\:g^{M2*}_{+--} \:pq |\Ukp|^2 \tbpm \ , \\ 
\partial_t^2 \tbqm &= - g^{M2}_{+--}\:g^{M1*}_{+--}\: pq |\Ukp|^2 \tbqm \ . 
\end{cases}
\end{align}
Similar to the analysis of large-scale dynamo action, the prefactors in front 
of the fluctuations are positive for classes D2 and D3 while they are
negative for classes D1 and D4. The growth of the small-scale magnetic 
field that is due to a large-scale velocity field is 
therefore also described by processes D2 and D3.
However, we observe an important difference compared to the large-scale dynamo: 
Now, at the largest wavenumber, $p$,   the magnetic fluctuation with the 
{\it same} helicity of the stretching velocity field has the larger growth rate, because 
\be
\label{eq:ss-ratio}
\left(\frac{g^{M1}_{++-}\:g^{M2*}_{++-}}{g^{M1}_{+-+}\:g^{M2*}_{+-+}}\right)^{1/2}= \frac{|k_0+p-q|}{|k_0-p+q|} > 1 \ ,
\ee
as can be seen from Equation \eqref{eqapp:gorder-ss} in Appendix \ref{app:stability}.
Therefore we expect the opposite helical sigature compared to the large-scale dynamo, that is to say, \\
(iii) {\it  the growth of the magnetic field at scales smaller than the characteristic scale of the flow 
will mainly have the same helicity as the flow.} \\
By comparison of the ratio between the two growth rates for
processes D2 and D3 we also observe that the small-scale dynamo 
operates more locally, because the growth rates diminish in both cases 
if the geometry of the triad is strongly nonlocal, i.e. 
for $ k_0<< p \simeq q$ 
we obtain $(g^{M1}_{++-}\:g^{M2*}_{++-})^{1/2} \to 0$ and 
$(g^{M1}_{+-+}\:g^{M2*}_{+-+})^{1/2} \to 0$ as can be seen from 
Equation \eqref{eq:ss-ratio}.  Our predictions for the small-scale dynamo are summarized in
Figure \ref{fig:ss-kinematic-triads}. \\

\noindent In conclusion,  there are four possible classes of 
triad-by-triad dynamo action, 
out of which two classes correspond to large-scale 
dynamo action, and two classes to small-scale dynamo action. 
For the large-scale dynamo one class has the same helical signature as the 
$\alpha$-effect \citep{Linkmann16}, this is class D3 given in Equation \eqref{eq:bminuskin1-ls} 
and depicted in Figure \ref{fig:ls-kinematic-triads} (b).
The other class, D2, which is described in Equation \eqref{eq:bpluskin2-ls} and 
depicted in Figure \ref{fig:ls-kinematic-triads} (a), has
the opposite helical signature. 
Most importantly, the $\alpha$-like dynamo has the higher growth rate. 
Conversely, the two classes corresponding to the small-scale dynamo are shown in Figures \ref{fig:ss-kinematic-triads}(a-b) 
and described in Equations \eqref{eq:bpluskin2-ss} and \eqref{eq:bminuskin1-ss}. At scales smaller than the 
characteristic scale of the mechanical equilibrium, class D2  
mainly amplifies magnetic field perturbations with the same sign of helicity as the flow, 
while class D3 amplifies magnetic field modes with the opposite sign of helicity as the flow, and 
we found that class D2 has the higher growth rate compared to D3. 
The combination of the two classes of dynamo action with the higher growth rate, i.e.     
the combination of the $\alpha$-like large-scale D3 with the small-scale D2, produces a
helical signature consistent with the Stretch-Twist-Fold  (STF) mechanism \citep{Vainshtein72,Childress95,Mininni11} of dynamo action in a 
positively helical flow: Negative magnetic helicity is generated 
at the large scales, while positive magnetic helicity is generated 
at the small scales. The main result from the theoretical analysis of 
triad-by-triad dynamo action can therefore be summarized in the following statement: \\ 
(iv) {\em The dominant linear instabilities present in the basic triadic dynamics lead to dynamo action 
with the same helical signature as the STF-mechanism.
  } 
We point out that since in full MHD all modes interact, it yet remains 
to be seen to what extent the MTI dynamos D2 and D3 
are representative of the 
full dynamics. We will address this point in Section \ref{sec:linear}. 
\\

\subsection{Inverse cascade of magnetic helicity} \label{sec:selfinteraction}
In a strongly magnetized flow the magnetic field interacts with itself due to its back-reaction on the flow by 
the Lorentz force. 
In order to assess this interaction, we consider a positively helical magnetic 
equilibrium $(\Bpp \ne 0, \Bpm =\Upm =\Upp= 0)$ that is subject to magnetic
and mechanical perturbations. 
Unlike for the kinematic dynamo, now the magnetic and mechanical
perturbations couple, i.e., the inverse cascade and the effect of the Lorentz
force are described by the same MTI system 
coupling the two effects.
The latter implies that the
distinction between homo- and heterochiral MTI systems is relevant if
both processes are to be analyzed. 
However, concerning the {\em existence} of linear
instabilities, the analysis can be simplified by restricting the focus on the
inverse cascade and leaving the effect of the Lorentz force on the flow aside.
Then it suffices to consider only one type of interaction 
by fixing the magnetic and mechanical perturbations at 
a given wavenumber each, and it is not necessary to further distinguish between homo- and
heterochiral systems. 
For simplicity we only consider magnetic perturbations at the 
smallest wavenumber $k$ in the triad and mechanical perturbations 
at the intermediate wavenumber $q$.  
The analysis of the 32 possible MTIs can then be reduced to 
the study of 4 individual MTIs by a similar argument as applied to the 
kinematic dynamo.
This allows us to briefly summarise the known results on the inverse 
cascade \citep{Linkmann16} and to outline the main points.  
The evolution equations 
for the magnetic and mechanical 
perturbations $\tbkp$, $\tbkm$, $\tuqp$ and $\tuqm$ are
\begin{align}
\label{eq:basic-triads-SI1}
IC1 \qquad &
\begin{cases}
\dt |\tbkp|^2 = k g^{M1}_{+++} \:\tbkp \Bpp \tuqp  + \mbox{c.~c.} \ ,  \\
\dt |\tuqp|^2 =  -g^{LF}_{+++} \:(k-p_0)\:\tuqp \Bpp \tbkp  + \mbox{c.~c.} \ ,  
\end{cases} \\
\nonumber \\
IC2 \qquad &
\label{eq:basic-triads-SI2}
\begin{cases}
\dt |\tbkp|^2 = k g^{M1}_{++-}\:\tbkp \Bpp \tuqm  + \mbox{c.~c.} \ ,  \\
\dt |\tuqm|^2 =  -g^{LF}_{++-}\:(k-p_0)\:\tuqm \Bpp \tbkp  + \mbox{c.~c.} \ , 
\end{cases} \\
\nonumber \\
IC3 \qquad &
\label{eq:basic-triads-SI3}
\begin{cases}
\dt |\tbkm|^2 = -k g^{M1}_{-++} \: \tbkm \Bpp \tuqp + \mbox{c.~c.} \ ,  \\
\dt |\tuqp|^2 = -g^{LF}_{-++}\: (-k-p_0) \: \tuqp \Bpp \tbkm + \mbox{c.~c.} \ ,  
\end{cases} \\
\nonumber \\
IC4 \qquad &
\label{eq:basic-triads-SI4}
\begin{cases}
\dt |\tbkm|^2 = -k g^{M1}_{-+-} \: \tbkm \Bpp  \tuqm + \mbox{c.~c.} \ , \\ 
\dt |\tuqm|^2 = - g^{LF}_{-+-} \: (-k-p_0)\: \tuqm \Bpp  \tbkm + \mbox{c.~c.} \ ,  
\end{cases} 
\end{align}
where again each equation describes the dynamics of one particular 
MTI system. 
Since this section is concerned with the dynamics of the magnetic 
field, we focus on the second-order evolution equations of the magnetic perturbations only:  
\begin{align}
\label{eq:bplusSI1} 
IC1 \qquad &
\partial_t^2 \tbkp  = -g^{M1}_{+++}\:g^{LF*}_{+++}\: k(k-p_0)\:|\Bpp|^2  \tbkp  \qquad  \qquad \Bpp \xrightarrow{\tuqp} \tbkp \ , \\ 
\nonumber \\
\label{eq:bplusSI2} 
IC2 \qquad &
\partial_t^2 \tbkp  = -g^{M1}_{++-}\:g^{LF*}_{++-} \: k(k-p_0)|\Bpp|^2  \tbkp \qquad  \qquad \Bpp \xrightarrow{\tuqm} \tbkp \ , \\ 
\nonumber \\
\label{eq:bminusSI1} 
IC3 \qquad &
\partial_t^2 \tbkm  = -g^{M1}_{-++}\:g^{LF*}_{-++} \: k(k+p_0) \, |\Bpp|^2 \tbkm \ , \\ 
\nonumber \\
\label{eq:bminusSI2} 
IC4 \qquad &
\partial_t^2 \tbkm  = -g^{M1}_{-+-}\:g^{LF*}_{-+-}  \: k(k+p_0) \, |\Bpp|^2 \tbkm \ . 
\end{align}
Again we consider under which circumstances linear instabilities occur. 
The products of coupling coefficients in Equations \eqref{eq:bplusSI1}-\eqref{eq:bminusSI2} are always positive,
hence we observe that the prefactors on RHS of the evolution equations for the negatively 
helical magnetic perturbations (Equation \eqref{eq:bminusSI1} and Equation \eqref{eq:bminusSI2}) are always negative, 
while the corresponding prefactors in the evolution equations for the positively helical magnetic 
perturbations (Equation \eqref{eq:bplusSI1} and Equation \eqref{eq:bplusSI2})  
are positive because we chose $k < p_0$. We immediately see that

\noindent (v) 
{\it a positively helical magnetic equilibrium at a given scale 
can only be unstable with respect to positively helical 
magnetic perturbations at larger scales} \citep{Linkmann16}. 

\noindent Since the magnetic perturbation and the equilibrium are of like-signed helicity, this linear instability can be identified 
with the inverse cascade of magnetic helicity, 
by which magnetic helicity of one sign is transported from 
smaller to larger scales.

We point out that the terms on the RHS of Equations \eqref{eq:bplusSI1}-\eqref{eq:bminusSI2} 
are essentially nonlinear, because their derivation
required the coupling of the momentum and induction equations through the
Lorentz force, as can be seen by the occurrence of the Lorentz coupling
factors $g^{LF}$ in Equations \eqref{eq:bplusSI1}-\eqref{eq:bminusSI2}. The 
coupling between the momentum and induction equations, which
is the only nonlinear contribution to the evolution of the magnetic field,
is therefore still present.
Equations \eqref{eq:bplusSI1} and \eqref{eq:bplusSI2} thus describe a nonlinear
contribution to the interscale energy transfer because  $\vec{b}$ acts on
$\vec{u}$, which is acting back on $\vec{b}$: $\bb \to \bu \to \bb$.  That is,
care has to be taken in the graphical representation and the physical
interpretation of Equations \eqref{eq:bplusSI1} and \eqref{eq:bplusSI2} as it may be
tempting to associate linear instabilities with purely magnetic energy transfer
due to advection by the velocity field: $\bb \to \bb$.

There are two classes of MTIs that have a linear instability associated with the inverse cascade 
of magnetic helicity, IC1 and IC2, where the IC1 describes the evolution of a positively magnetic field 
in a positively helical flow because the catalyzer mode $\uqp$ is positively helical, while 
the IC2 describes the evolution of a positively helical magnetic field in a 
negatively helical flow. By comparison of the respective growth rates, 
we can assess in which situation the inverse cascade of magnetic helicity is most efficient.
According to the discussion in Appendix \ref{app:stability}, we obtain 
\be
\label{eq:ic-ratio}
\left(\frac{g^{M1}_{++-}\:g^{LF*}_{++-}}{g^{M1}_{+++}\:g^{LF*}_{+++}}\right)^{1/2}= \frac{|k+p_0-q|}{|k+p_0+q|} < 1.
\ee
that is, IC1 leads to a larger
growth rate for the large-scale magnetic perturbation than IC2.
Therefore we conclude that 

\noindent (vi) {\it  the inverse cascade of magnetic helicity 
is more efficient in a helical flow where magnetic and kinetic helicity are of the same sign 
than in a helical flow where magnetic and kinetic helicity are of opposite sign.} 

\begin{figure}[h]
\center
\includegraphics[scale=0.25]{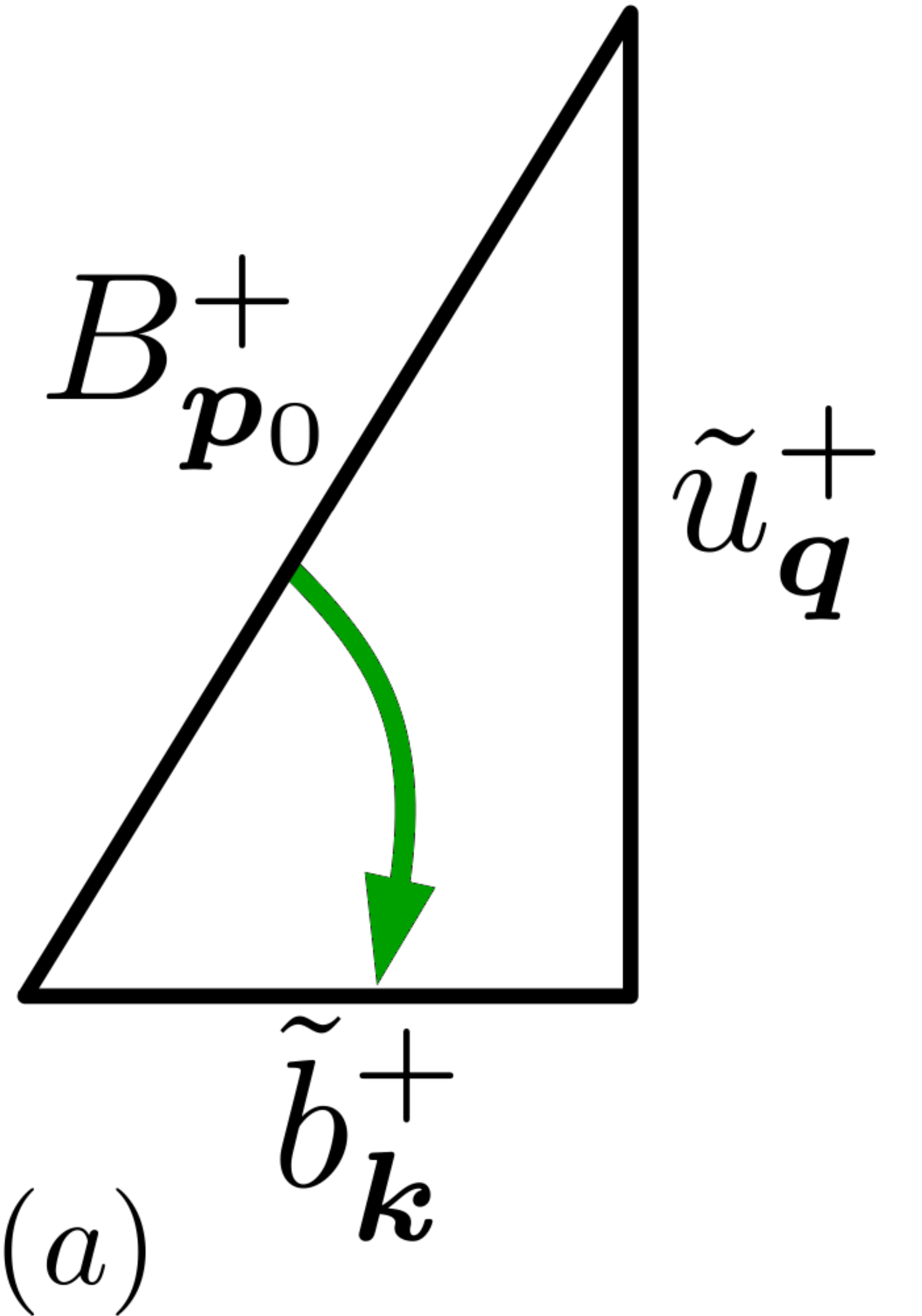} 
\includegraphics[scale=0.25]{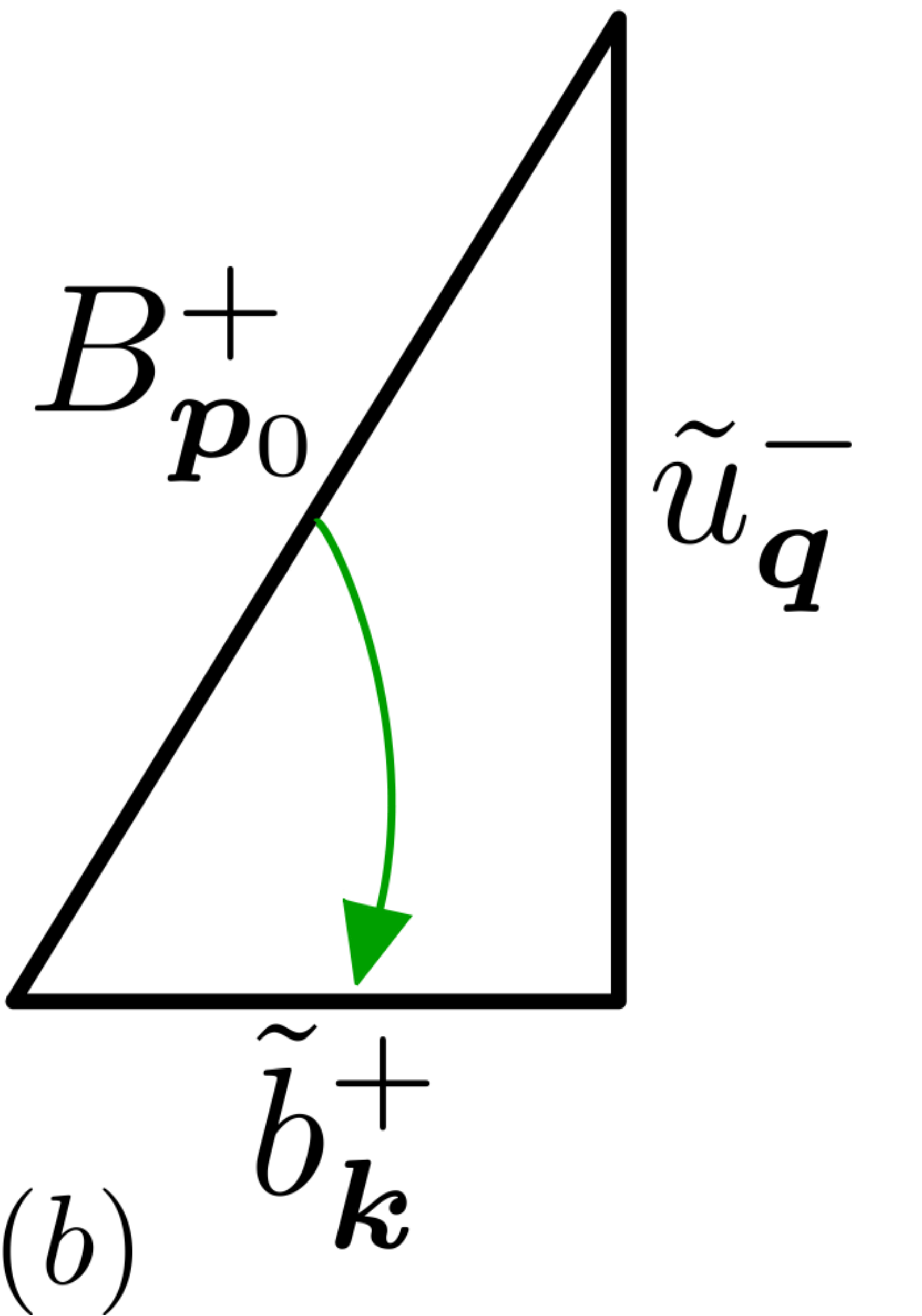} 
\caption{Summary of inverse transfer processes $IC1: \Bpp \xrightarrow{\tuqp} \tbkp$ (a) and $IC2: \Bpp \xrightarrow{\tuqm} \tbkp$ (b).
The thickness of the arrows indicates the qualitative difference in the intensity of the respective energy transfers.}
\label{fig:IC-triads}
\end{figure}

\noindent These results are summarized in Figure \ref{fig:IC-triads}.
Furthermore, we note that IC2-transfers are more local than IC1-transfers because for 
strongly nonlocal triads $k<<p_0\simeq q$ 
the combination of coupling factors 
$g^{M1}_{+++}\:g^{LF*}_{+++}$ corresponding to IC2 tends to zero, as can be seen from Equation \eqref{eq:ic-ratio}.
\\

\section{Numerical results} \label{sec:simulations} 
The analysis carried out in the previous sections is based on a simplified single-triad
dynamics, which cannot be the end of the story \citep{Moffatt14}. During the non-linear evolution
of the whole set of triads in the full MHD equations different instabilities
that are due to different triad shapes and helical content of the corresponding MTIs will superpose and
interact. Therefore, it is not trivial to predict the behavior of the full
system and the typical transfer that will dominate in the fully coupled MHD regime.
A paradigmatic example is given by the case of the Navier-Stokes equations in
absence of the magnetic field \citep{Biferale12,Biferale13,Biferale13a}. 
There, we know that one set of MTIs (formed by velocity fields with the same
helicity sign) is characterized by an inverse energy cascade. 
Once all of them are coupled together by the whole Navier-Stokes equation we typically have a
forward energy transfer, albeit a switch to an inverse cascade
is observed in presence of strong rotation \citep{Smith96,Mininni09a} or in shallow 
fluid layers \citep{Nastrom84,Lautenschlager98,Smith99,Celani10,Xia11}.\\
Concerning the full MHD case of interest here,
 we carried out a series of DNSs of Equations (\ref{eq:Fmomentum}) and (\ref{eq:Finduction}) using a
fully de-aliased pseudospectral code and up to  $512^3$ collocation points in a triply
periodic domain of size $L=2\pi$. We stir the velocity field with a random Gaussian forcing,
\[\langle \vec{f}_u(\bk,t) \vec{f}_u^*(\bq,t') \rangle = f_u \delta(\bk-\bq) \delta(t-t')
\hat Q(\bk),\] where $\hat Q(\bk)$ is a projector assuring incompressibility
and $f_u$ is nonzero in a given band of Fourier modes (concentrated either at large or at small scales)
 $ k_u^f \in [k_{\rm min}:k_{\rm max}]$. Moreover, by  decomposing the forcing in its helical modes,
$ \vec{f}_u = \vec{f}^+_u + \vec{f}^-_u$, we can further control the
 chirality of the  mechanical injection. 
The same kind of forcing is also used for the magnetic field (when applied). 
Further details concerning the implementation of the helical projection can be found in 
the papers by \citet{Biferale12,Biferale13a}.
 \\
 In order to test the predictions of the analysis presented in
Section \ref{sec:theory} and to refine the understanding of the interaction of a
magnetic field in a helical flow, we carried out two sets of numerical
experiments. \\
In the first set of simulations we studied the evolution of the
magnetic field initially seeded at small-scales and without any injection of magnetic energy, i.e. only the velocity field is forced. These
simulations  are labeled as linear in reference to the character of
the initial magnetic growth. 
We studied three different
underlying velocity configurations:  a chiral  kinetic forcing at large
scales, leading to a turbulent helical  velocity field  (R1-D); a chiral forcing at small scales,
leading to a laminar helical velocity field  (R2-D) and a chiral forcing
 at small scales combined with a  strong decimation of the velocity field 
on only positively helical modes, leading to a
 turbulent helical velocity field in an inverse-cascade regime  (R3-D).
  The latter case  allows us to assess the growth of a small-scale magnetic field 
in a strongly helical turbulent flow and to select only velocity 
modes with the same helical sign. The label (D) stands for dynamo. \\
The second set of simulations is denoted as 
nonlinear because we  add a small-scale forcing on the magnetic
component also.   In this second series of simulations we started from the velocity configuration  
 (R1-D) described above,  and  we investigated different
subcases by changing the magnetic conditions:
a magnetic forcing with the same sign of helicity as the mechanical forcing 
 (R1-IC) and with opposite sign (R2-IC).  The effects of a
small-scale magnetic forcing are also studied starting from the decimated setup
described by the class of simulations (R3-D), i.e. where the velocity
field is constrained to evolve only on one set of helical modes. 
Now we inject magnetic helicity with the same or opposite sign with
respect to the kinetic helicity (R3-IC)  and (R4-IC)
respectively. The letters (IC) stand for inverse (magnetic helicity)
cascade.

\subsection{Large- and small-scale dynamo}
\label{sec:linear}
In this section we report results from the  numerical experiments
where we let the magnetic field evolve freely in different types of flows. 
The initial conditions are always generated from a stationary
simulation of a nonconducting fluid, and the initial magnetic seeding is always at
small scales to be amplified by kinematic dynamo action. The simulations are evolved 
beyond the kinematic regime into the nonlinear dynamo regime. 
The
numerical experiments discussed in this section differ in the range of scales, 
$k_u^f$, of the applied mechanical force, $\vec{f}_u$,  and in
the projection onto different helical sectors of the velocity field.
Detailed information about this series of simulations is given in
Table~\ref{tbl:simulations-linear}.\\

\begin{figure*}[h]
\center
{\bf Magnetic field}  \\
\includegraphics[scale=0.7]{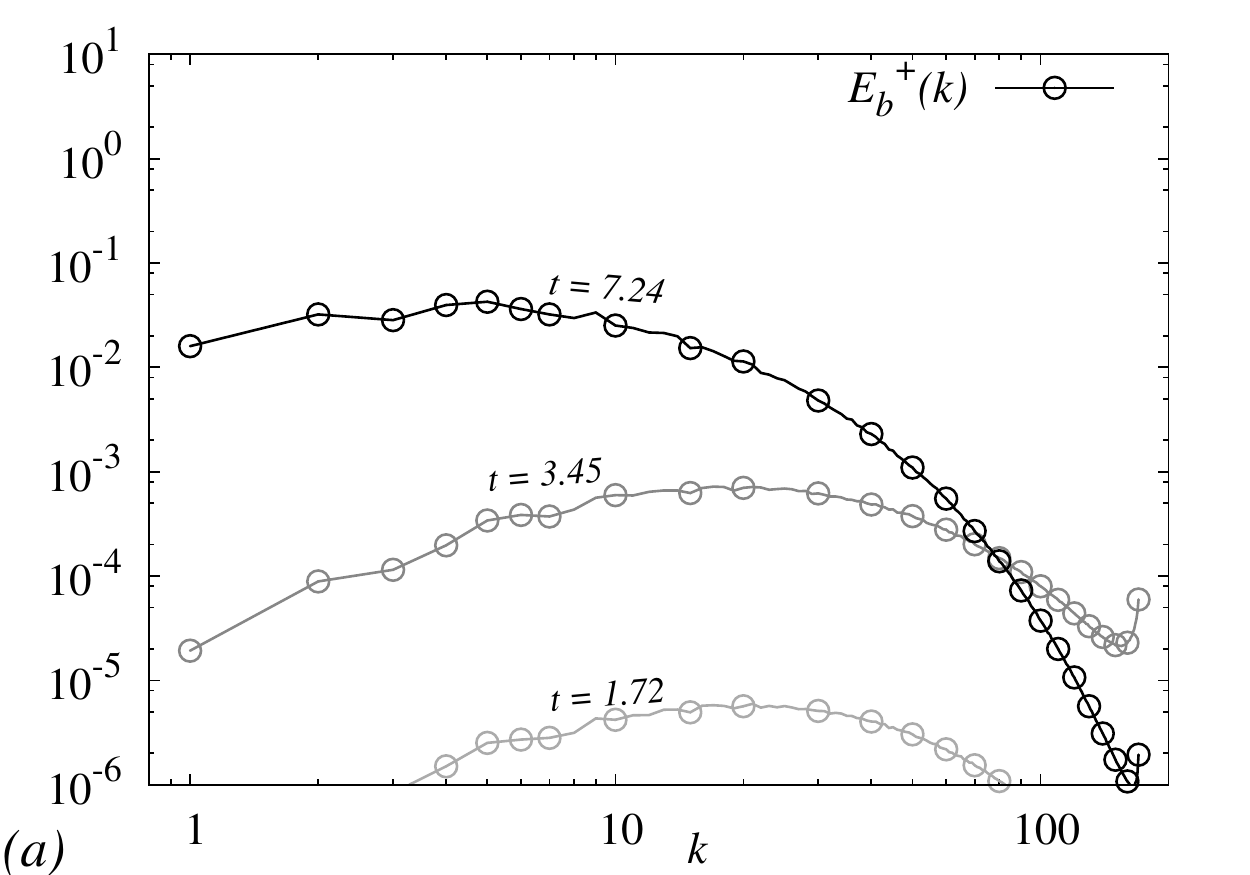} 
\includegraphics[scale=0.7]{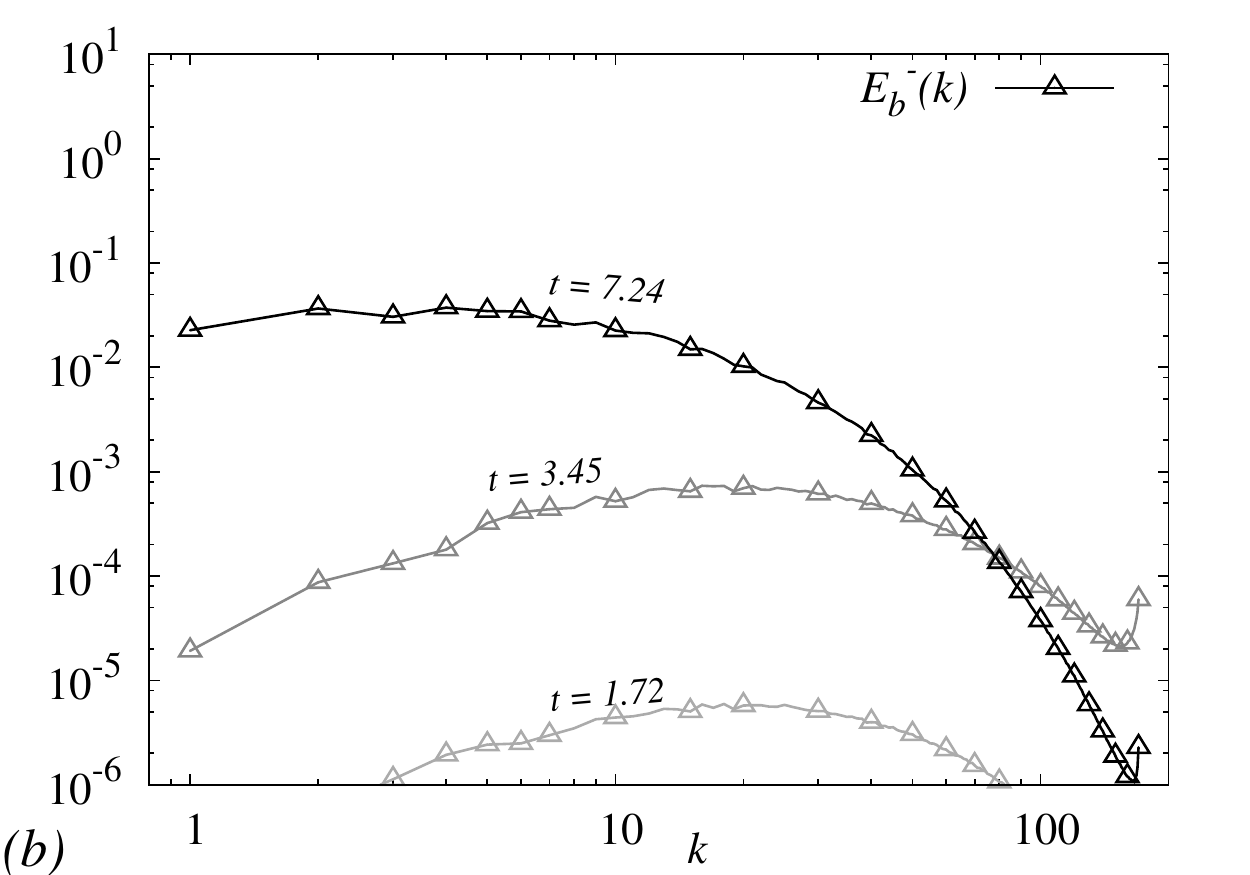} 
{\bf Velocity field}  \\
\includegraphics[scale=0.7]{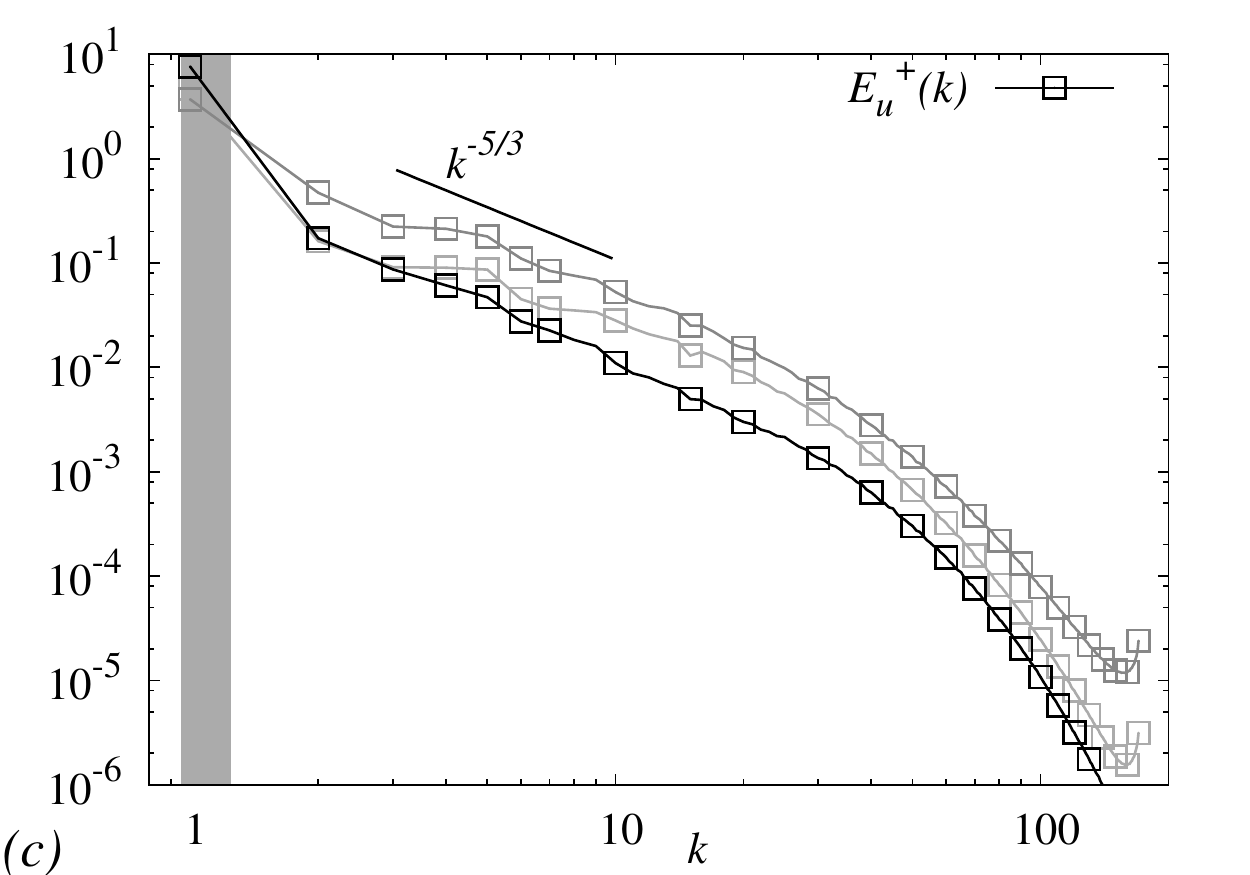} 
\includegraphics[scale=0.7]{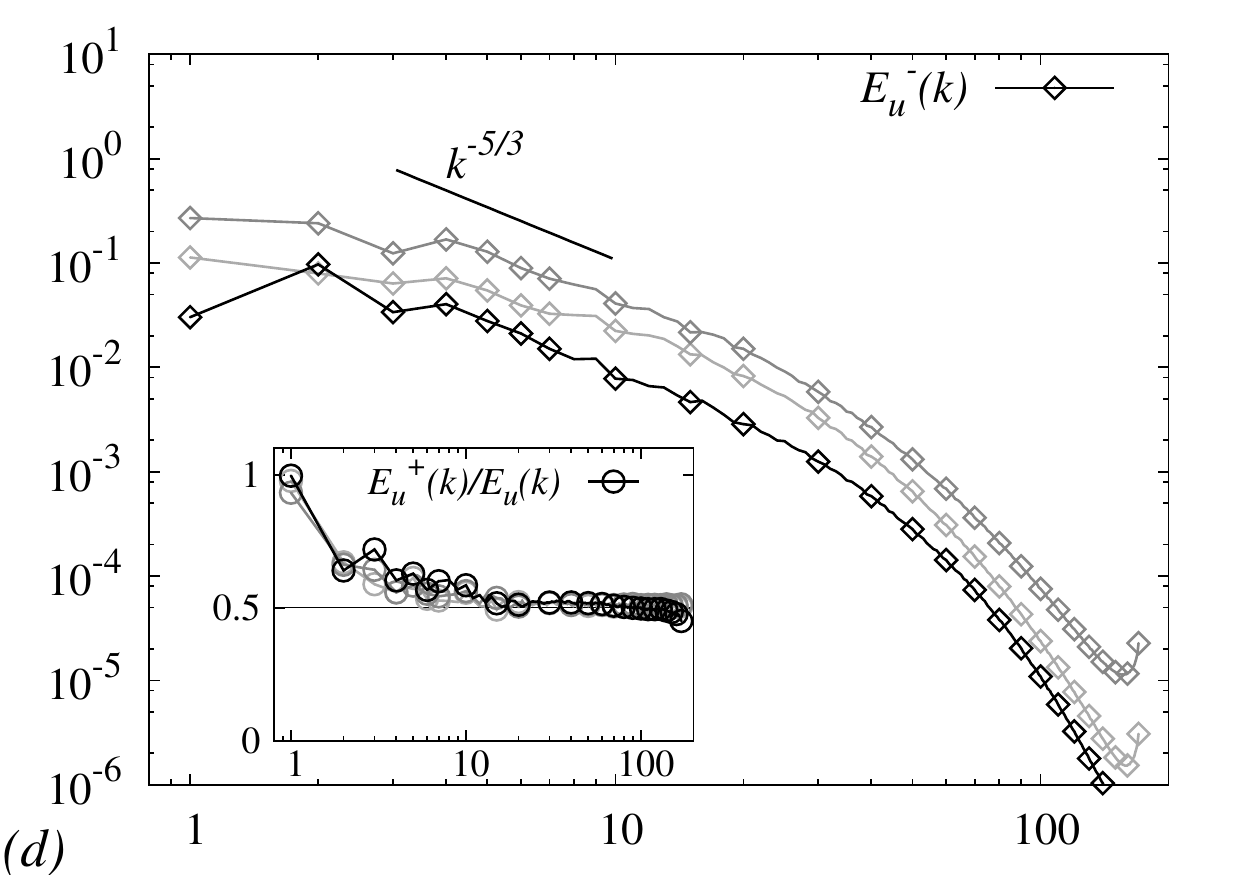} 
\caption{
Run R1-D.
 Panels (a) and (b) show a log-log plot of the magnetic energy spectra $E^+_b(k)$ and $E^-_b(k)$ against $k$ at different times. 
Panels (c) and (d) are the same as (a) and (b), but for the 
kinetic energy spectra $E^+_u(k), E^-_u(k)$.  Time is expressed in units of the forcing-scale turnover time, $T$ 
(see Table~\ref{tbl:simulations-linear}). The gray color-coding corresponds to different
instants during the time evolution, while the dark rectangular area corresponds to the band of forced wavenumbers. 
The inset of panel (d) shows the ratio of the kinetic energy in the positively helical modes with 
respect to the total kinetic energy, $E^+_u(k)/E_u(k)$; the solid horizontal line marks the value $0.5$ 
which corresponds to the mirror symmetric case. 
}
\label{fig:symmetric-dynamo}
\end{figure*}

\begin{figure*}[h]
\center
{\bf Magnetic field}  \\
\includegraphics[scale=0.7]{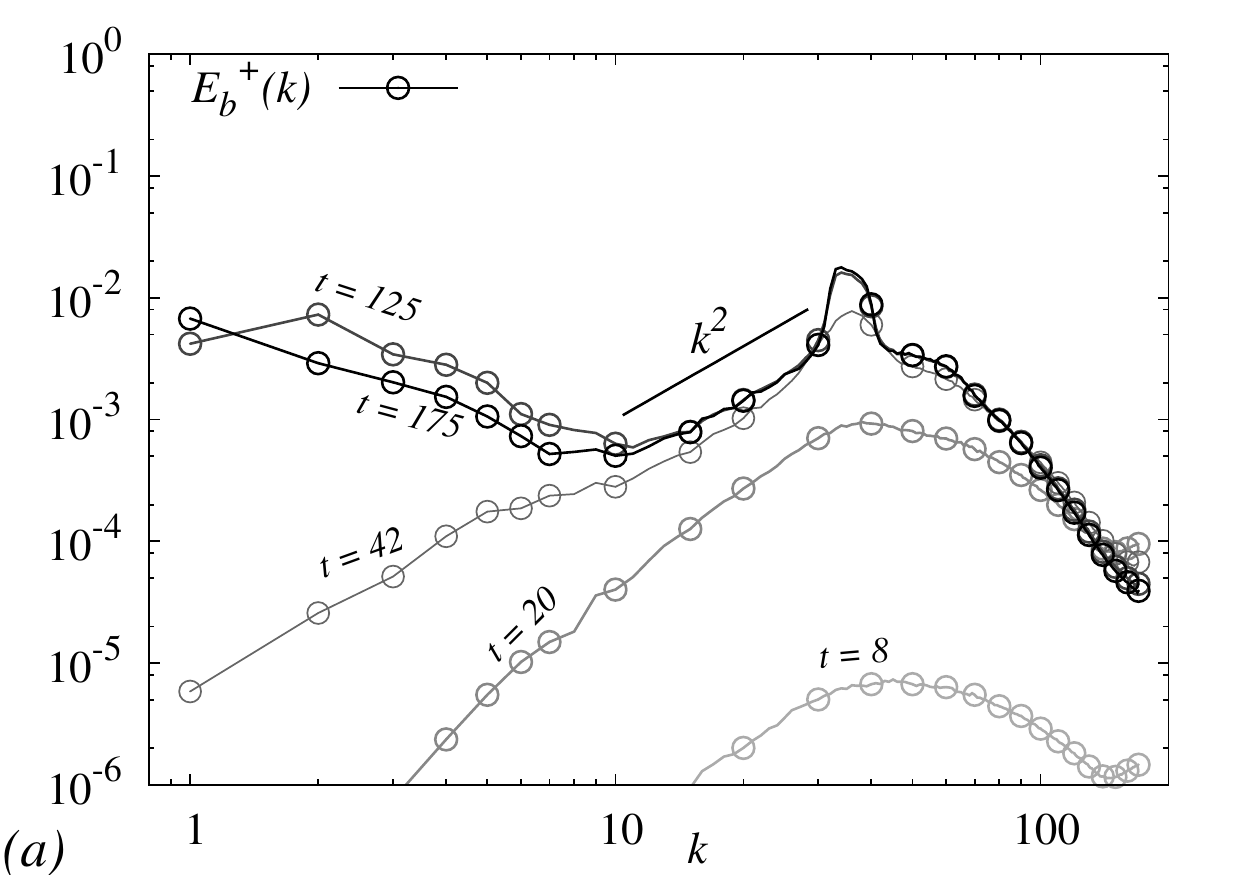} 
\includegraphics[scale=0.7]{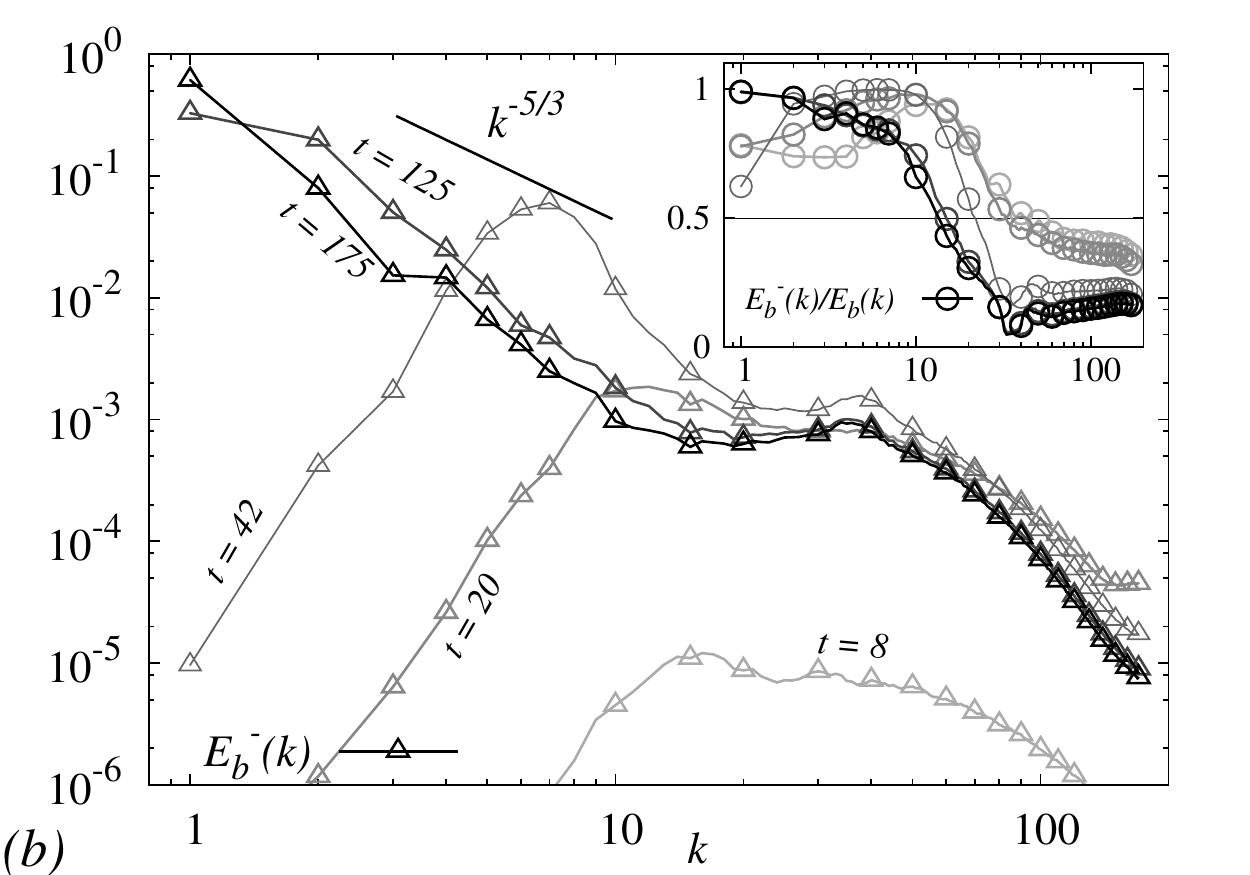} 
{\bf Velocity field}  \\
\includegraphics[scale=0.7]{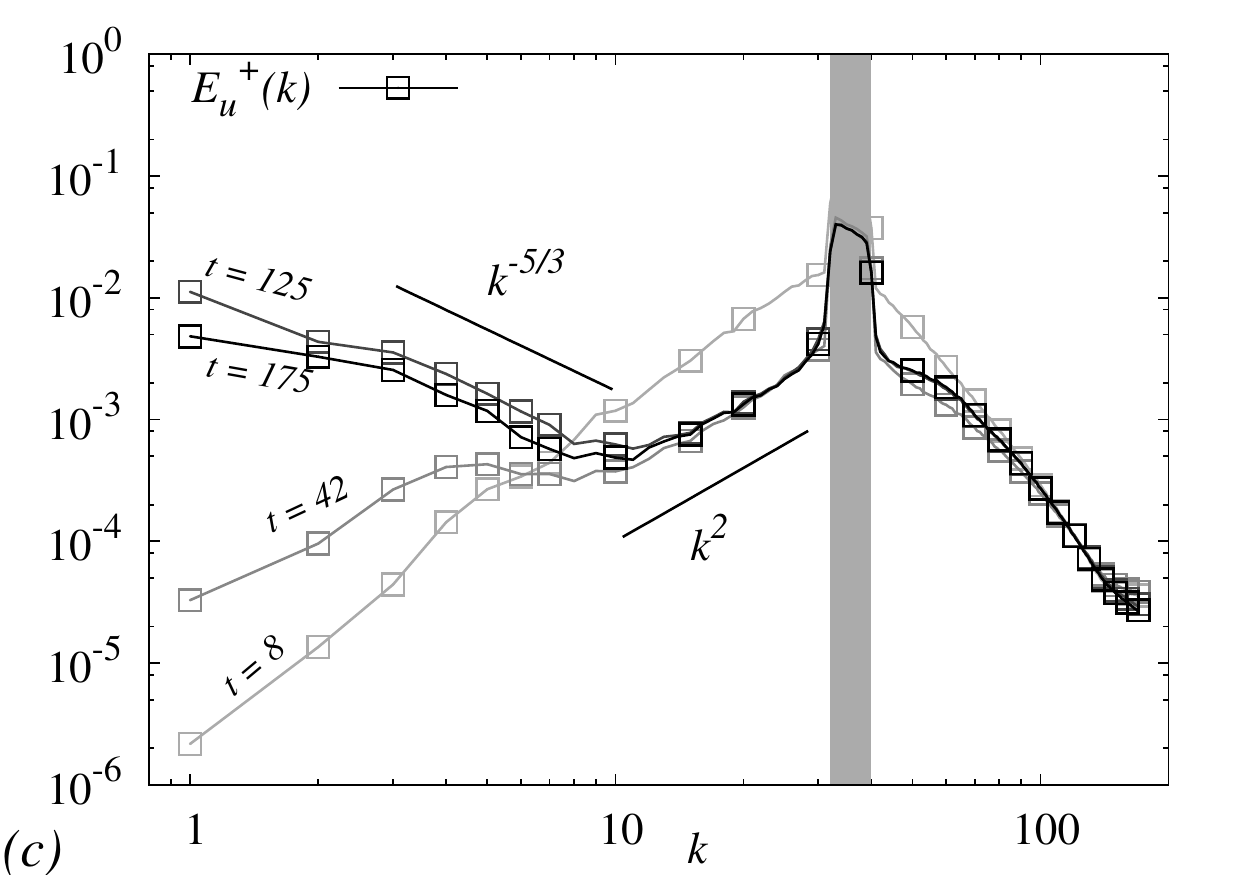} 
\includegraphics[scale=0.7]{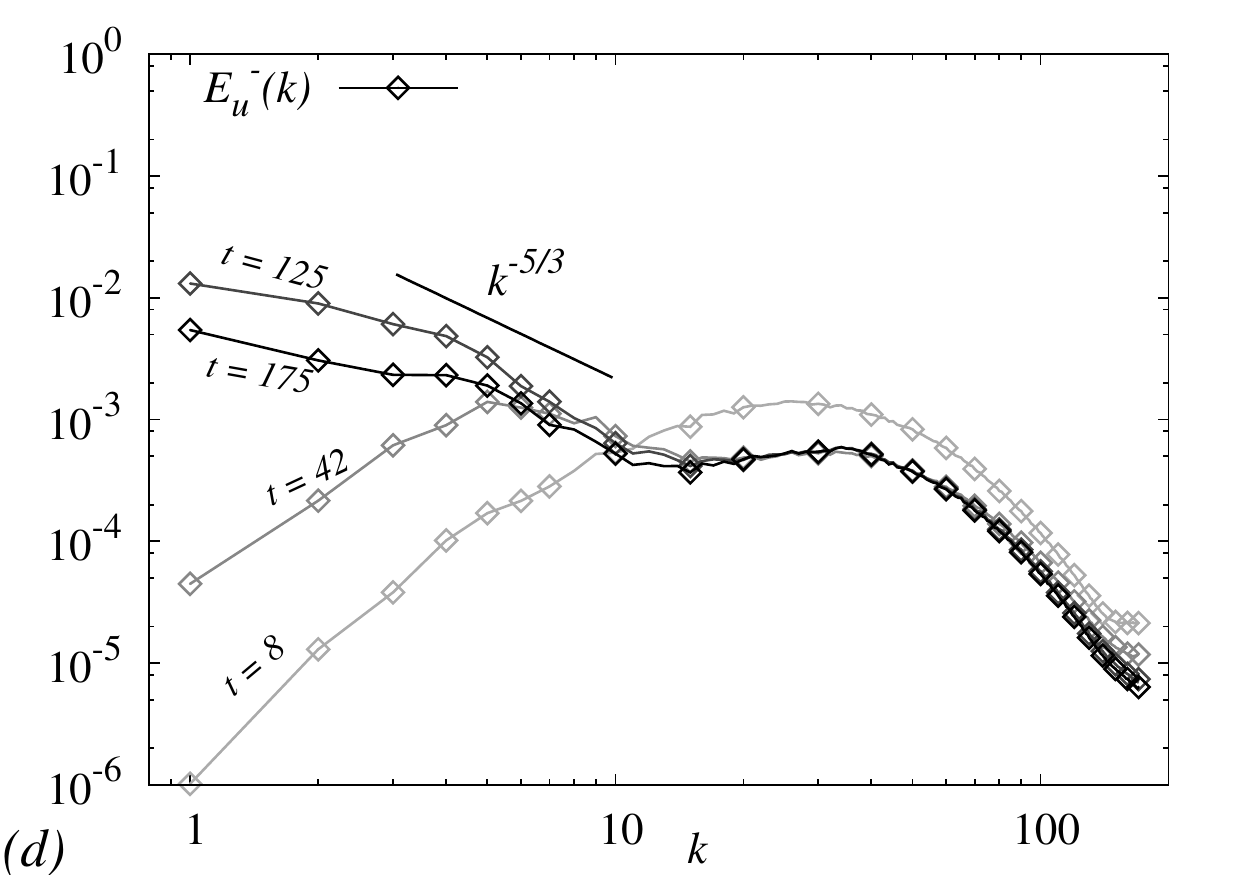} 
\caption{
Run R2-D.
Panels (a) and (b) show a log-log plot of the magnetic energy spectra $E^+_b(k)$ and $E^-_b(k)$ against $k$ at different times. 
The inset of panel (b) shows the ratio of the magnetic energy in the positively helical modes with 
respect to the total magnetic energy, $E^+_b(k)/E_b(k)$; the solid horizontal line marks the value $0.5$ 
which corresponds to the mirror symmetric case. 
Panels (c) and (d) are the same as (a) and (b) but for the 
kinetic energy spectra $E^+_u(k)$ and $E^-_u(k)$.  
}
\label{fig:dynamo}
\end{figure*}

\noindent
R1-D:  In this case we evolve the  velocity fields with a positively
helical mechanical force,  $\vec{f}_u = \vec{f}_u^+$, applied at  large scales,
$ k_u^f \in [0.25:1.25]$. As can be seen in
Figures \ref{fig:symmetric-dynamo}(a-b), the growth of $\bb^+$ and $\bb^-$ both
at small and large scales is symmetric despite the helical large-scale forcing.
We explain this lack of helical asymmetry in the magnetic field growth by
observing that the flow is nearly mirror symmetric for $ k > k_u^f$, as shown
in Figures \ref{fig:symmetric-dynamo}(c-d).  This is not surprising, it is well
known that fully homogeneous and isotropic turbulence tends to quickly recover
small-scale mirror symmetry even in presence of a large-scale helicity
injection 
\citep{Chen03a,Chen03b,Mininni10,Sahoo15a,Sahoo16,Gledzer15,Deusebio14,Kessar15,Stepanov15,Stepanov15a}. 
To quantify the rate of
recovery we show in the inset of Figure \ref{fig:symmetric-dynamo}(d) the ratio
$E^+_u(k)/(E^+_u(k)+E^-_u(k))$. As a result, unless other mechanisms, e.g.
rotation or convection, enhance preferentially one small-scale helical 
component, the
magnetic seed evolves initially on a non-helical turbulent flow, and no
difference is expected between the growth rate of the positively and
negatively helical magnetic field components. \\
In case 
R2-D the velocity field is subjected to a small-scale helical force, $\vec{f}_u = \vec{f}_u^+$ with 
$k_u^f \in [32:40]$. With small-scale injection of kinetic energy, the forward energy cascade cannot develop and the flow is laminar. 
Now, from Figures \ref{fig:dynamo}(a-b) we observe an asymmetric growth of
positively and negatively helical magnetic field  consistent with the
analytical predictions from the MTI systems D2 and D3 discussed  in
Sections \ref{sec:lsdynamo} and \ref{sec:ssdynamo} and depicted in
Figures \ref{fig:ls-kinematic-triads} and \ref{fig:ss-kinematic-triads}. The
curves are always color-coded with darker colors indicating later times, 
with the time evolution of magnetic and kinetic energies shown in  
Figure \ref{fig:dynamo_timeseries}(a) on a linear-logarithmic scale in order to indicate 
when the kinematic stage of the dynamo ends and the evolution of the magnetic field 
becomes nonlinear. Time is expressed in 
units of forcing-scale turnover time $T=L_f/u_{rms}$, where $u_{rms}$ refers to the root mean square (rms) 
value during the kinematic stage of the dynamo (see Table \ref{tbl:simulations-linear}),
and the color-coded arrows in Figure \ref{fig:dynamo_timeseries}(a) correspond to the 
color-coded spectra shown in Figures \ref{fig:dynamo}(a-b). 
In Figures \ref{fig:dynamo}(c-d) the evolution of the kinetic energy spectra is also
shown. As can be seen from Figures \ref{fig:dynamo}(a-b), at scales larger
than the forcing scale the negatively helical modes grow faster than the
positively helical modes, while the opposite is true at scales smaller than the
forcing scale. This is also quantified in the inset of Figure 
\ref{fig:dynamo}(b) where the ratio $E^-_b(k)/(E^+_b(k)+E^-_b(k))$ is plotted
at different times.  The above empirical observation  is precisely the
expected helical signature of an STF-dynamo operating in a positively helical
flow as explained in  Sections \ref{sec:lsdynamo} and \ref{sec:ssdynamo} and summarized 
in statement (iv) in Section \ref{sec:ssdynamo}.
%
%

Interestingly, the STF-type dynamo continues beyond the kinematic regime, as 
can be seen in Figures \ref{fig:dynamo}(a-d), where at later times the magnetic field is not
negligible and has a clearly visible feedback on the evolution of the kinetic energy.
As indicated in Figure \ref{fig:dynamo_timeseries}(a), the two curves representing 
$t/T=125$ and $t/T=175$ shown as dark lines in Figures \ref{fig:dynamo}(a-b) 
correspond to snapshots of the magnetic field during nonlinear evolution, while 
the two earlier snapshots at $t/T=8$ and $t/T=20$
shown as light gray curves correspond to the kinematic stage of the dynamo. 
As can be seen from Figures \ref{fig:dynamo}(a-b), at the largest resolved scale 
$E_b^-(k)$ grows better than $E_b^+(k)$, even in the nonlinear regime. 
However, we also observe a saturation effect, and at later times, the 
difference in the growth of positively and negatively helical sectors diminishes.
At that late stage, both magnetic and kinetic energy spectra have a slope
qualitatively compatible with $k^{-5/3}$-scaling at low wavenumbers, while 
at intermediate wavenumbers the positively helical components $E_b^+(k)$ and $E_u^+(k)$ 
have a $k^2$ slope. \\  

\begin{figure*}[h]
\center
\hspace{2.5cm} {\bf Magnetic field} \hspace{6cm} {\bf Velocity field} \\
\hspace{-0.3cm}
\includegraphics[scale=0.5]{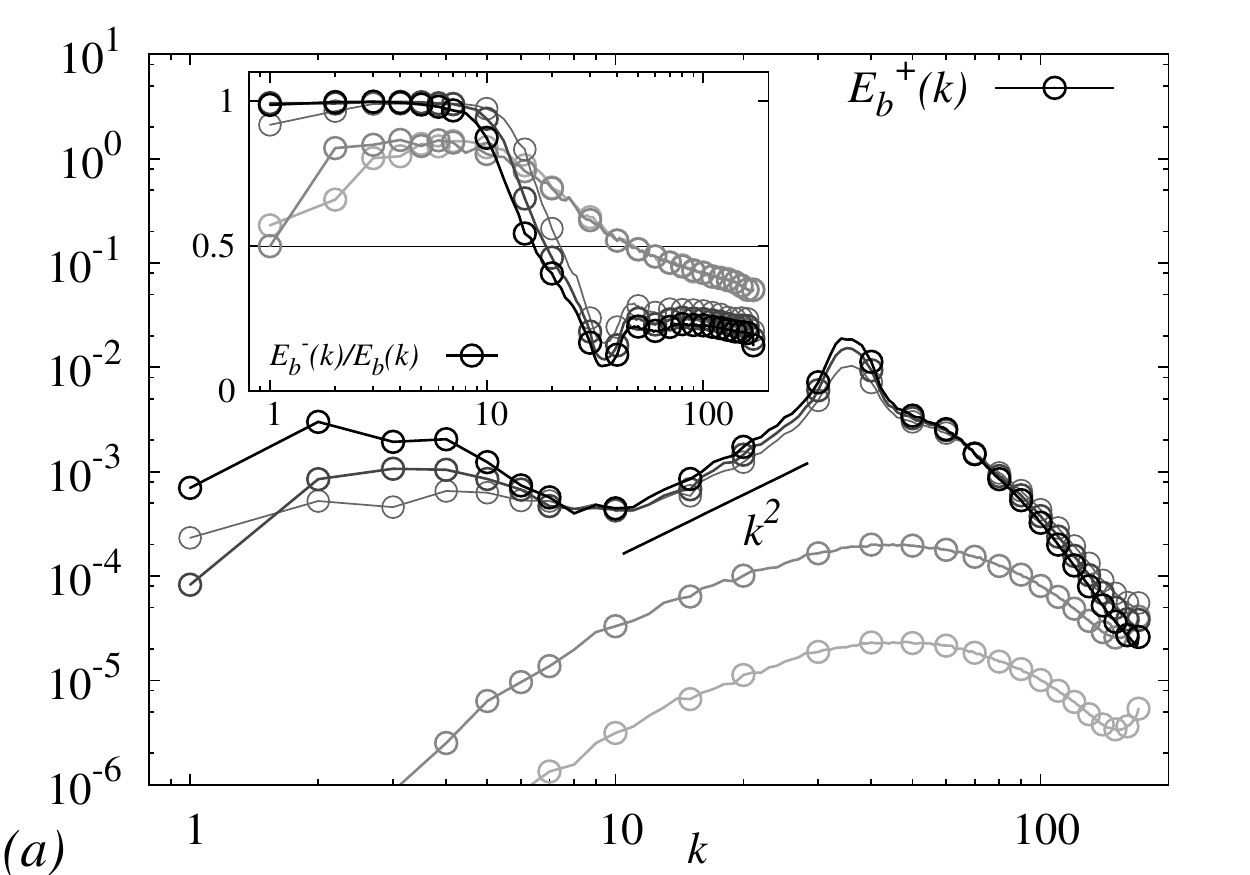} 
\hspace{-0.6cm}
\includegraphics[scale=0.5]{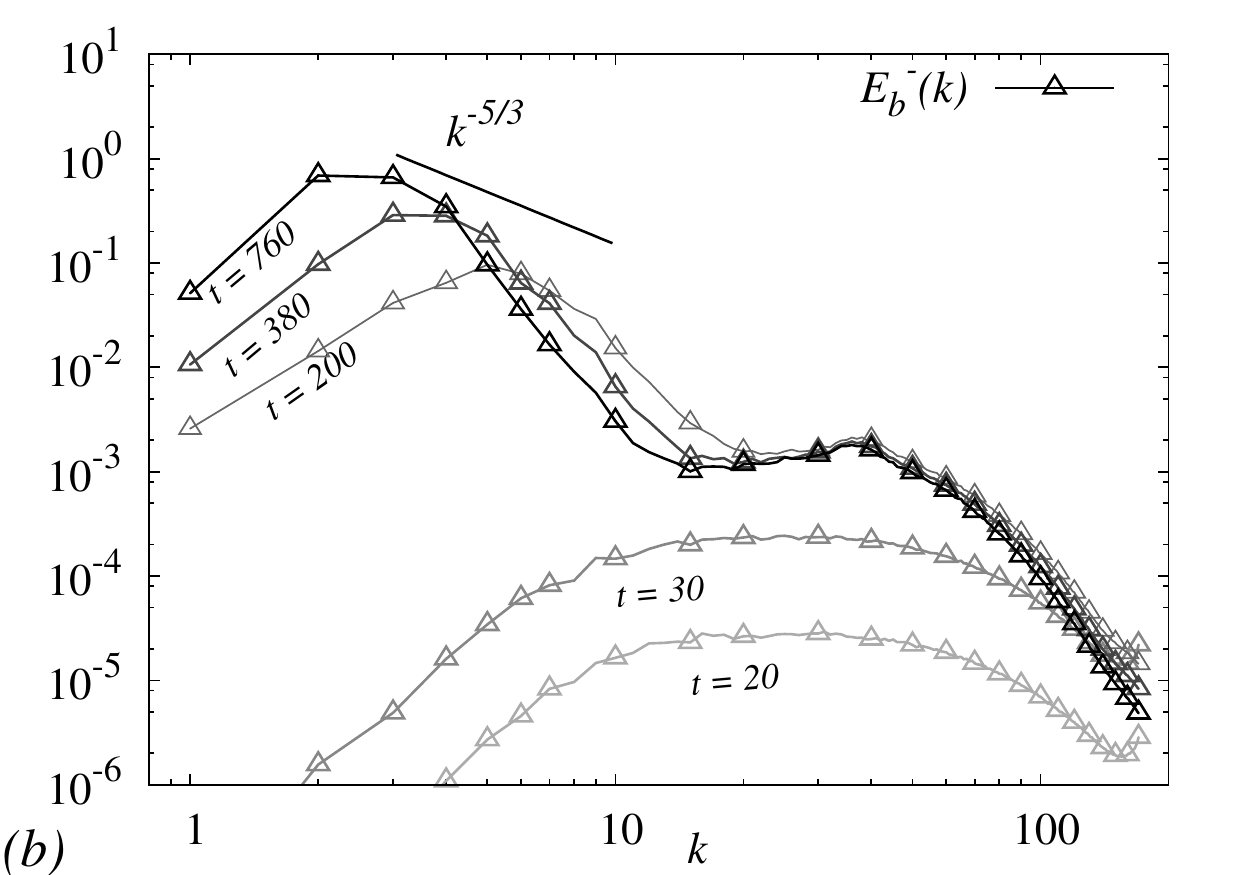} 
\hspace{-0.6cm}
\includegraphics[scale=0.5]{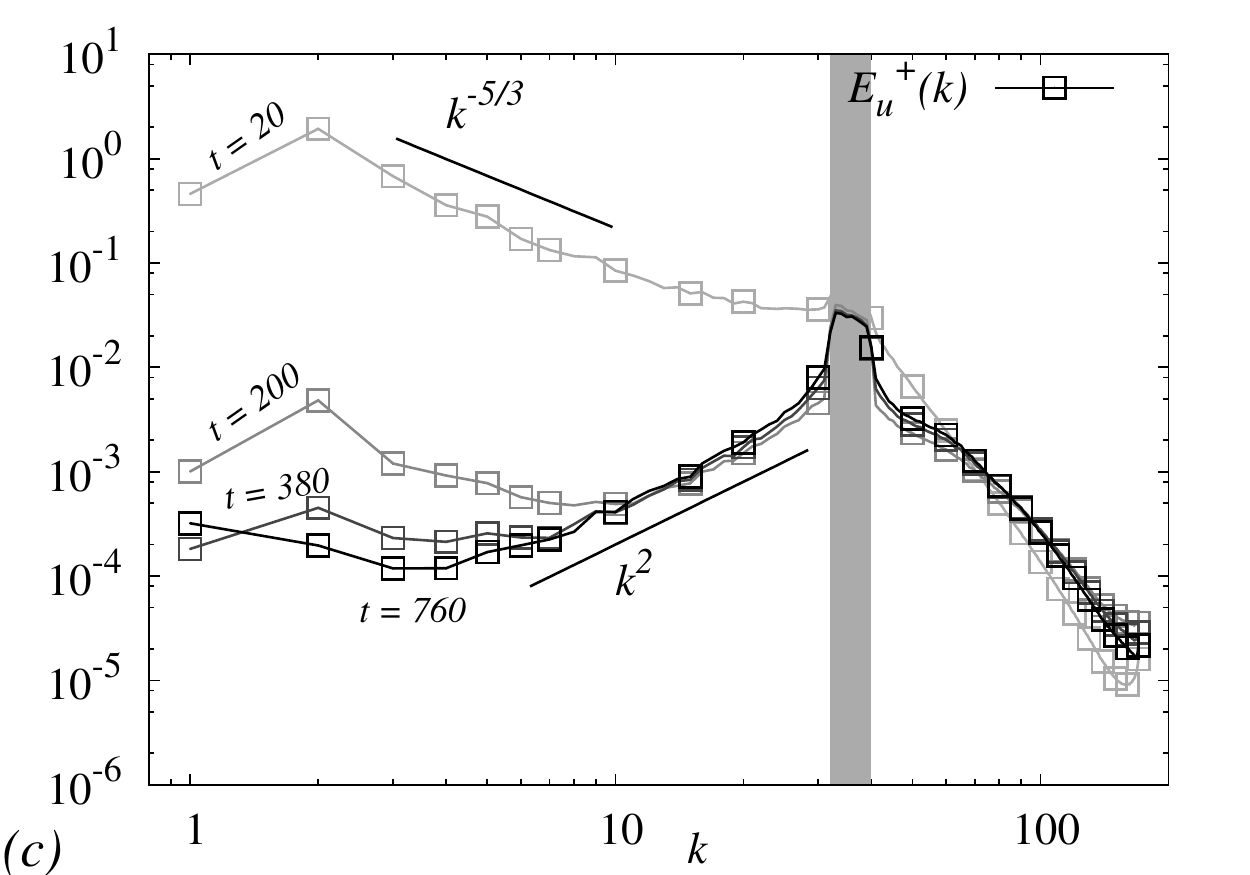} 
\caption{
Run R3-D.
Panels (a) and (b) show a log-log plot of the magnetic energy spectra $E^+_b(k)$ and $E^-_b(k)$ against $k$ at different times. 
The inset of panel (a) shows the ratio of the magnetic energy in the positively helical modes with 
respect to the total magnetic energy, $E^+_b(k)/E_b(k)$; the solid horizontal line marks the value $0.5$ 
which corresponds to the mirror symmetric case.  
Panel (c) is the same as (a) and (b), but for the 
kinetic energy spectrum $E_u(k) = E^+_u(k)$.  
}
\label{fig:upbpbm}
\end{figure*}

\begin{figure*}[h]
\center
\hspace{0.5cm} {\bf R2-D} \hspace{7.5cm} {\bf R3-D} \\
\includegraphics[scale=0.7]{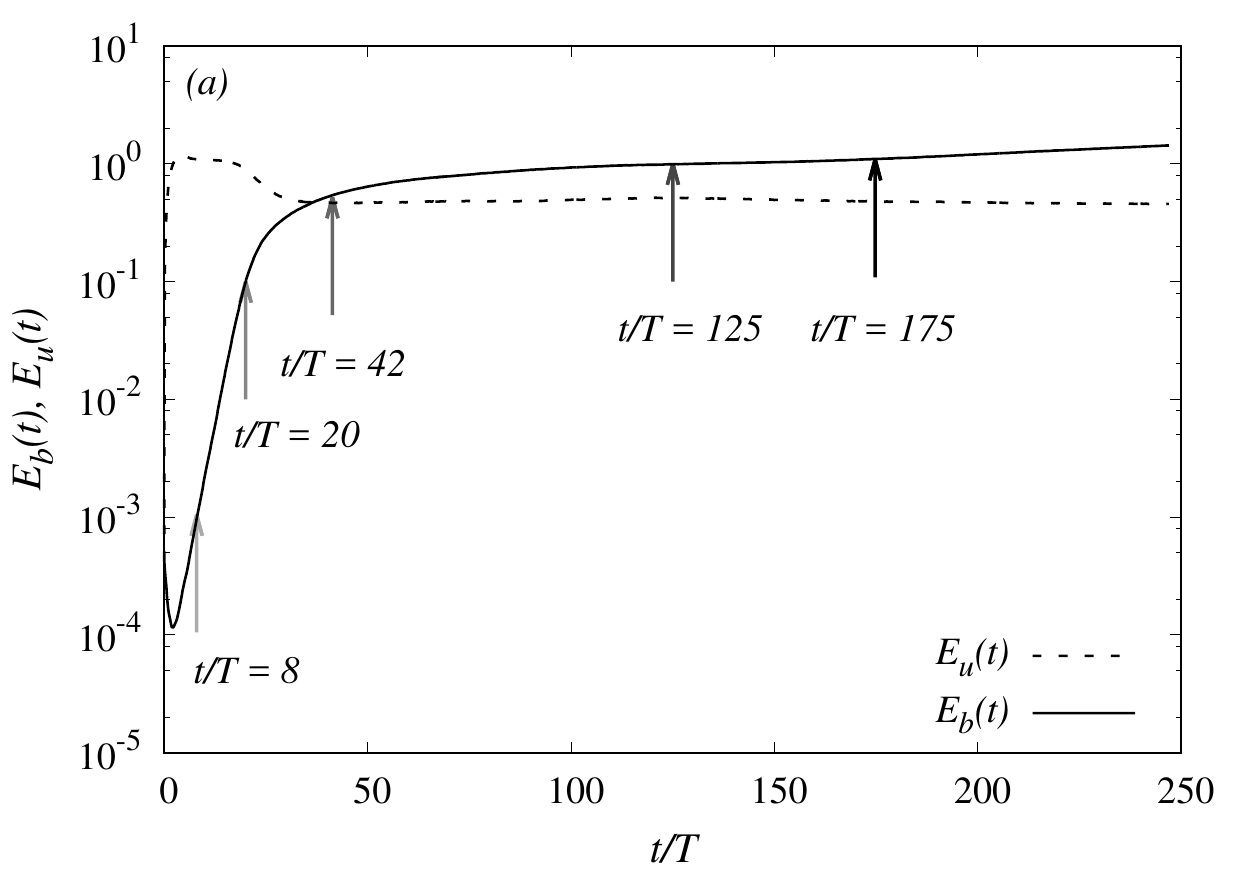} 
\includegraphics[scale=0.7]{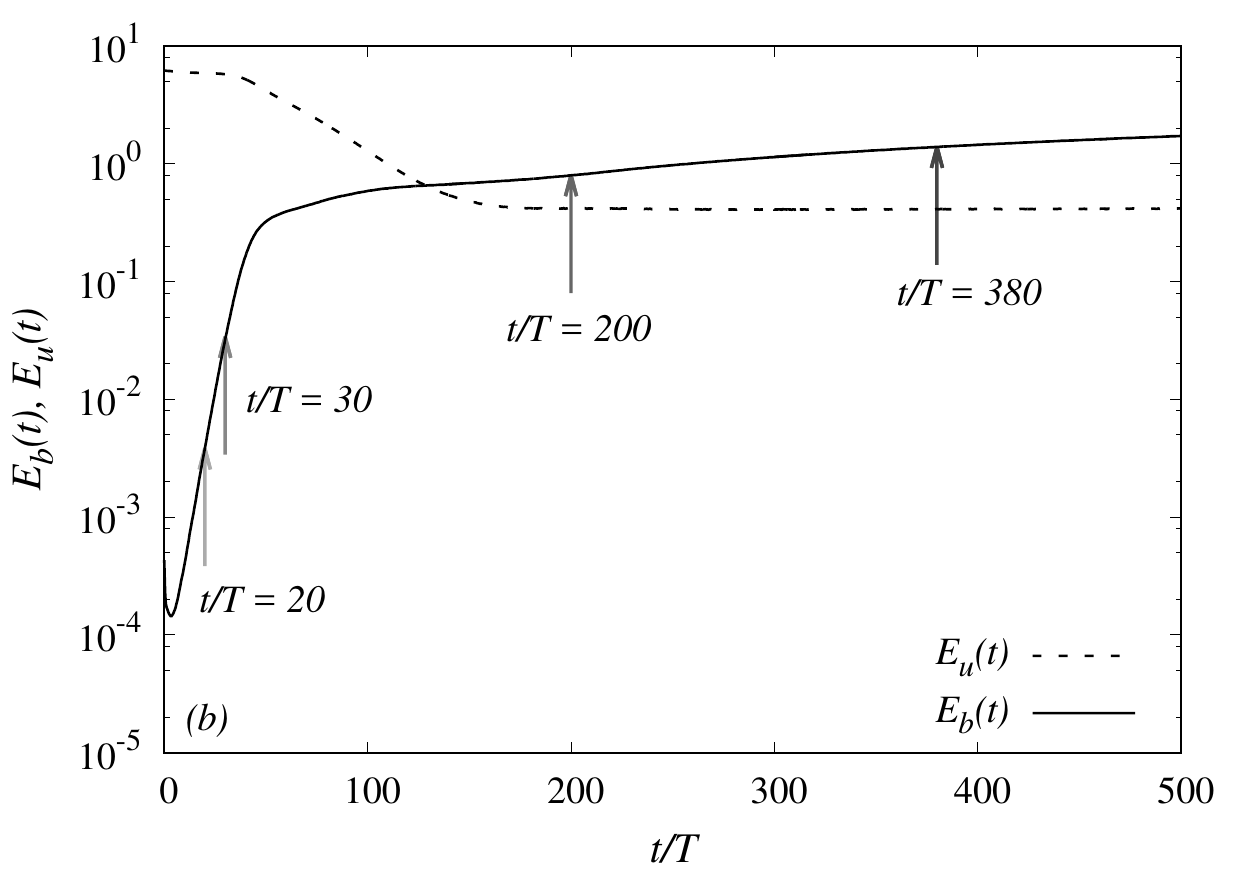} 
\caption{
Time series of the total magnetic (solid line) and kinetic (dashed line) energies on a linear-logarithmic scale for runs 
R2-D (panel (a)) and R3-D (panel (b)).
The arrows in panels (a) and (b) 
indicate the time corresponding to the magnetic energy spectra
shown in Figures~\ref{fig:dynamo}(a-b) and \ref{fig:upbpbm}(a-b), respectively. 
Time is given in units of forcing-scale turnover time $T$, as specified in 
Table \ref{tbl:simulations-linear}. In order to improve the readability of the figure showing 
clearly the initial exponential growth, the time evolution in panel (b) is only given up to $t/T = 500$. 
}
\label{fig:dynamo_timeseries}
\end{figure*}

\noindent The growth of the magnetic field in the non-standard case given by the setup
R3-D is shown in Figures \ref{fig:upbpbm}(a-b).
Here the velocity field is again forced at small scales, but at difference from
the case R2-D initially in an inverse-cascade regime, i.e., with fully
turbulent helical modes at all scales. This is achievable due to the fact
that we have constrained the Navier-Stokes velocity evolution only on positive helical
modes \citep{Biferale12,Biferale13a}. The interest in studying this case is twofold.
First, we wish to understand how robust the inverse kinetic transfer is 
under magnetic perturbations. Second, we aim to study the evolution of a
small-scale magnetic field in a fully helical turbulent flow. From
Figures \ref{fig:upbpbm}(a-b), we see that the magnetic growth is very similar to
the case R2-D, further confirming the theoretical triad-by-triad analysis.
Interestingly, the magnetic field growth has a dramatic effect
on the inverse kinetic energy cascade, as demonstrated in Figure \ref{fig:upbpbm}(c) where
the $k^{-5/3}$-slope of $E^+_u(k)$ is immediately destroyed. In other words, the magnetic 
field grows at the expense of the velocity field modes, strongly
perturbing the phase-correlation that leads to the inverse transfer in absence of
magnetic perturbations. Eventually a $k^2$-scaling for $E_u(k) = E_u^+(k)$
develops at late times for intermediate $k$, as in the laminar case R2-D.
The time evolution of the magnetic energy is shown in  
Figure \ref{fig:dynamo_timeseries}(b) on a linear-logarithmic scale, with color-coded arrows 
indicating that the curves at $t/T = 200$, $t/T = 380$ and $t/T = 760$
shown in Figure \ref{fig:upbpbm} correspond to the nonlinear stage of the dynamo, while the 
curves at $t/T = 20$ and $t/T=30$ show snapshots in the kinematic stage. Similar to case R2-D, 
we observe that the STF-type dynamo continues to be active in the nonlinear regime, albeit 
showing slower magnetic field growth and less difference between the growth of the 
positive and negatively helical sectors. \\

\noindent
In summary, we find in both cases 
R2-D and R3-D that the linear dynamics appear to be quite strong, 
as the magnetic field evolution still bears the same helical signature 
even in the nonlinear dynamo regime. As can be expected, this effect diminishes 
with time as the overall magnetic field growth tends to saturate. 
In relation to the predictions from the linear stability analysis, we conclude
that the results from the triadic dynamics describe the evolution of the magnetic
field well during the kinematic stage of the dynamo and also during the 
onset of nonlinear evolution, even in the presence of a large number of interacting modes. 
Eventually, the validity of the predictions breaks down due to saturation. \\

\begin{table}[hb]
  \caption{
Details of the linear (dynamo) simulations. 
$N$: number of collocation points along each axis in a periodic cube of size
$L=2\pi$; 
$\nu$: kinematic viscosity;
$\eta$: magnetic resistivity;
$k_u^f$: range of forced wavenumbers for velocity field; 
$T$: forcing-scale turnover time ($L_f/u_{\rm rms}$), where $L_f = 4\pi/(k^f_{min}+k^f_{max})$
and $u_{rms}$ is measured in the kinematic stage;
$\varepsilon$: kinetic dissipation rate;
$Re_{\lambda}$: Taylor-Scale Reynolds number.
}
  \label{tbl:simulations-linear}
\begin{tabular*}{\linewidth}{@{\extracolsep{\fill} }  c  c  c  c  c  c  c  c  c  c  c  c c }
    \toprule
   RUN  &  $N$   & helical modes  & $\nu=\eta$ & $k_u^f$ & $\fup$ & $\fum$ & $T$ & $u_{rms}$ & $\varepsilon$ & $Re_{\lambda}$ &  \\ \colrule
   R1-D  & $512$ & $\Up$, $\Um$, $\Bp$, $\Bm$ & $0.002$  & $[0.25,1.25]$ & $5$  & $0$ & 2.9 & 2.9 & 2.5 & 230 &  nonhelical dynamo, turbulent flow \\ 
   R2-D  & $512$ & $\Up$, $\Um$, $\Bp$, $\Bm$ & $0.002$  & $[32,40]$ &  $5$  & $0$ & 0.12 & 1.5 & 3.5 & 15 &   large-scale dynamo, laminar flow \\ 
  \hline
   R3-D & $512$ & $\Up$, $\Bp$, $\Bm$ & $0.002$    & $[32,40]$ &  $5$  & $-$ & 0.05 & 3.5 & 6.2 & 140 & helical dynamo, turbulent flow \\
    \botrule
  \end{tabular*}
\end{table}

\subsection{Inverse cascade of magnetic helicity} \label{sec:nonlinear}
In order to assess the dynamics of strongly magnetized flows and in particular of the inverse cascade of 
magnetic helicity, we carried out a series of simulations subjecting the system also to small-scale electromagnetic forces with $k_b^f \in [32:40]$.
Similar to the previous section, we distinguish the simulations according to the
characteristic scale of the mechanical force and the helical content of the
velocity field. Full details of this series of numerical experiments are contained in 
Table~\ref{tbl:simulations-nonlinear}. 
\\

\begin{table}[hb]
  \caption{Details of the nonlinear (inverse cascade) simulations. 
$N$: number of collocation points along each axis in a periodic cube of size
$L=2\pi$; 
$\nu$: kinematic viscosity;
$\eta$: magnetic resistivity;
$k_u^f$: range of forced wavenumbers for velocity field; 
$k_b^f$: range of forced wavenumbers for magnetic field; 
$T = L_f/u_{\rm rms}$: forcing-scale turnover time,
where $u_{rms}$ is measured in steady state before applying 
${\bm f}_b$;
$\varepsilon$: kinetic dissipation rate;
$Re_{\lambda}$: Taylor-Scale Reynolds number.
}
  \label{tbl:simulations-nonlinear}
\begin{tabular*}{\linewidth}{@{\extracolsep{\fill} } c  c  c  c  c  c  c  c  c  c  c  c  c  c  c  c  c  c c }
    \toprule
   RUN  &  $N$   & helical modes             & $\nu=\eta$ & $k_u^f$   & $k_b^f$   & $\fup$ & $\fum$ & $\fbp$ & $\fbm$
   & $T$ & $u_{rms}$ & $\varepsilon$ & $Re_{\lambda}$  \\ \colrule
   R1-IC  & $512$ & $\Up$, $\Um$, $\Bp$, $\Bm$ & $0.002$    & $[0.25,1.25]$ & $[32,40]$ & $1$  & $0$ & $25$ & $0$ & 2.9 &2.9 & 2.5 & 230   \\
   R2-IC & $512$ & $\Up$, $\Um$, $\Bp$, $\Bm$ & $0.002$    & $[0.25,1.25]$ & $[32,40]$ & $1$  & $0$ & $0$ & $25$ & 2.9 & 2.9& 2.5 & 230  \\
\colrule                                                                          
    R3-IC & $512$ & $\Up$,  $\Bp$, $\Bm$       & $0.002$    & $[32,40]$   & $[32,40]$ & $5$  & $-$ & $25$ & $0$ & 0.05 & 3.5& 6.2 & 140 \\
    R4-IC & $512$ & $\Up$,  $\Bp$, $\Bm$       & $0.002$    & $[32,40]$   & $[32,40]$ & $5$  & $-$ & $0$ & $25$ & 0.05 & 3.5& 6.2 & 140  \\
    \botrule
  \end{tabular*}
\end{table}

\subsubsection{Full velocity field}
\begin{figure}[h]
\center
{\bf Magnetic field}  \\
\includegraphics[scale=0.7]{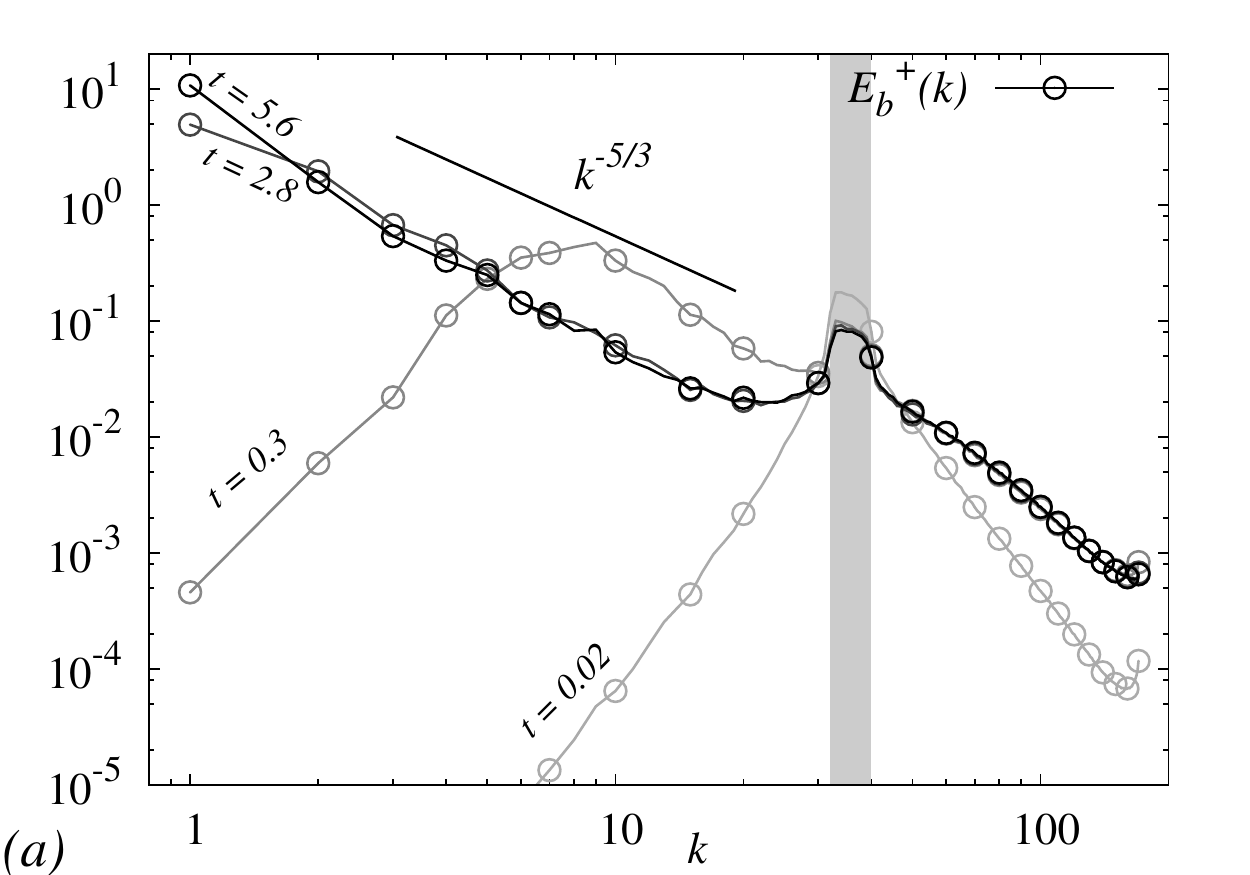} 
\includegraphics[scale=0.7]{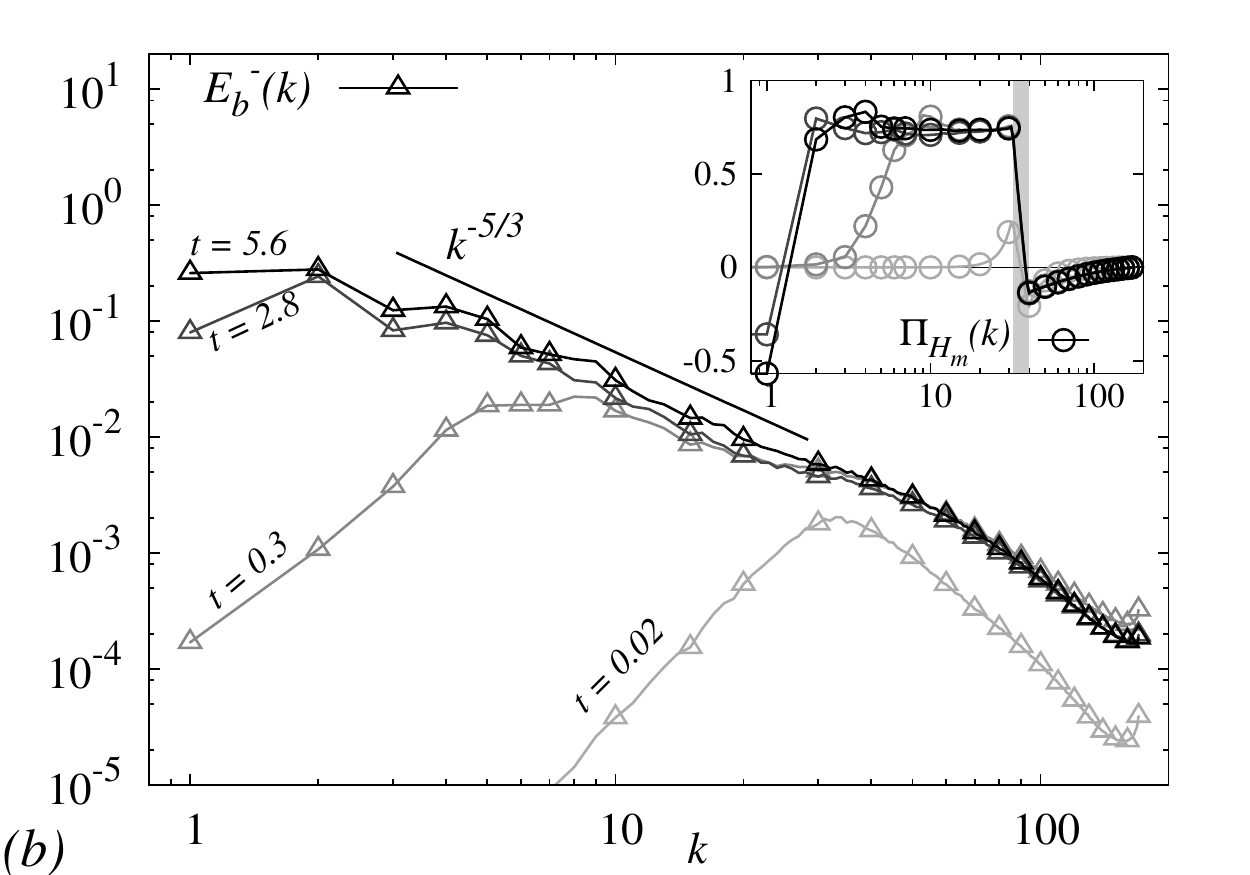} 
{\bf Velocity field}  \\
\includegraphics[scale=0.7]{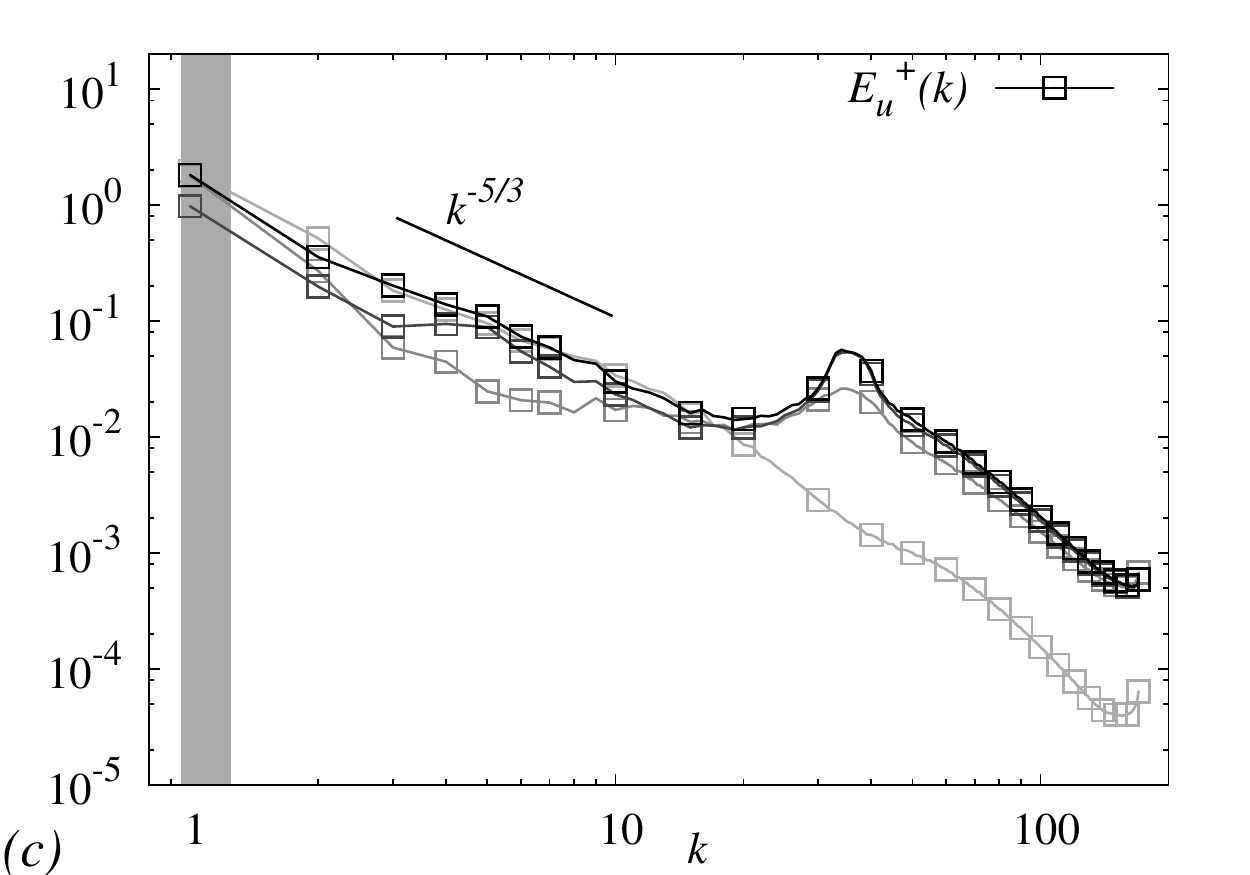} 
\includegraphics[scale=0.7]{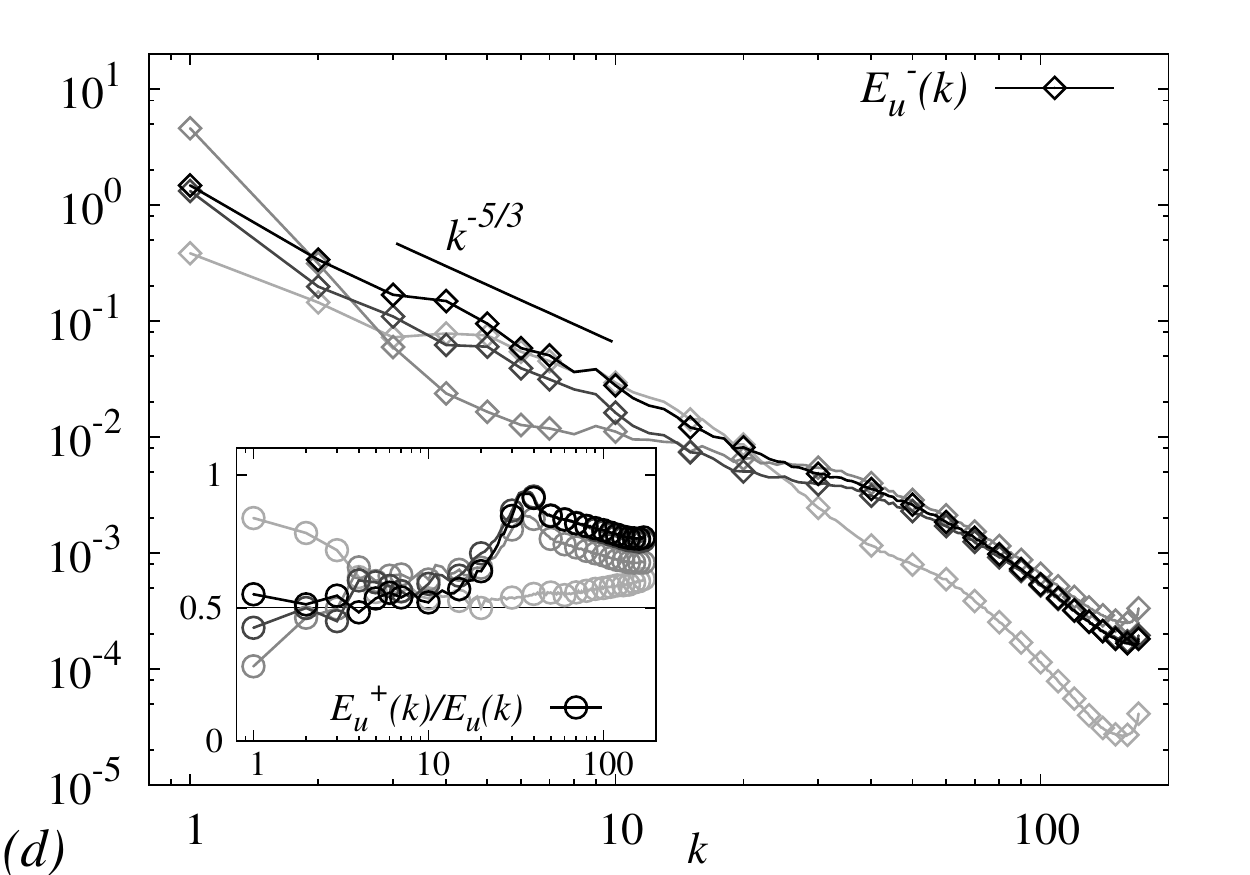} 
\caption{
Run R1-IC.
Panels (a) and (b) show a log-log plot of the magnetic energy spectra $E^+_b(k)$ and $E^-_b(k)$ against $k$ at different times. 
The inset of panel (b) shows the flux of magnetic helicity $\Pi_{H_m}(k)$.
Panels (c) and (d) are the same as (a) and (b), but for the 
kinetic energy spectra $E^+_u(k)$ and $E^-_u(k)$.  
The inset of panel (d) shows the ratio of the kinetic energy in the positively helical modes with 
respect to the total kinetic energy, $E^+_u(k)/E_u(k)$; the solid horizontal line marks the value $0.5$ 
which corresponds to the mirror symmetric case. 
}
\label{fig:r2ic-spectra}
\end{figure}

\begin{figure}[h]
\center
{\bf Magnetic field}  \\
\includegraphics[scale=0.7]{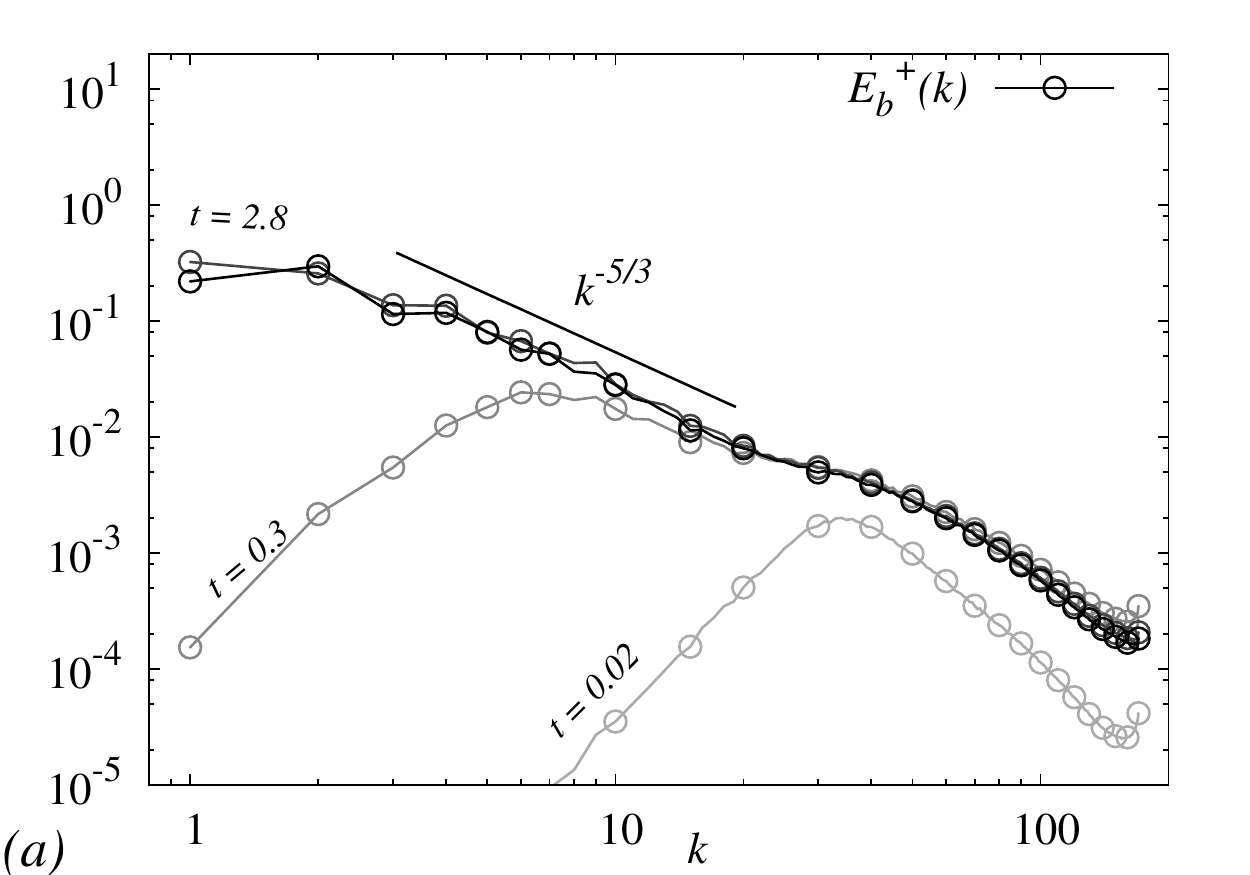} 
\includegraphics[scale=0.7]{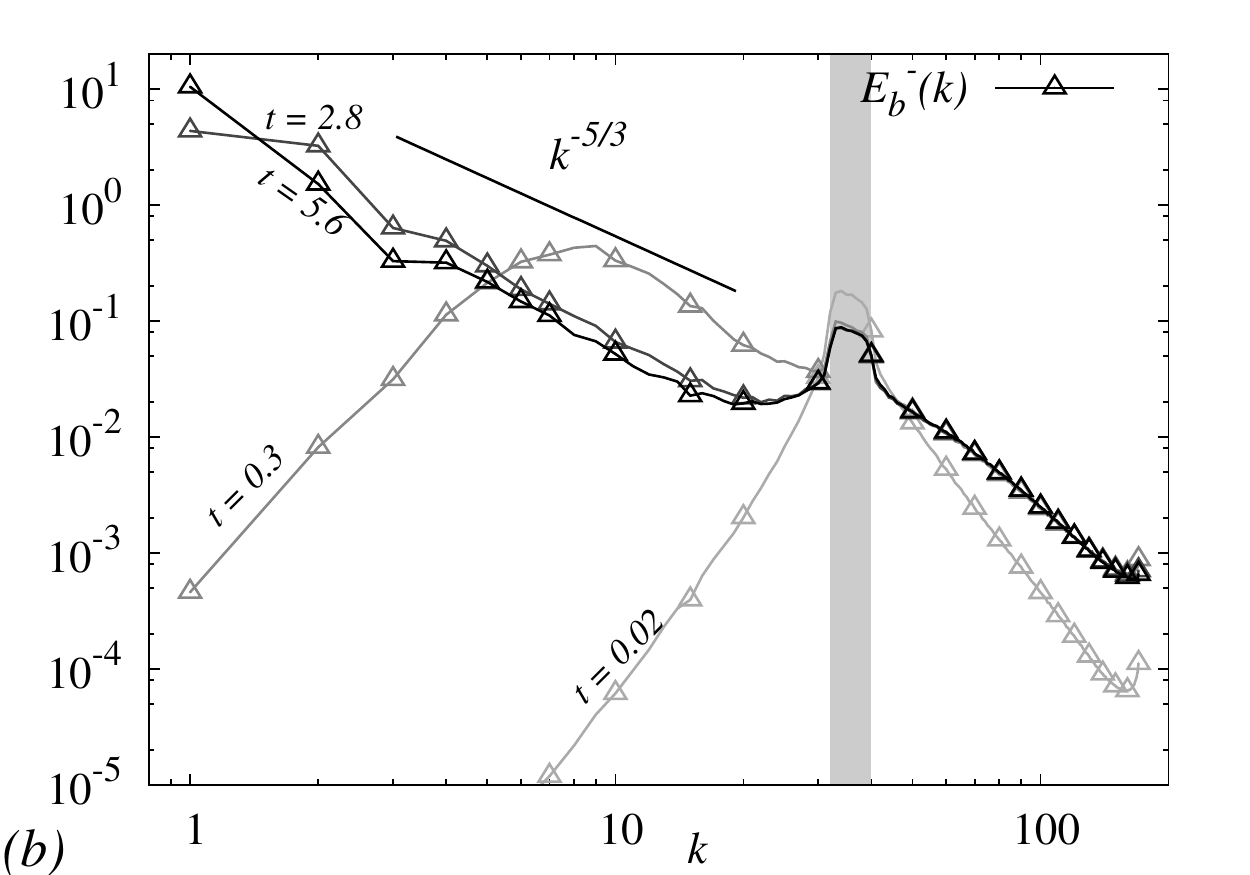} 
{\bf Velocity field} \\
\includegraphics[scale=0.7]{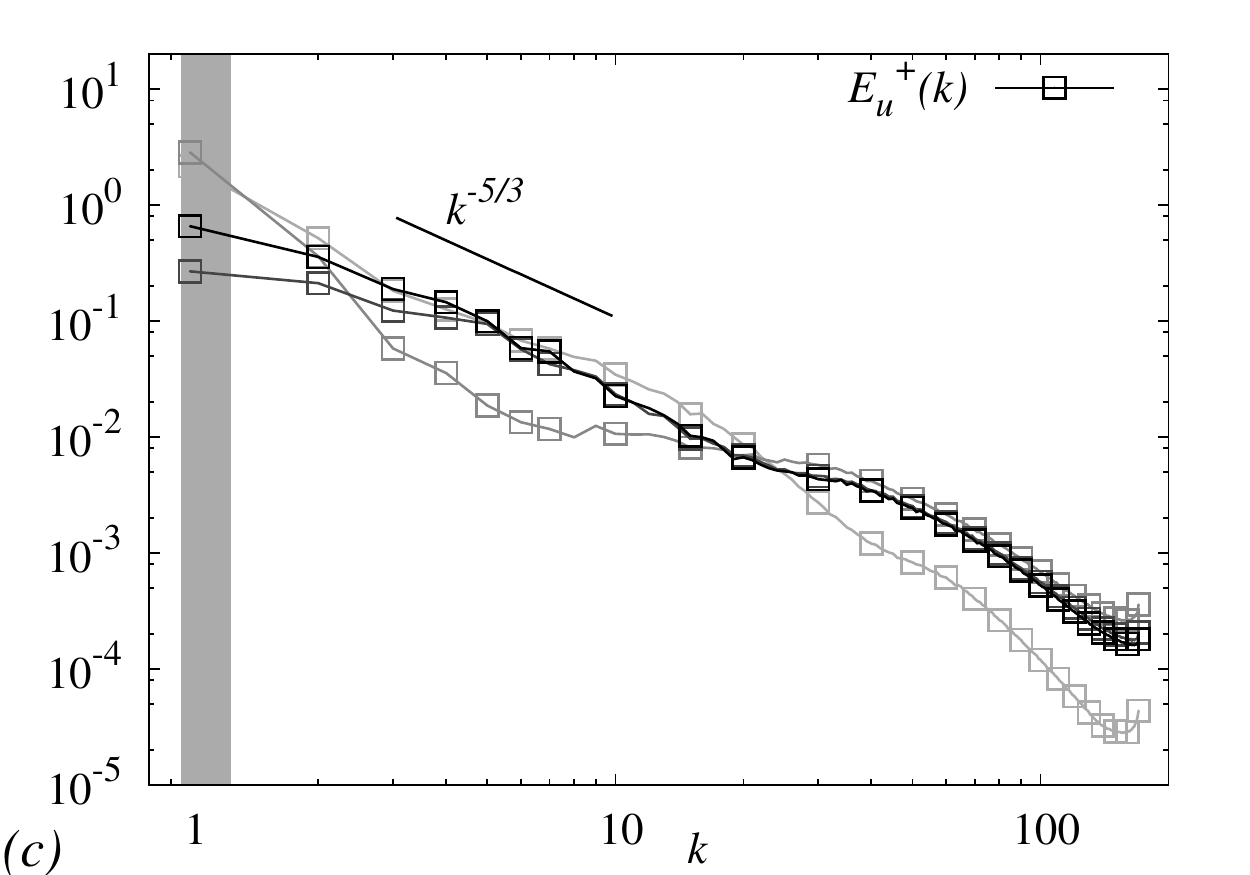} 
\includegraphics[scale=0.7]{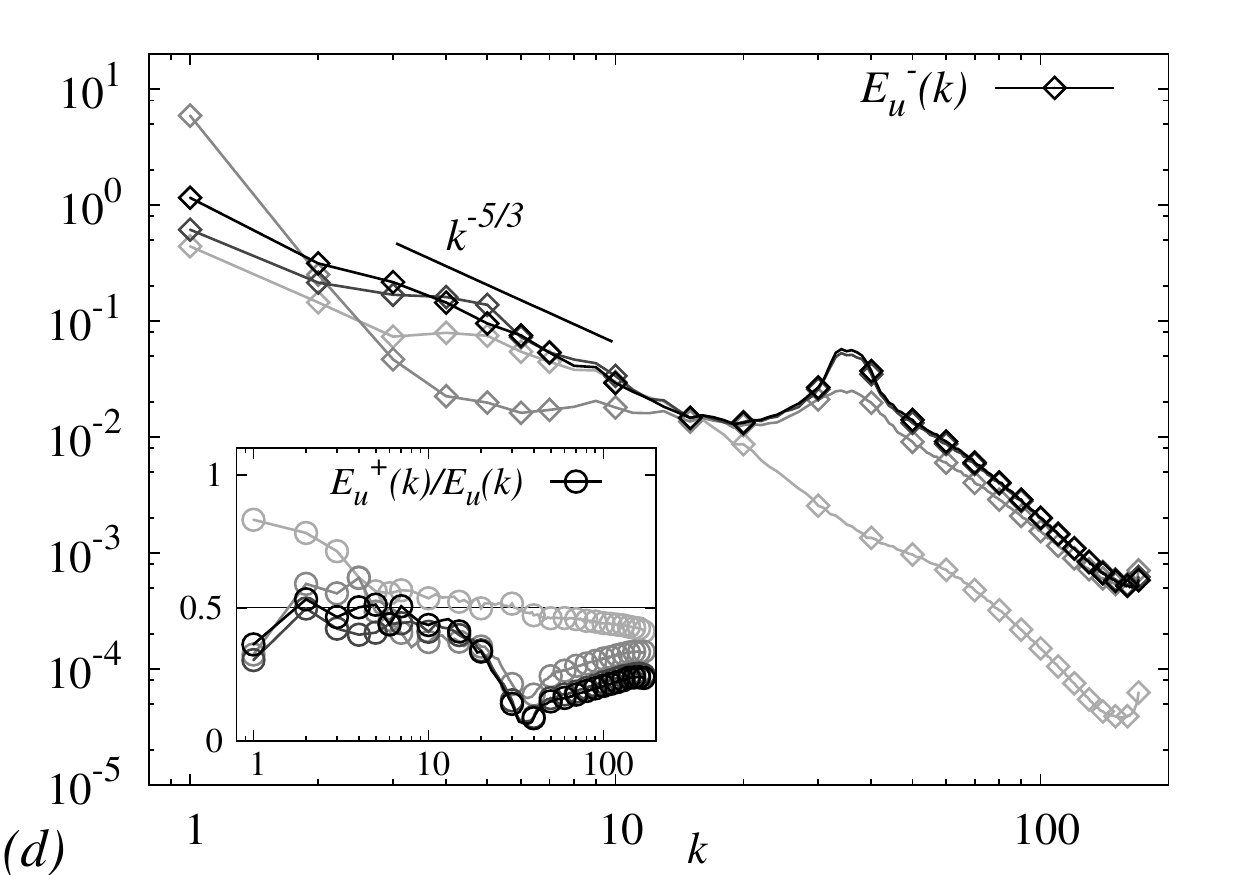} 
\caption{
Run R2-IC.
Panels (a) and (b) show a log-log plot of the magnetic energy spectra $E^+_b(k)$ and $E^-_b(k)$ against $k$ at different times. 
Panels (c) and (d) are the same as (a) and (b), but for the 
kinetic energy spectra $E^+_u(k)$ and $E^-_u(k)$.  
The inset of panel (d) shows the ratio of the kinetic energy in the positively helical modes with 
respect to the total kinetic energy, $E^+_u(k)/E_u(k)$; the solid horizontal line marks the value $0.5$ 
which corresponds to the mirror symmetric case. 
}
\label{fig:r3ic-spectra}
\end{figure}

\noindent
In cases R1-IC and R2-IC the force is either positively helical,
$\vec{f}_b = \vec{f}_b^+$, or  negatively helical, $\vec{f}_b = \vec{f}_b^-$.
Results for the time evolution of these two configurations  are shown in
Figures~\ref{fig:r2ic-spectra} and \ref{fig:r3ic-spectra}, respectively.
From a comparison of Figure~\ref{fig:r2ic-spectra}(a) with  
Figure~\ref{fig:r3ic-spectra}(b), it is clear that the evolution of 
the magnetic energy spectra 
is fully dominated by the signature of the
magnetic forcing, i.e., there is always an inverse cascade of the same magnetic
helical component injected at small scales. 
An important indicator for a cascade process is given by the flux of the cascading quantity. 
As can be seen in the inset of Figure~\ref{fig:r2ic-spectra}(b), the magnetic helicity flux
\be
\Pi_{H_m}(k)=\sum_{k'=k_{min}}^k \sum_{|\bk|=k'} \ba_{\bk}^* \cdot i\bk \times \sum_{\bk+\bp +\bq =0} \bu_{\bp} \times \bb_{\bq} + \mbox{c.c.} \ ,  
\ee
is constant and positive in the region $ 1 < k < k_b^f$, indicating indeed an 
inverse cascade of positive magnetic helicity. 
At difference from the dynamo
case, the large-scale magnetic helicity has the same sign as the
small-scale kinetic helicity. 
This is possible because of the strong intensity of
the small-scale magnetic field which is always far from a kinematic dynamo
regime. This is in agreement with statement (v) of the theoretical Section
\ref{sec:selfinteraction}. 
When we compare Figure~\ref{fig:r2ic-spectra}(c) 
with Figure~\ref{fig:r3ic-spectra}(d) it is important to
note that the feedback of the magnetic field via the Lorentz force on the
velocity field leaves a small-scale helical signature on the flow itself. This 
is quantified in the insets of Figures~\ref{fig:r2ic-spectra}(d) and \ref{fig:r3ic-spectra}(d). On
the other hand, the magnetic feedback on the large-scale velocity field tends
to make it helically neutral, recovering the large-scale mirror symmetry as shown in
the insets of Figures~\ref{fig:r2ic-spectra}(d) and \ref{fig:r3ic-spectra}(d). \\
In summary, we find that the injection of magnetic helicity leads to
large-scale magnetic field growth mainly of the forced helical sector of the
magnetic field, as expected from the inverse cascade of magnetic helicity and
consistent with the theoretical results summarized in statement (v) of
Section~\ref{sec:selfinteraction}.  We also find that the conversion of magnetic
to kinetic energy due to the action of the Lorentz force at  small-scales 
proceeds preferentially between magnetic and velocity field modes with the same sign of
helicity.

\subsubsection{Decimated helical velocity field}
\begin{figure}[h]
\center
\hspace{2.5cm} {\bf Magnetic field} \hspace{6cm} {\bf Velocity field} \\
\hspace{-0.3cm}
\includegraphics[scale=0.5]{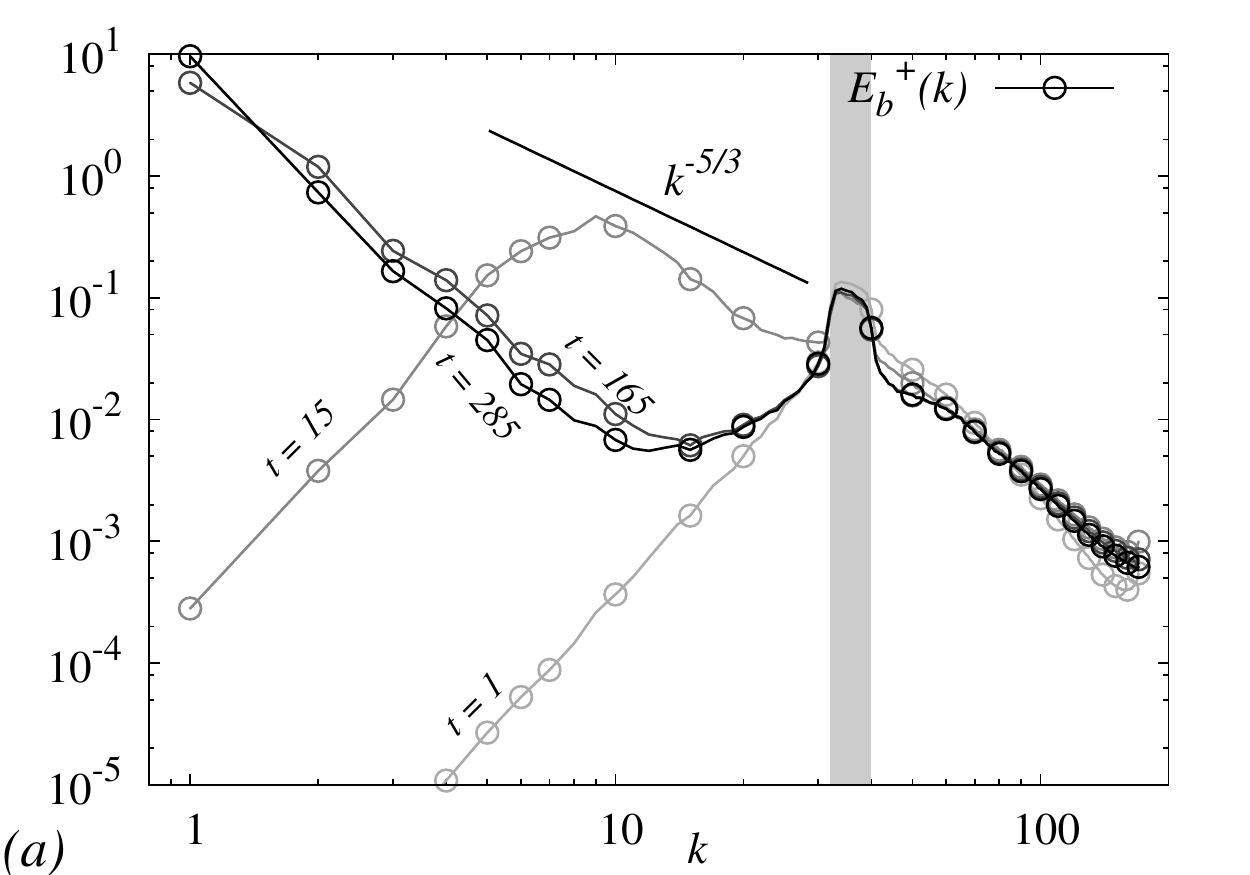} 
\hspace{-0.6cm}
\includegraphics[scale=0.5]{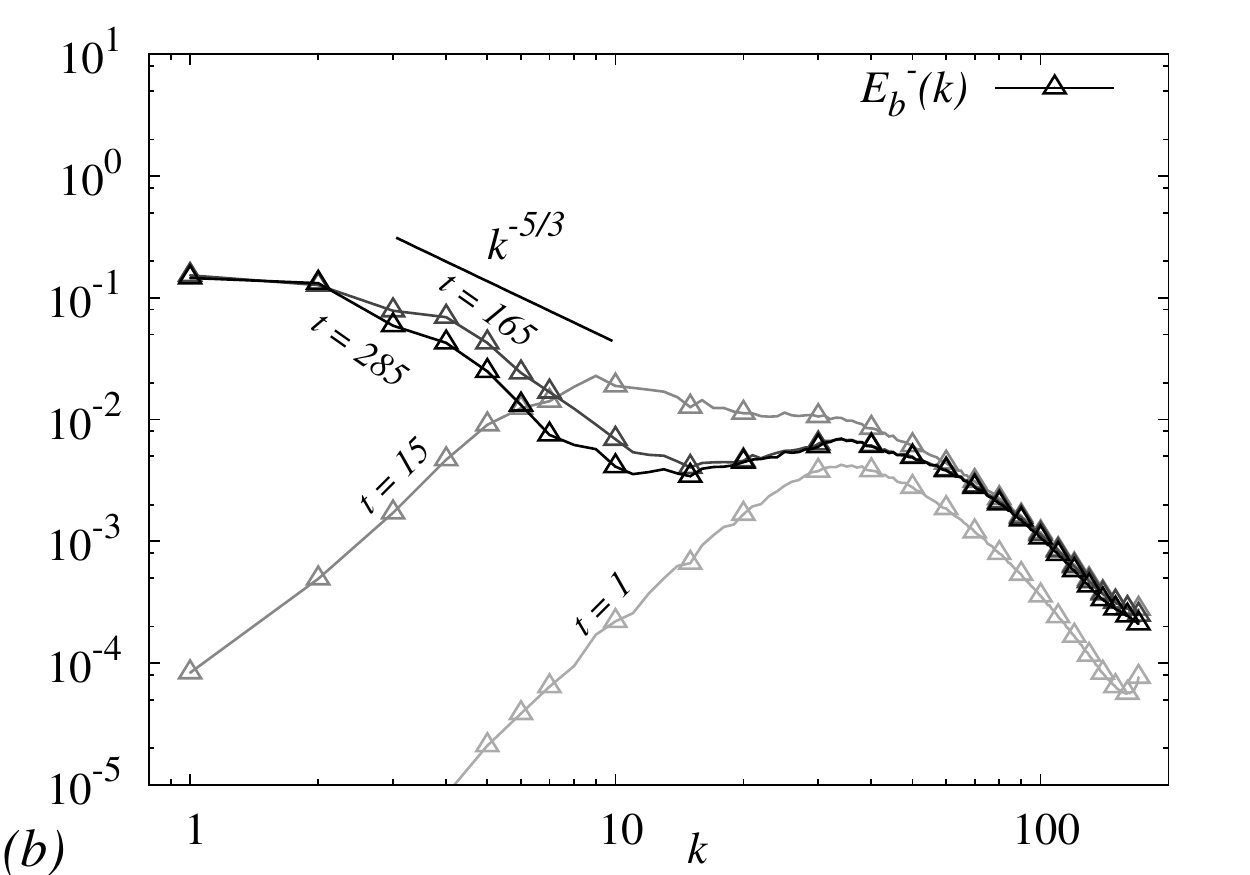} 
\hspace{-0.6cm}
\includegraphics[scale=0.5]{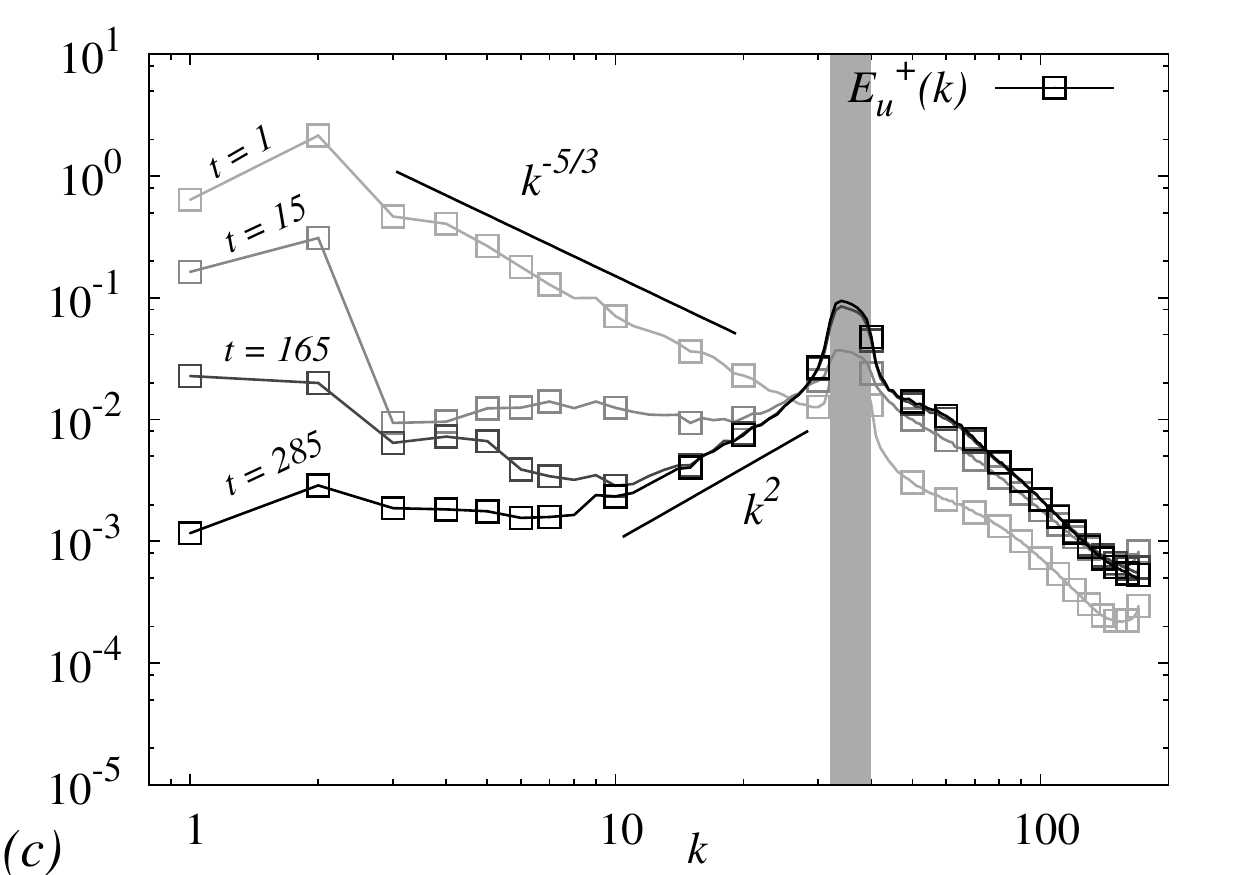} 
\caption{
Run R3-IC.
Panels (a) and (b) show a log-log plot of the magnetic energy spectra $E^+_b(k)$ and $E^-_b(k)$ against $k$ at different times. 
Panel (c) is the same as (a) and (b), but for the 
kinetic energy spectrum $E_u(k) = E^+_u(k)$. 
}
\label{fig:r5ic-spectra}
\end{figure}

\begin{figure}[h]
\center
\hspace{2.5cm} {\bf Magnetic field} \hspace{6cm} {\bf Velocity field} \\
\hspace{-0.3cm}
\includegraphics[scale=0.5]{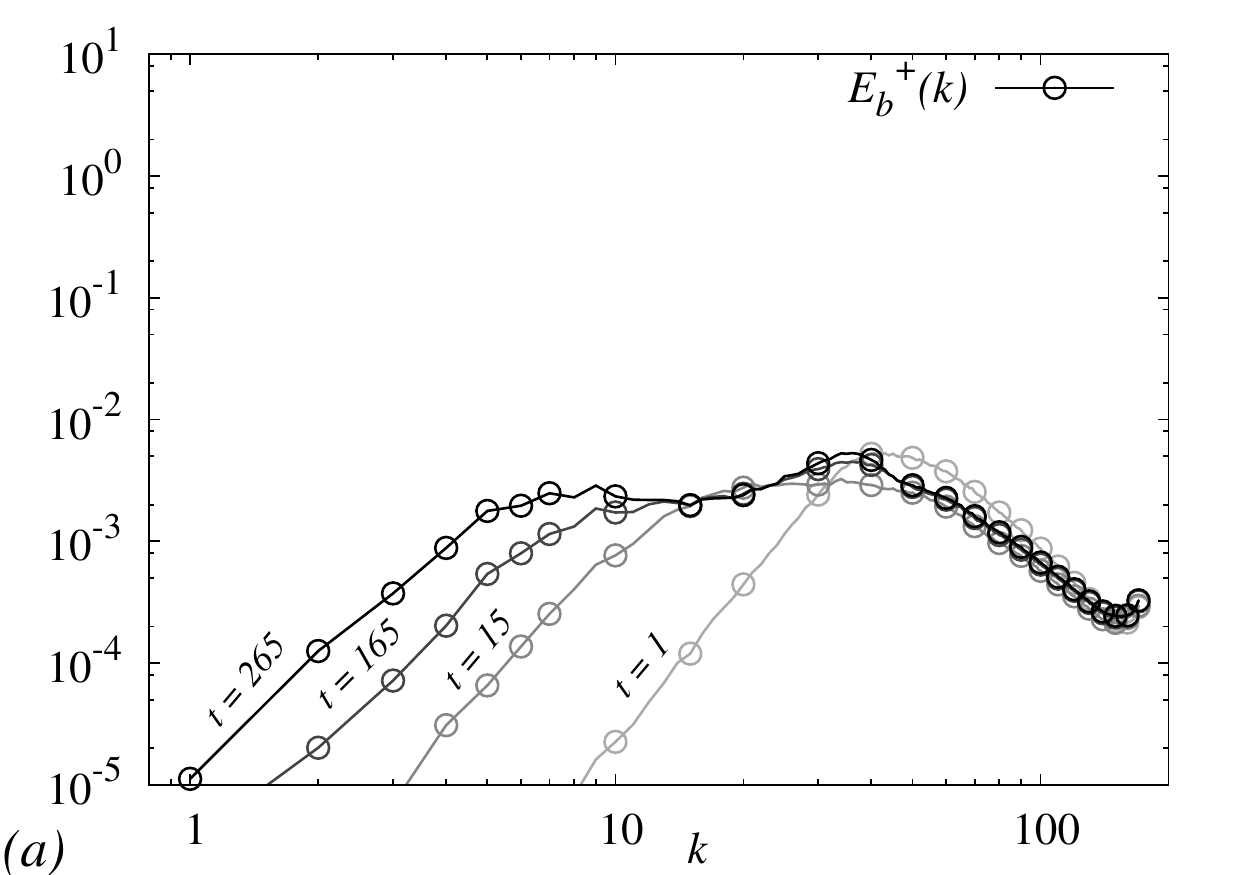} 
\hspace{-0.6cm}
\includegraphics[scale=0.5]{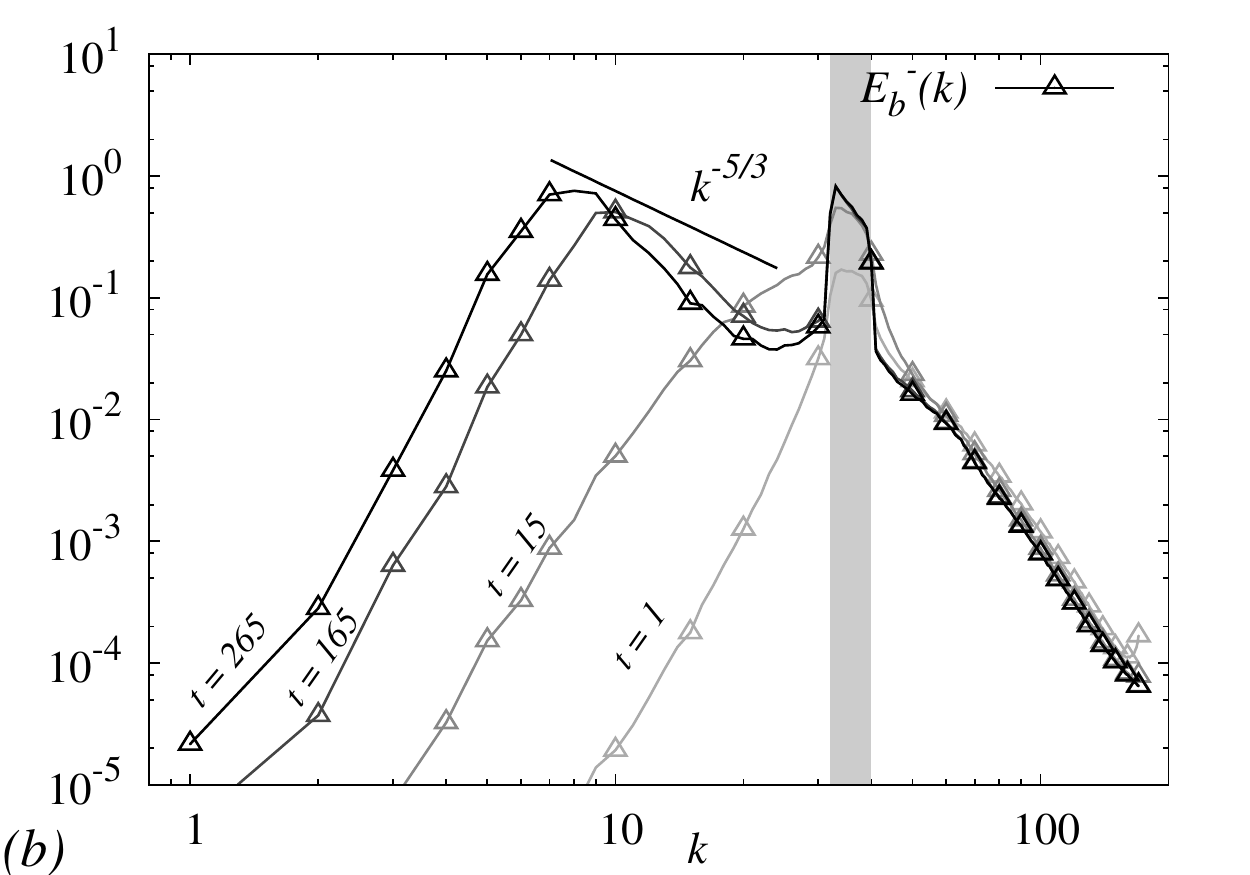} 
\hspace{-0.6cm}
\includegraphics[scale=0.5]{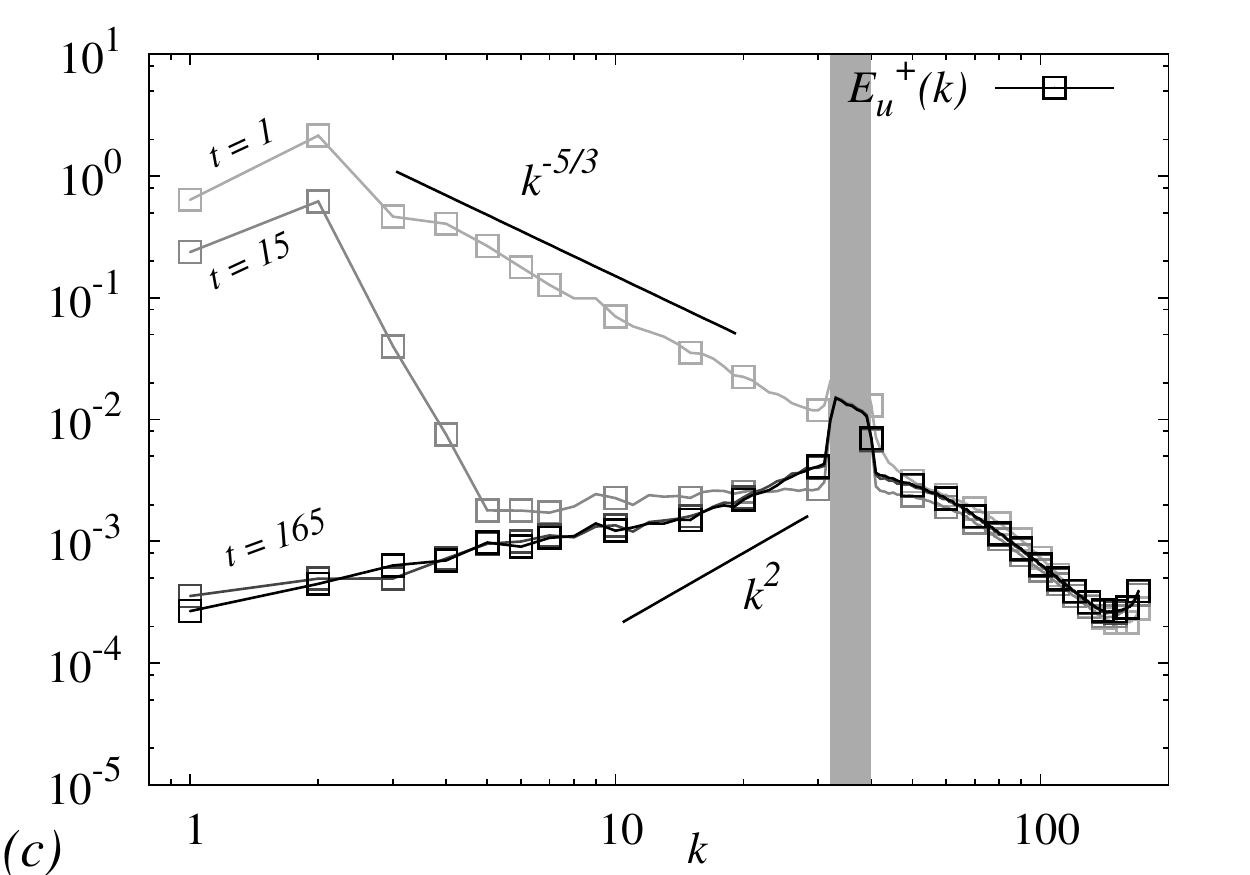} 
\caption{
Run R4-IC.
Panels (a) and (b) show a log-log plot of the magnetic energy spectra $E^+_b(k)$ and $E^-_b(k)$ against $k$ at different times. 
Panel (c) is the same as (a) and (b), but for the 
kinetic energy spectrum $E_u(k) = E^+_u(k)$.  
}
\label{fig:r6ic-spectra}
\end{figure}

\begin{figure}[h]
\center
\includegraphics[scale=0.285]{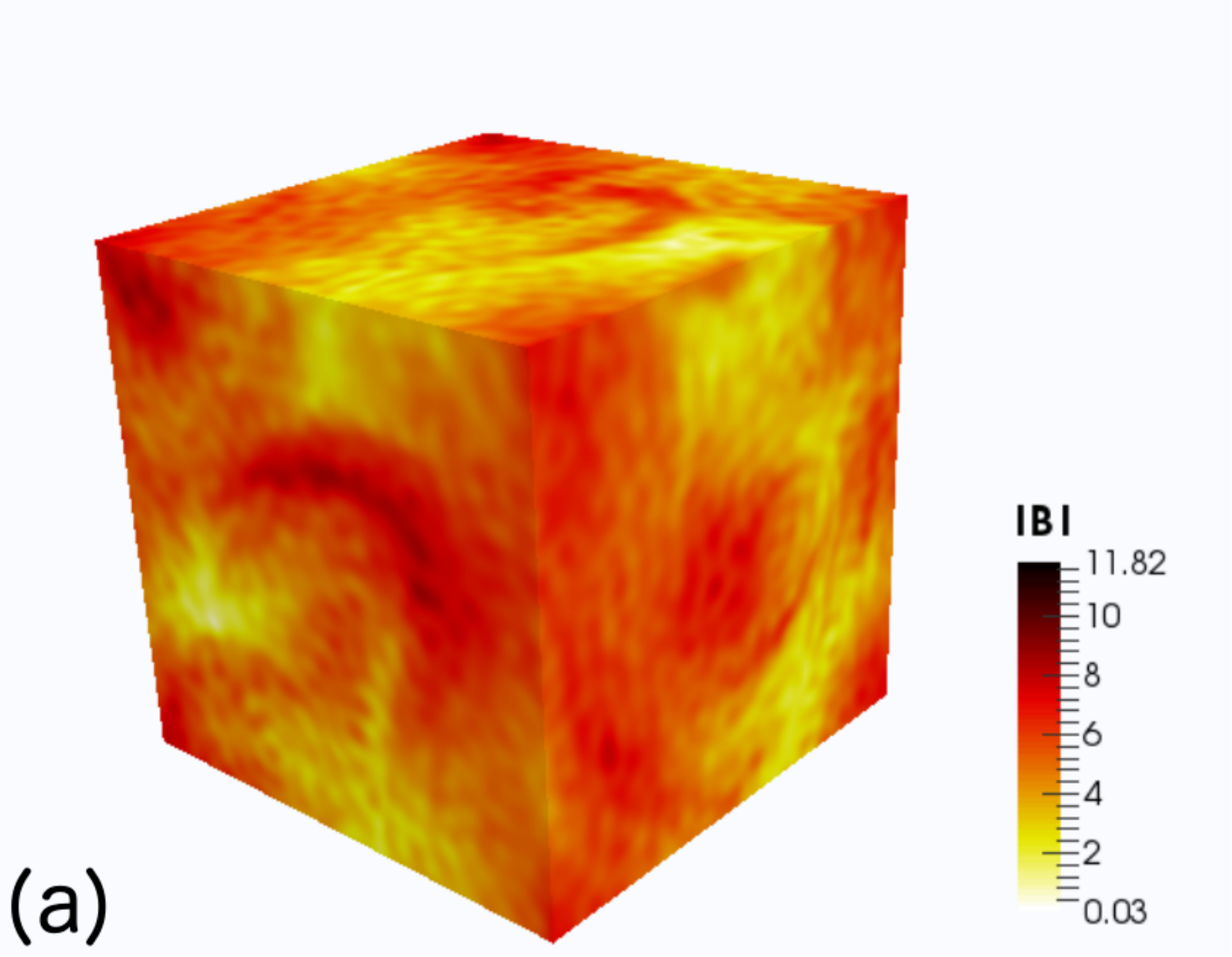} 
\hspace{-0.4cm}
\includegraphics[scale=0.285]{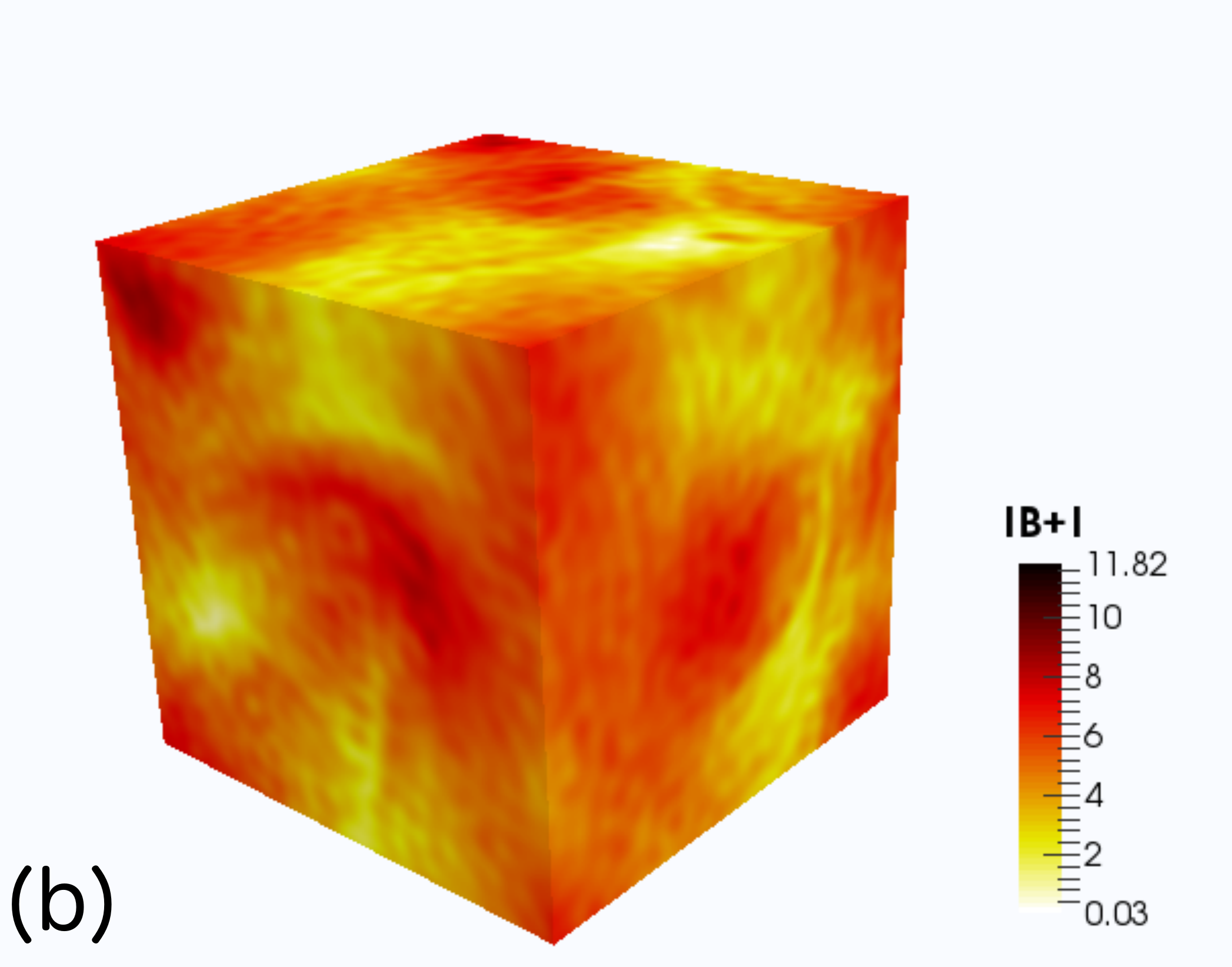} 
\hspace{-0.4cm}
\includegraphics[scale=0.285]{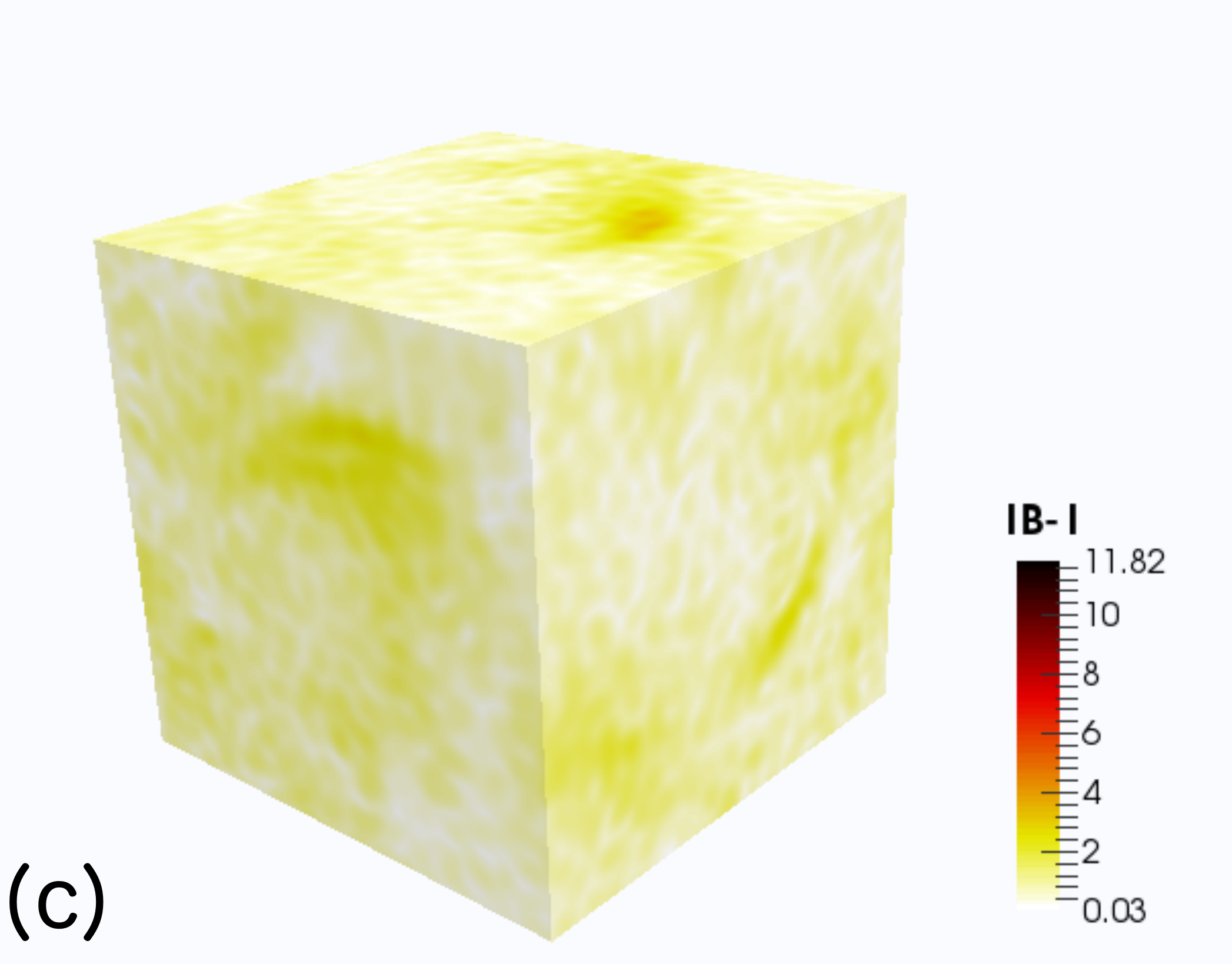} 
\caption{ Visualizations of the magnetic field magnitude for run R3-IC at $t/T = 7.88$, which
corresponds to the latest snapshot in time shown in Figure~\ref{fig:r5ic-spectra}. Panel (a) shows
the magnitude of the full magnetic field $|\bb|$, panel (b) shows the positively helical
component $|\bb^+|$ and panel (c) the negatively helical component $|\bb^-|$.   
A sharp Fourier filter has been applied to all fields to keep only 
modes at wavevectors $\bk$ with $|\bk| \leqslant 20$ in order to remove the 
small-scale contribution from the forcing. 
}
\label{fig:r5ic-visu}
\end{figure}

We now focus on the evolution of a strong magnetic field in a flow where the
velocity is constrained to have only positive helical modes, such that we
expect to see differences in the evolution of positively and negatively helical
magnetic modes due to the asymmetry of the advecting flow.  Again we consider
two cases which only differ in the helical content of the electromagnetic
force, which allows us to examine the effect of kinetic helicity on the
dynamics of the inverse magnetic helicity cascade.  In case R3-IC, the
base flow,  $\vec{u}=\vec{u}^+$ , and the electromagnetic force
$\vec{f}_b=\vec{f}_b^+$, have the same sign of helicity, while in case 
R4-IC the force $\vec{f}_b=\vec{f}_b^-$ is of opposite helicity compared to
the flow.  The time evolution of the helical magnetic energy spectra is shown
in Figure~\ref{fig:r5ic-spectra}(a-b) for case R3-IC and in
Figure~\ref{fig:r6ic-spectra}(a-b) for case R4-IC.  We observe a pronounced
large-scale magnetic field growth in the forced helical sector in both cases,
as expected from the theoretical results on linear instabilities of the triadic
dynamics summarized by statement (v) in Section~\ref{sec:selfinteraction}.  The
large-scale growth of $E_b^+(k)$ in case R3-IC and $E_b^-(k)$ in case
R4-IC are associated with the inverse cascade of magnetic helicity.
Visualizations of the magnitude of the full magnetic field $\bb$ and of the helical components $\bb^+$ and $\bb^-$ for case R3-IC 
are shown in Figure~\ref{fig:r5ic-visu}, where the formation of large-scale
positively helical structures is clearly visible.
The visualized data have been filtered with a sharp Fourier 
filter to keep only modes at wavevectors $\bk$ with $|\bk| \leqslant 20$ 
in order to remove the small-scale contribution from the 
forcing. 
Nevertheless, the inverse cascade is more efficient for R3-IC compared
to R4-IC. Indeed as can be seen from  Figure~\ref{fig:r5ic-spectra}(a),
$E_b^+(k)$ grows more efficiently than $E_b^-(k)$ in 
Figure~\ref{fig:r6ic-spectra}(b). 
This confirms the prediction from the stability
analysis concerning the inverse cascade of magnetic helicity summarized in
statement (vi) in Section~\ref{sec:selfinteraction}. 
This shows that the kinetic helicity has a profound effect on the efficiency
of nonlinear interactions distributing magnetic helicity across the scales: 
the inverse cascade of magnetic helicity in a helical flow is more
efficient and more nonlocal if kinetic and magnetic helicity are of the same
sign.  \\

\section{Conclusions} \label{sec:conclusions}
We studied the dynamics of helical triad interactions in 
homogeneous MHD turbulence both analytically and numerically. 
We have shown that the helical Fourier decomposition of the full MHD equations
is a key tool to better disentangle different inertial transfer processes in 
the fully coupled dynamics. 
First, we extended the set of helical triad interactions 
\citep{Lessinnes09,Linkmann16} 
to the most general MTI systems, and we clarified in which cases the   
stability analysis of the subset of triadic interactions 
carried out in \citep{Linkmann16} is sufficient to capture all possible linear 
instabilities. 
We further analysed two cases of 
astrophysical interest concerning the emergence of large scale magnetic fields,
i.e. dynamo action and the inverse cascade of magnetic helicity, by extending the results
of  \citet{Linkmann16} to provide qualitatively testable predictions on the helical contents of 
 the resulting magnetic growth at large scales and small scales. 
Subsequently, we carried out two series of suitably designed numerical
experiments in order to test the theoretical results.  In
Sections~\ref{sec:lsdynamo} and  \ref{sec:ssdynamo} we clarified which of the
linear instabilities identified by \citet{Linkmann16} is the leading one and
which global helical signature should be expected for the magnetic field. 
In Section~\ref{sec:selfinteraction} we focused on linear instabilities which can be associated
with the inverse cascade of magnetic helicity: a helical magnetic
equilibrium at a given scale can only be unstable with respect to like-signed
helical magnetic perturbations at larger scales \citep{Linkmann16}. 
 We  found that the level of kinetic helicity affects the growth 
rates  associated with the inverse magnetic helicity cascade: the
inverse cascade of magnetic helicity is more efficient in a helical flow where
magnetic and kinetic helicity are of the same sign. 
We point out that the perturbation problem
considered for strongly magnetised flows was restricted to 
large-scale magnetic perturbations only, and it did not include 
the effect of the Lorentz force on the flow. In principle, 
these two dynamical effects are intimately related. A more 
refined analysis that distinguishes between nonlinear 
dynamo and inverse cascade effects, for instance, therefore 
requires a simultaneous analysis of the Lorentz force,
which in turn requires a distinction between homo- and heterochiral 
MTI systems. 
Similarly, the possible influence of the characteristic scale
of the flow on the local or non-local nature 
of the inverse cascade would also require a study of the general MTI 
system. 
Due to the structure of the chosen equilibria 
our analysis did not consider the effect of non-negligible 
cross-helicity on the evolution of the magnetic field.    
Let us remark that the effects induced by an equilibrium solution with a
non-trivial cross-helicity can also be handled analytically 
as shown by \citet{Linkmann16}. 
\\

\noindent
All theoretical results have been derived from a linear stability 
analysis of the basic triadic
structure of the MHD equations where only three modes interact.  However, in
any physical MHD configuration all modes interact, therefore it is not immediately 
obvious if the theoretical results obtained from the single MTI systems 
correctly predict the behavior of the full dynamics \citep{Moffatt14}. 
Therefore we carried out two series of numerical experiments, where series 
D discussed in Section~\ref{sec:linear} corresponds to dynamo simulations and 
series IC in Section~\ref{sec:nonlinear} to
simulations with an inverse magnetic helicity cascade. The numerical results confirm 
the theoretical predictions presented in the previous paragraph. 
Our dynamo results agree qualitatively with the dynamo simulations by \citet{Brandenburg01}
except for some quantitative difference in the dimensionless growth rates   
that is due to different forcing strategies and scales.

\noindent
It is important to note that the triadic dynamo instabilities analyzed here do not result
from any further modeling assumptions such as scale separation or the
first-order smoothing approximation \citep{Moffatt78,Krause80,Brandenburg05}. 
Instead, they are present in the basic dynamics of the MHD equations restricted to a small number
of degrees of freedom.   Furthermore, the triadic $\alpha$-type dynamo
instabilities found here may not have the same limitations as the classical
$\alpha$-effect of mean-field electrodynamics that originate from the more efficient growth of the
small-scale magnetic field compared to the large-scale magnetic field
\citep{Vainshtein92}.  In the latter case, one considers the
mean-field induction equation of the $\alpha$-dynamo
$
\partial_t \vec{B}_0 = \alpha \nabla \times \vec{B}_0 \ ,
$
where the magnetic field $\vec{B}$ has been decomposed into a large-scale mean
and a small-scale fluctuating part $\vec{B} = \vec{B}_0 + \bb$, and the
coefficient $\alpha$ is given by 
$
\alpha = \frac{1}{3}\Big(-\langle \bu \cdot \vec{\omega} \rangle + \langle \bb \cdot \vec{j} \rangle \Big) \ , 
$
with the angled brackets denoting an appropriate average \citep{Brandenburg05}
and $\vec{j} = \nabla \times \bb$ the current density.
If the growing small-scale magnetic field $\bb$ and the small-scale flow $\vec{u}$ have
like-signed helicities, 
then the coefficient $\alpha$ decreases, thus
quenching the  growth rate of the large-scale magnetic field $\vec{B}_0$. Because
the evolution of the small-scale magnetic field $\bb$ is faster than that of
large-scale magnetic field $\vec{B}_0$, the coefficient $\alpha$ could be
quenched before leading to large-scale magnetic field growth. Therefore
concerns have been raised in the literature about whether the classical
$\alpha$-effect is efficient enough to generate large-scale magnetic fields.
This is especially problematic at high magnetic Reynolds numbers, where it eventually
leads to catastrophic $\alpha$-quenching \citep{Cattaneo96}.  Our analysis is concerned with instabilities of the ideal MHD
equations, and the corresponding growth rates do not depend on the magnetic
Reynolds number $Rm$. Hence the dynamo instabilities we found are
in principle present even at large $Rm$. 
A process similar to $\alpha$-quenching and/or 
saturation can also be studied within our approach, but it may not be catastrophic because the 
the growth rates are independent of $Rm$. The small-scale instabilities we 
found in Section~\ref{sec:ssdynamo} preferentially lead to a growing small-scale 
magnetic field with the same sign of helicity as the flow. Eventually, 
the small-scale magnetic field will back-react on the flow, and the linear instability 
leading to the $\alpha$-like large-scale dynamo may be removed.       
At this point we cannot be more precise, as a rigorous assessment of dynamo 
quenching is outside the scope of the stability analysis
carried out here. This requires mixed equilibria, while only purely mechanical or electromagnetic equilibria
were analyzed here. However, we point out that dynamo quenching and the effect of the 
Lorentz force can also be assessed by a similar kind of stability analysis. 
This analysis, which requires a different set of equilibria and is technically 
more complex, is currently in progress and will be reported elsewhere. \\

\noindent
The $\alpha$-effect has further limitations due to its intrinsic scale
separation.  \citet{Boldyrev05} showed by considering the Kazantsev model
\citep{Kazantsev68} that fast-growing eigenmodes exist at all scales, which are
not included in the $\alpha$-dynamo due to the required scale separation.
Although the Kazantsev model assumes the velocity field to have Gaussian
statistics and is as such is not applicable to turbulence, the important point
is that a correct description of dynamo action should involve all scales. We
point out that scale separation is not necessary for the derivation of the
triad-by-triad dynamo instabilities. 
Nevertheless,
information about local and nonlocal dynamics can be obtained by varying the
shape of the wavevector triad, and we find that strongly nonlocal triads lead
mostly to $\alpha$-type dynamo action. 
A recent numerical investigation into 
the efficiency of the kinematic dynamo depending on the energy injection scale
showed that intermediate-scale forcing results in the most efficient dynamo \citep{Sadek16}.
This may be qualitatively interpreted by noting that the triadic dynamo growth
rates depend on the geometry of the triad, i.e., the locality and nonlocality
of the triadic interactions. The growth rate first increases with decreasing
equilibrium (or forcing) scale, but this trend is reversed once the scale
separation becomes very large, which suggests an intermediate
range of forcing scales where a triadic dynamo may indeed be most efficient.
\\

\noindent
Beyond the confirmation of the theoretical results further interesting
observations can be made from the numerical work.  
First, the instability of a
mechanical helical equilibrium associated with large-scale kinematic dynamo
action appears to persist in the nonlinear regime, as shown in Figures~\ref{fig:dynamo} and \ref{fig:upbpbm} in 
Section~\ref{sec:linear}.  This suggests that a very
strong magnetic field is necessary in order to quench the dynamo. Second, as shown in 
Figures~\ref{fig:r2ic-spectra}(c) and \ref{fig:r3ic-spectra}(d) in Section~\ref{sec:nonlinear}, the transfer of
magnetic to kinetic energy due to the feedback of the Lorentz force on the flow at the small scales
is sensitive to the sign of magnetic helicity: The velocity field modes with
the same sign of helicity as the magnetic field increase in intensity. 
On the theoretical side, we found an important difference between the 
large-scale magnetic field growth
due to dynamo instabilities compared to instabilities of the inverse cascade
type, which occur due to nonlinear self-interaction in strongly magnetized
flows.  Dynamo action produced large-scale magnetic fields of opposite helicity
compared to the small-scale flow while the inverse cascade of magnetic helicity
was most efficient in the helical magnetic field sector with the same sign of
helicity as the flow. This may lead to transitional behavior with increasing
small-scale magnetic field strength, eventually changing the helical signature
of the large-scale magnetic field.  Furthermore, it suggests the possible existence of a
dynamo quenching mechanism at the basic triad level. \\ 

\section*{Acknowledgements}
\noindent We thank V.~Dallas for helpful discussions. 
The research leading to these results has received funding from the European
Union's Seventh Framework Programme (FP7/2007-2013) under grant agreement No.
339032. M.L.~and M.M.~acknowledges support from the Scottish Universities Physics Alliance
and the UK Engineering and Physical Sciences Research Council (EP/K503034/1 and EP/M506515/1). 
A.B.~is funded by a STFC Consolidated Grant. 
The theoretical work was planned and carried out by M.L., while
G.S.~designed and carried out all numerical simulations and 
performed the data analysis.
L.B.~contributed to the theoretical work and designed the numerical
simulations. Everyone contributed to the interpretation of the numerical
results. The paper was written by M.L.~and L.B.
\appendix
\section{Coupling coefficients}
\label{app:stability}
The results in Sections~\ref{sec:lsdynamo}-\ref{sec:selfinteraction} were derived by  
analyzing the structure of the coupling factors 
$g^{IN}$, $g^{LF}$, $g^{M1}$ and $g^{M2}$, which quantify the strength
and locality of a given combination of helical modes. 
We begin by stating the definition of a generic 
coupling factor associated to a given triadic interaction 
\begin{align}
\label{eqapp:gfactor}
g^X_{s_k s_p s_q} &= -\frac{1}{2} {\hsk} \cdot \left( {\hsp} \times {\hsp} \right)
         = s_k s_p s_q e^{i\beta}\frac{Q}{4kpq}(s_kk +s_pp +s_qq) \ , 
\end{align}
where $Q^2=2(k^2p^2+p^2q^2+q^2k^2)-k^4-p^4-q^4 \geqslant 0$ depends on the
shape of the triad and $\beta = \beta(s_k,s_p,s_q,k,p,q)$ 
is a real number determined by the orientation of the triad \citep{Waleffe92}
and the superscript $X$ stands for any of the identifiers 
$IN$, $LF$, $M1$ and $M2$.
As can be seen, the coupling factor only depends on the geometry of the 
wavevector triad and not on the type of interaction.
This immediately implies that
the product of the coupling factors that appear in the
second-order evolution equations of the perturbations of a helical mechanical or magnetic 
equilibrium in Section~\ref{sec:summary-theory} are always positive, 
as only combinations with the same helical content occur
\begin{align}
\label{eqapp:gsquare}
g^{X1}_{s_ks_p s_q}g^{X2*}_{s_ks_ps_q} &= \left(s_k s_p s_q e^{i\beta}\frac{Q}{4kpq}(s_kk +s_pp +s_qq)\right)
                                      \left(s_q s_k s_p e^{i\beta}\frac{Q}{4kpq}(s_qq +s_kk +s_pp)\right)^* \nonumber \\
                                   &= \left(\frac{Q}{4kpq}\right)^2(s_kk +s_pp +s_qq)^2
                                    = |g^{X1}_{s_k s_p s_q}|^2 =|g^{X2}_{s_k s_p s_q}|^2 \ ,
\end{align}
where $X1$ and $X2$ label the different interactions, and we note 
that an inertial coupling factor $g^{IN}$ never couples to any of the
others.  
Equation \eqref{eqapp:gsquare} holds for any permutation of $s_k, s_p$ and $s_q$. That is, the product of
two coupling factors in a given interaction always equals the modulus square of one of 
the factors. 
The second step in the analysis required an ordering of the product of coupling factors.
Since the term $(Q/4kpq)^2$ in eq.~\eqref{eqapp:gsquare} is independent of 
the combination of helical modes, the relative ordering of the coupling coefficients 
is determined by the term $(s_kk +s_pp +s_qq)^2$ in eq.~\eqref{eqapp:gsquare}, 
i.e., it depends on the helicities and the ordering of wavenumbers in a given triad. 
In this paper we chose without loss of generality the wavenumber ordering $k\leqslant q \leqslant p$ and obtain
\be
(k +p +q)^2 \geqslant (-k +p +q)^2 \geqslant (k +p -q)^2 \geqslant (k -p +q)^2 \ ,
\ee
which results in
\be
\label{eqapp:gorder}
  |g^X_{+++}|  \geqslant |g^X_{-++}| \geqslant |g^X_{++-}| \geqslant  |g^X_{+-+}| \ ,
\ee
where the subscripts in this equation correspond to the different possible helicity combinations
for $g^{X}_{s_ks_p s_q}$, that is, $s_k$ is always the left subscript, $s_p$ the middle subscript and 
$s_q$ the right subscript. 
The ordering is invariant under reflections, i.e., the same ordering holds 
if $+$ and $-$ are interchanged.
Finally, for the two large-scale dynamo classes D2 and D3 
we obtain for the ratio of the prefactors in   
eqs.~\eqref{eq:bpluskin2-ls} and \eqref{eq:bminuskin1-ls} 
\be
\label{eqapp:gorder-ls}
\frac{g^{M1}_{++-}g^{M2*}_{++-}}
{g^{M1}_{-++}g^{M2*}_{-++}}
 = \frac{|g^{M1}_{++-}|^2}{|g^{M1}_{-++}|^2}= \frac{|k+p_0-q|^2}{|-k+p_0+q|^2} < 1 \ . 
\ee 
Similarly, for the small-scale dynamo we obtain for we obtain for the ratio of the prefactors in
eqs.~\eqref{eq:bpluskin2-ss} and \eqref{eq:bminuskin1-ss} 
\be
\label{eqapp:gorder-ss}
\frac{g^{M1}_{++-}g^{M2*}_{++-}}
{g^{M1}_{+-+}g^{M2*}_{+-+}}
 = \frac{|g^{M1}_{++-}|^2}{|g^{M1}_{+-+}|^2}= \frac{|k_0+p-q|^2}{|k_0-p+q|^2} > 1 \ .
\ee 
Finally, for the ratio of the prefactors in the inverse cascade processes in 
eqs.~\eqref{eq:bplusSI1} and \eqref{eq:bplusSI2} we obtain
\be
\label{eqapp:gorder-ic}
\frac{g^{M1}_{+++}g^{LF*}_{+++}}
{g^{M1}_{++-}g^{LF*}_{++-}}
= \frac{|g^{M1}_{+++}|^2}{|g^{M1}_{++-}|^2} = \frac{|k+p_0+q|^2}{|k+p_0-q|^2} > 1 \ . 
\ee

\bibliography{apj_refs}

\end{document}